\documentclass[12pt,epsf]{article}
\usepackage{verbatim}
\usepackage{anyfontsize}
\usepackage{dsfont}
\usepackage{tikz}
\usepackage{sidecap}
\sidecaptionvpos{figure}{t}
\usepackage{subfigure}
\usepackage{enumitem}
\usepackage[english]{babel}
\usepackage{enumitem}  

\usepackage{amssymb,amsmath}
\usepackage{graphicx, xcolor, varwidth}
\usepackage{setspace}
\usepackage[permil]{overpic}
\usepackage{cite}
\usepackage{mathtools}

\DeclareSymbolFont{matha}{OML}{txmi}{m}{it}
\DeclareMathSymbol{v}{\mathord}{matha}{118}

\usepackage[framemethod=default]{mdframed}
\newmdenv[skipabove=7pt,
skipbelow=7pt,
rightline=false,
leftline=false,
topline=false,
bottomline=false,
backgroundcolor=blue!10,
linecolor=blue,
innerleftmargin=5pt,
innerrightmargin=5pt,
innertopmargin=5pt,
innerbottommargin=5pt,
leftmargin=0cm,
rightmargin=0cm,
linewidth=4pt]{bBox}

\colorlet{darkblue}{blue!70!black}
\colorlet{darkgreen}{green!70!black}

\usepackage[colorlinks=true,urlcolor=darkblue,linktocpage=true,linkcolor=darkblue,citecolor=darkblue]{hyperref}
\hypersetup{ 
colorlinks=true, 
linkcolor=blue, 
citecolor=magenta, 
}

\numberwithin{equation}{section}
\DeclareMathSymbol{v}{\mathord}{matha}{118}
\newcommand{\be}{\begin{equation}}
\newcommand{\ee}{\end{equation}}
\newcommand{\bea}{\begin{eqnarray}}
\newcommand{\eea}{\end{eqnarray}}
\newcommand{\bear}{\begin{eqnarray}}
\newcommand{\eear}{\end{eqnarray}}
\newcommand{\beas}{\begin{eqnarray*}}
\newcommand{\p}{\partial}
\newcommand{\eeas}{\end{eqnarray*}}
\newcommand{\ba}{\begin{array}}
\newcommand{\ea}{\end{array}}

\def\ba#1\ea{\begin{align}#1\end{align}}
\def\bs#1\es{\begin{split}#1\end{split}}

\renewcommand{\r}{\rho}
\newcommand{\br}{\bar{\rho}}

\newcommand{\pd}[2][1]{\ifnum#1=1 \frac{\partial}{\partial {#2}} \else
  \frac{\partial^#1}{\partial {#2}^{#1}}\fi}
\newcommand{\dpd}[2][1]{\ifnum#1=1 \dfrac{\partial}{\partial {#2}} \else
  \frac{\partial^#1}{\partial {#2}^{#1}}\fi}
\newcommand{\td}[2][1]{\ifnum#1=1 \frac{d}{d{#2}} \else
  \frac{d^#1}{d{#2}^{#1}}\fi}

\newcommand{\rr}{\text{\boldmath$\rho$}}

\newcommand{\x}{\xi}

\renewcommand{\(}{\left(}
\renewcommand{\)}{\right)}

\newcommand{\nbox}{{\,\lower0.9pt\vbox{\hrule \hbox{\vrule height 0.2 cm \hskip 0.19 cm \vrule height 0.2 cm}\hrule}\,}}

\newcommand{\ie}{{\it i.e.,}\ }

\def\O{{\cal O}}

\newcommand{\bh}{\bar{h}}

\newcommand{\bz}{\bar{z}}



\begin{document}
\begin{spacing}{1.3}
\begin{titlepage}

\begin{center}
{\Large 
\vspace*{6mm}

Swampland Conditions for Higher Derivative Couplings\\
 from CFT

}

\vspace*{6mm}

Sandipan Kundu

\vspace*{6mm}

\textit{Department of Physics and Astronomy,
\\ Johns Hopkins University,
Baltimore, Maryland, USA\\}

\vspace{6mm}

{\tt \small kundu@jhu.edu}

\vspace*{6mm}
\end{center}

\begin{abstract}
There are effective field theories that cannot be embedded in any UV complete theory. We consider scalar effective field theories, with and without dynamical gravity, in $D$-dimensional anti-de Sitter (AdS) spacetime with large radius and derive precise bounds (analytically) on the coupling constants of higher derivative interactions $\phi^2\Box^k\phi^2$ by only requiring that the dual CFT obeys the standard conformal bootstrap axioms. In particular, we show that all such coupling constants, for even $k\ge 2$, must satisfy positivity, monotonicity, and log-convexity conditions in the absence of dynamical gravity. Inclusion of gravity only affects constraints involving the $\phi^2\Box^2\phi^2$ interaction which now can have a negative coupling constant. Our CFT setup is a Lorentzian four-point correlator in the Regge limit. We also utilize this setup to derive constraints on effective field theories of multiple scalars. We argue that similar analysis should impose nontrivial constraints on the graviton four-point scattering amplitude in AdS.

\end{abstract}

\end{titlepage}
\end{spacing}

\vskip 1cm
\setcounter{tocdepth}{2}  
\tableofcontents

\begin{spacing}{1.3}


\section{Introduction}

By now it is well-known that not all effective field theories (EFTs) can be UV completed. One famous example is the EFT of a massless scalar with higher derivative interaction  
\be\label{intro:adams}
S=\frac{1}{2}\int d^D x \(-(\p\phi)^2+\mu \phi^2\Box^2 \phi^2+\cdots\)\ ,
\ee
which does not admit a UV completion for $\mu<0$ \cite{Adams:2006sv}. Conceptually, this represents a substantial departure from our traditional understanding of Wilsonian EFTs. In fact, this remarkable result led to the proof of the 4D $a$-theorem establishing irreversibility of unitary renormalization group flows between conformal fixed points \cite{Komargodski:2011vj}. More generally this constraint is related to the idea of ``swampland" of EFTs that cannot be obtained as a low energy approximation of a consistent theory of quantum gravity (oftentimes string theory) \cite{Vafa:2005ui,Ooguri:2006in,Brennan:2017rbf,Palti:2019pca,vanBeest:2021lhn,ArkaniHamed:2006dz}.

The constraint on the EFT (\ref{intro:adams}) is not actually an accident, but part of a general feature of low energy EFTs with higher derivative interactions. There is a growing body of literature with similar precise bounds on IR couplings of an EFT from UV consistency \cite{deRham:2017avq,deRham:2017zjm,Chandrasekaran:2018qmx,Zhang:2018shp,Bi:2019phv,Remmen:2019cyz,Remmen:2020vts,Zhang:2020jyn,Yamashita:2020gtt,Fuks:2020ujk,Remmen:2020uze,Caron-Huot:2020cmc,Bellazzini:2020cot,Tolley:2020gtv,Gu:2020ldn,Arkani-Hamed:2020blm,Li:2021cjv}. 
All these bounds have one thing in common that they do not depend on the details of  the UV completion. However, these types of bounds are generally derived under the assumption that the $2\rightarrow 2$ scattering amplitude obeys (i) analyticity (in the usual regime), (ii) partial wave unitarity, (iii) crossing symmetry, and (iv) Regge boundedness conditions even in the UV.  Such S-matrix based arguments can be unsatisfying since some of these assumptions (even though well-motivated) have not yet been rigorously established.\footnote{In recent years, significant progress has been made  both in analytical and numerical approaches to the S-matrix bootstrap \cite{Paulos:2016but,Caron-Huot:2016icg,Caron-Huot:2016owq,Paulos:2017fhb,Cordova:2018uop,Guerrieri:2018uew,Homrich:2019cbt,EliasMiro:2019kyf,Karateev:2019ymz,Correia:2020xtr,Bose:2020shm,Guerrieri:2020bto,Guerrieri:2020kcs,Kaplan:2020ldi,Bose:2020cod,Huang:2020nqy,Hebbar:2020ukp,Sinha:2020win,Tourkine:2021fqh,Haldar:2021rri,He:2021eqn}. }

Another technical challenge of these EFT arguments is to incorporate dynamical gravity mainly because of the graviton pole in the $2\rightarrow 2$ scattering amplitude. Recently, an elegant framework has been introduced in \cite{Caron-Huot:2021rmr} that bypasses this problem by studying scattering amplitudes at finite impact parameter (see \cite{Noumi:2021uuv} for related discussions). Under the same assumptions about  the $2\rightarrow 2$ scattering amplitude, this framework leads to non-trivial and rigorous two sided bounds on coupling constants of higher derivative interactions in $D>4$ dimensions. The analysis necessarily requires that the $2\rightarrow 2$ scattering amplitude ${\mathcal A}(s,t)<|s|^2$ for large $s$ (at fixed $t<0$).  However, it is unclear whether this Regge boundedness condition is valid in the presence of dynamical gravity.\footnote{Note that the Froissart bound \cite{Froissart:1961ux,Martin:1962rt,CHAICHIAN1992151} does not hold without a mass gap in the theory. Hence, the Regge boundedness condition ${\mathcal A}(s,t)<|s|^2$ is subtle whenever there are massless states in the theory, even in the absence of gravity. For example, the same issue persists even for the 4-point scattering amplitude of  the dilaton  that led to the proof of the 4D $a$-theorem in \cite{Komargodski:2011vj}. However, in that case, the Regge boundedness follows from conformal invariance of the UV fixed point \cite{Komargodski:2011xv,Luty:2012ww}. On the other hand, the same argument for the 4-point dilaton amplitude in 6D imposes a weaker condition ${\mathcal A}(s,t)<|s|^3$ \cite{Elvang:2012st,Heckman:2021nwg}.} Nevertheless, these bounds provide compelling evidence in favor of the expectation that all higher derivative interactions must have order one coupling constants in the units of the UV cut-off scale. The main motivation of this paper is to derive similar bounds on EFTs in anti-de Sitter (AdS) spacetime where such loopholes can be avoided.

In this paper, we will address a closely related question: what scalar EFTs in AdS$_D$ cannot be embedded into a UV theory that is dual to a CFT$_{D-1}$ obeying the usual CFT axioms? We will provide a partial answer to this question by leveraging the huge advancement in constraining the space of consistent CFTs from well-established conformal bootstrap axioms (for a review see \cite{Poland:2018epd}). The main logic of our argument parallels recent developments in constraining EFTs in AdS (with or without dynamical gravity) from rigorous analysis in the dual CFT \cite{Hofman:2008ar,Hartman:2015lfa,Afkhami-Jeddi:2016ntf,Paulos:2016fap,Paulos:2016but,Kulaxizi:2017ixa,Costa:2017twz,Afkhami-Jeddi:2017rmx,Paulos:2017fhb,Cordova:2017zej,Meltzer:2017rtf,Afkhami-Jeddi:2018own,Afkhami-Jeddi:2018apj,Homrich:2019cbt,Kaplan:2019soo,Haldar:2019prg,Conlon:2020wmc,Komatsu:2020sag,Kundu:2020bdn,Hijano:2020szl,Alday:2021odx}. For example, the sign constraint on the $\phi^2\Box^2 \phi^2$ coupling in the EFT (\ref{intro:adams})   can be  derived in AdS  from the conformal bootstrap \cite{Hartman:2015lfa}. 

The main advantage of our AdS argument is that the bounds follow directly from the conformal bootstrap axioms which are, both conceptually and technically, well-understood, at least at the level of four-point correlators. We will derive the $\phi^2\Box^2 \phi^2$ constraint (with and without dynamical gravity) as a special case of an infinite set of similar constraints on higher derivative interactions of the form $\phi^2 \Box^k \phi^2$ from a simple CFT setup.

We consider a scalar EFT in AdS$_D$ with an effective action\footnote{We are ignoring $\phi^3$, $\phi^4$, $\phi^2\Box \phi^2$, and all other higher derivative interactions that cannot be written as $\phi^2\Box^k \phi^2$  since these interactions, as well as presence of other low spin $(J\le 1)$ fields, will not affect the final bounds. However, these interactions can sometimes create obstruction to a flat space limit, especially at low spacetime dimensions. We will discuss this in section \ref{sec:flat}.} 
\begin{align}\label{intro:EFT}
S&=S_{EH}\\
&+\frac{1}{2} \int d^{D} x \sqrt{-g} \left(- g^{\mu\nu}\nabla_\mu \phi \nabla_\nu\phi -m^2 \phi^2+\mu \sum_{k=2,3,4,\cdots} \left(\frac{\lambda_k}{n_k(\Delta) M^{2(k-2)}}\right)\phi^2 \Box^k \phi^2\right)+\cdots \ , \nonumber
\end{align}
where, $S_{EH}$ is the Einstein-Hilbert action with  a negative cosmological constant and $M$ is the mass-scale of new physics. We allow for the possibility that the scalar field has a mass $0\le m^2\ll M^2$. First, let us explain our convention. We have defined  a positive coupling constant $\mu\ge 0$ which has the dimension $1/M^{D}$. The coupling constants $\lambda_k$ are dimensionless and normalized by introducing (dimensionless and known) $\O(1)$ positive numerical factors $n_k(\Delta)$.\footnote{The numerical factor $n_k(\Delta)>0$ is defined in (\ref{eq:gamma}) as ratios of $\Gamma$-functions. Note that $n_2(\Delta)=n_4(\Delta)=1$. Moreover, in the large AdS radius limit with finite and non-zero $m$, this factor $n_k(\Delta)= 1$ for all finite $k$.  For large AdS radius, the numerical factor $n_k(\Delta)$ is non-trivial (\ie $n_k(\Delta)\neq 1$) only in the massless limit (or for $k\gg m R_{\rm AdS}$).} The choice of this particular normalization makes the final bounds rather simple. Moreover, without loss of generality, we will assume that $\lambda_2$ is order one, however, to begin with we do not assume that the other coupling constants are order one as well. Our goal is to derive necessary conditions (analytically) for the tree level EFT (\ref{intro:EFT}) to have a UV completion.\footnote{We assume that the EFT (\ref{intro:EFT}) is weakly coupled such that $G_N$ (if non-zero) and $\mu$ are small and of the same order in the units of the cut-off scale $M$.} In particular, we will  impose precise constraints on $\lambda_k$ coupling constants, irrespective of the details of the UV physics, from the requirement that the dual CFT satisfies the bootstrap axioms. 

There are non-trivial constraints on the EFT (\ref{intro:EFT}) even when gravity is non-dynamical $(G_N=0)$. So, first we focus on this simpler case. The EFT (\ref{intro:EFT}) in AdS with large radius $R_{\rm AdS}M\gg 1$ enjoys a dual CFT description  in $D-1$ spacetime dimensions. Specifically, it was shown by  \cite{Heemskerk:2009pn} and subsequent authors \cite{Heemskerk:2010ty,Fitzpatrick:2010zm,Penedones:2010ue,ElShowk:2011ag,Fitzpatrick:2011ia,Fitzpatrick:2011hu,Fitzpatrick:2011dm,Fitzpatrick:2012cg,Goncalves:2014rfa,Alday:2014tsa,Hijano:2015zsa,Aharony:2016dwx}, that the scalar EFTs  in AdS$_D$ of the form (\ref{intro:EFT}) are in one-to-one correspondence with perturbative solutions of crossing symmetry in CFT$_{D-1}$. This interacting dual CFT has a scalar primary operator $\O$ which is dual to the AdS field $\phi$ with dimension $\Delta$ given by $m^2 R_{\rm AdS}^2=\Delta(\Delta-D+1)$. Since there is no dynamical gravity, the stress tensor of the dual CFT must decouple from the low energy spectrum. This implies that we are in the limit of large central charge $c_T\rightarrow \infty$ with $R_{\rm AdS}M\equiv \Delta_{\rm gap}$ fixed (but large).\footnote{The central charge $c_T$ is the overall coefficient of the CFT stress tensor two-point function.} Of course, the dual CFT should be thought of as an ``effective" CFT which is embedded in some bigger CFT satisfying the usual CFT axioms. We utilize this dual description to study CFT Regge correlators associated with the EFT (\ref{intro:EFT}). At the leading order in $\mu$, these CFT Regge correlators grow in a very specific way within the regime of validity of the EFT (\ref{intro:EFT}). In fact, this type of Regge growth is known to be highly constrained by the argument of \cite{Kundu:2020gkz} (see section 6). In this paper, we revisit these bounds on the Regge growth of CFT correlators and show that they impose precise constraints on the EFT (\ref{intro:EFT}). In particular, in the limit of large $R_{\rm AdS}M$ we conclude that the coupling constants $\lambda_k$, irrespective of the rest of the theory,   must obey the following conditions for the EFT (\ref{intro:EFT}) (with $\mu\ge 0$)  to be embedded into a UV theory that is dual to a CFT obeying the CFT axioms:\footnote{All bounds obtained in this paper are valid in spacetime dimensions $D\ge 4$. We also expect that  our analysis is valid even for $D=3$ as long as $0\le m^2\ll M^2$ and the field $\phi$ has shift symmetry or $\mathbb{Z}_2$ symmetry.} 
\begin{itemize}
\item {{\bf Positivity}-- For all even $k\ge 2$
\be\label{intro:condition1}
 \lambda_k >0\ .
\ee}
\item  {{\bf Monotonicity}-- $\lambda_k$ as a function of even $k$ decreases monotonically\footnote{It should be noted that this condition, unlike other two conditions, depends on our exact definition of the cut-off scale $M$. For an arbitrary definition of $M$, there must always exist a rescaling $M\rightarrow XM$ with order one $X$ which makes the EFT consistent with the condition (\ref{intro:condition2}). } 
\be\label{intro:condition2}
\lambda_{k+2}\le \lambda_k
\ee
 for all even $k\ge 2$. }
\item  {{\bf Log-Convexity}-- $\lambda_k$, for even $k$, satisfies a global log-convexity condition and hence for any even $k_1$, $k_2$, and $k_3$  
\be\label{intro:condition3}
\frac{1}{k_2-k_1}\ln \frac{\lambda_{k_1}}{\lambda_{k_2}}\ge \frac{1}{k_3-k_1}\ln \frac{\lambda_{k_1}}{\lambda_{k_3}}\ , \qquad k_3>k_2>k_1\ge 2\ .
\ee
}
\end{itemize}
We emphasize that these constraints follow directly from analyticity, positivity, and crossing symmetry of CFT four-point correlators -- properties that are contained in the conformal bootstrap axioms. The above conditions, among other things, imply that all higher derivative interactions $\phi^2\Box^k \phi^2$ with even $k$ must have order one coupling constants in the units of the UV cut-off scale $M$. However, we believe   that (\ref{intro:condition1})-(\ref{intro:condition3}) are necessary conditions but they are far from being sufficient. For example, it is expected that similar bounds exist even for odd $k$. Whereas, our setup does not impose any restriction on the odd $\lambda_k$ couplings other than all $\lambda_k$ couplings  in AdS, even or odd, with $k\ge 3$ must vanish when $\lambda_2=0$.\footnote{This can  be alternatively stated as $\lambda_2> 0$ with $\mu\ge 0$. This condition is more subtle in the exact flat space limit, as we explain later. In flat space $\lambda_2=0$ does not necessarily requires  $\lambda_k=0$ for odd $k$. For an example see \cite{Heckman:2021nwg}. }

\begin{figure}[h]
\centering
\includegraphics[scale=0.4]{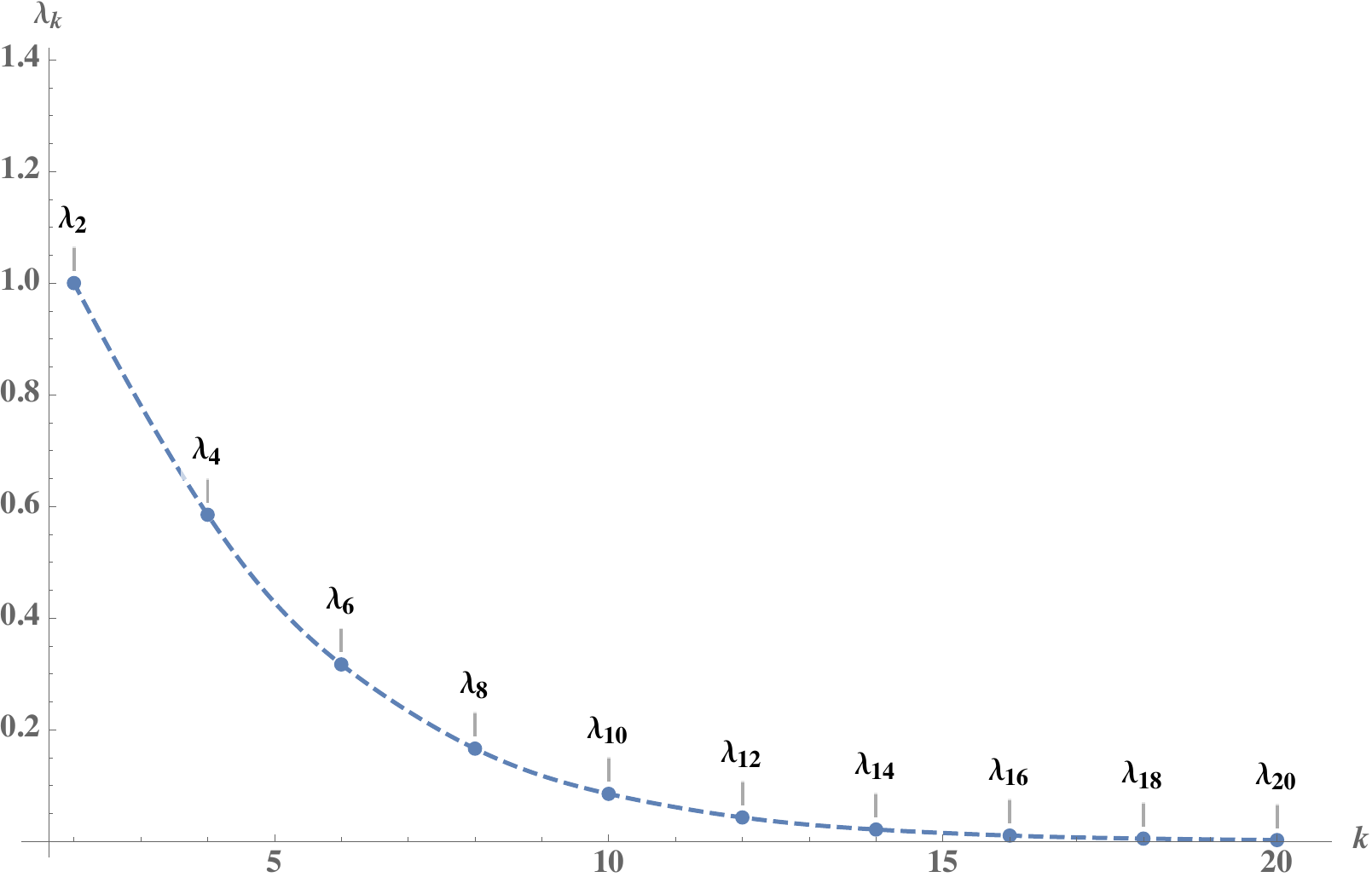}
\caption{ \label{fig:intro} \small The coupling constants $\lambda_k$ for even $k\ge 2$, without dynamical gravity, must obey the conditions (\ref{intro:condition1})-(\ref{intro:condition3}). There is always some choice of the scale $M$ (and $\mu$) for which the coupling constants have this generic structure. When gravity is included, only bounds on $\lambda_2$ become weaker.}
\end{figure}

Of course, the next key step is to include dynamical gravity ($G_N\neq 0$). In our setup, the inclusion of gravity is a rather trivial generalization. Now the central charge of the dual CFT is large $c_T\gg \Delta_{\rm gap}\gg 1$ but finite. The bulk graviton contributes only to the leading growing term of the Regge correlator of the dual CFT. This immediately implies that the constraints (\ref{intro:condition1})-(\ref{intro:condition3}) remain unchanged for all even $k\ge 4$. On the other hand, all conditions involving $\lambda_2$ now receive corrections from gravity. For example, gravity allows for the $\phi^2\Box^2\phi^2$ interaction to have a negative coupling constant 
\be\label{intro:bound}
\mu \lambda_2 > -\pi N_D(\Delta) G_N R_{\rm AdS}^2 \ ,
\ee
where $N_D(\Delta)$ is a positive order one numerical factor given in appendix \ref{App:N}. It is however unclear how to extract a precise bound from (\ref{intro:bound}) in the flat space limit. The AdS bound (\ref{intro:bound}), as we will explain,  suggests that in the flat space limit $\mu \lambda_2 M^D > -\varepsilon$, where $\varepsilon$ is some small positive number.  This is certainly consistent with the results of \cite{Caron-Huot:2021rmr}, however, we do not have a precise definition of $\varepsilon$. Nevertheless, this raises an interesting conceptual question of whether, and in what sense, the 4D $a$-theorem is valid in the presence of dynamical gravity.

Finally, we will generalize our analysis for EFTs of multiple scalar fields in AdS. The main motivation for this generalization is to demonstrate that there are other tools available when we go beyond a single scalar field. For example, the same CFT consistency conditions of \cite{Kundu:2020gkz} now also impose two-sided bounds on odd $k\ge3$ higher derivative interactions involving multiple fields (see \cite{Li:2021cjv} for similar bounds on flat space multi-field EFTs). Furthermore, for multiple fields there are interference effects that are also constrained by the CFT axioms leading to an infinite set of non-linear bounds among various higher derivative coupling constants.\footnote{Note that the same interference effects were utilized in \cite{Kundu:2020bdn} to derive non-linear bounds on the dilaton-axion effective action associated with 4D RG flows with global symmetry breaking. } These additional tools will certainly be useful for bounding the four-graviton scattering amplitude in AdS by using the dual CFT description.

Our CFT setup, from the dual gravity perspective, is probing local high energy scattering deep in the bulk. Since the local high energy scattering is insensitive to the spacetime curvature, on physical grounds we expect that AdS bounds obtained in this paper persist even in the flat space limit (other than the caveat mentioned after equation (\ref{intro:bound})). Indeed, we checked that weakly coupled string amplitudes satisfy all the conditions derived in this paper. However, there is one obvious but important issue that we must address. Any strict AdS inequality $A>0$ must be regarded as $A\ge 0$ in the flat space limit since the $A=0$ case can no longer  be ruled out due to finite curvature effects. 

An important feature of our AdS bounds is that they differ significantly for massive and massless scalars, especially when we take the flat space limit. In particular, when we take the  large $R_{\rm AdS}$ limit (with fixed $m$), our bounds agree completely with the ones obtained from the flat space dispersive sum-rules under the same set of assumptions about the four-point amplitude as mentioned in the beginning.\footnote{This is true even when we take $m\rightarrow 0$ after taking the large radius limit. See section \ref{sec:mass}.} On the other hand, when we take $m=0$ first and then  $R_{\rm AdS}\rightarrow \infty$, two sets of bounds differ significantly. This perhaps indicates that the Regge boundedness condition of the flat space amplitude can break down in the presence of massless states (see section \ref{sec:mass}). 

At this stage, one may wish to compare our bounds with the flat space bounds of \cite{Caron-Huot:2021rmr}. Indeed, there is some overlap between these two sets of bounds. Of course, from our CFT setup, it is not immediately clear how to obtain any constraints on odd $k$ coupling constants for a single scalar field. Such bounds will perhaps require a more sophisticated CFT analysis. Nevertheless, we observe that our constraints, in the flat space limit, are consistent with the bounds of \cite{Caron-Huot:2021rmr}. 

We have analyzed the EFT (\ref{intro:EFT}) at tree level. We note that the CFT consistency conditions  of \cite{Kundu:2020gkz} that we have utilized in this paper apply even when we include corrections from EFT loops. In fact, the CFT consistency conditions  of \cite{Kundu:2020gkz} (see section 6) hold even for arbitrary external CFT operators with or without spins (and not necessarily local or primary). So, it is  a straightforward exercise to extend our analysis to derive bounds on the graviton four-point scattering amplitude in AdS by studying Regge correlators of the stress tensor operator in the dual CFT. It would be interesting to compare such bounds with similar classical bounds of \cite{Chowdhury:2019kaq} from ``Classical Regge Growth" (CRG) conjecture and EFT bounds of \cite{Bern:2021ppb,Guerrieri:2021ivu}  from unitarity and crossing. We will have to leave this question for the future.

The rest of the paper is organized as follows. In section \ref{sec:EFTAdS} we begin by explaining our general setup.  In section \ref{sec:review} we review the bounds of \cite{Kundu:2020gkz} on certain CFT Regge correlators and explain how they follow from the conformal bootstrap axioms.  We use this CFT constraints in section \ref{sec:EFT_AdS} to derive bounds (\ref{intro:condition1})-(\ref{intro:condition3}) on the scalar EFT in AdS without dynamical gravity. Then in section \ref{sec:mass} we discuss some implications of these constraints for massless external scalars in the flat space limit. Section \ref{sec:gravity} studies the consequences of the inclusion of gravity. In section \ref{sec:twofields} we extend our analysis to impose bounds on the EFT of two scalar fields. Section \ref{sec:conclusions} contains our conclusions and final comments. Some additional aspects of our analysis are included in several appendices. In particular, in appendix \ref{app:rindler} we demonstrate how Rindler positivity in CFT follows from unitarity and crossing symmetry.

\section{Scalar EFT in AdS}\label{sec:EFTAdS}
We consider an EFT of a  single massless or massive scalar field in AdS with higher derivative interactions. We start with the following low energy effective action with four-point  interactions  
\begin{align}\label{action}
S&=\frac{1}{16\pi G_N} \int d^{D} x \sqrt{-g}\(R+\frac{(D-1)(D-2)}{R_{\rm AdS}^2}\)\nonumber\\
&+\frac{1}{2} \int d^{D} x \sqrt{-g} \left(- g^{\mu\nu}\nabla_\mu \phi \nabla_\nu\phi -m^2 \phi^2+\alpha_3 \phi^3+\sum_{k=0}^\infty \mu_k\ \phi^2 \Box^k \phi^2\right)+\cdots \ ,
\end{align}
where, $\alpha_3$ and $\mu_k$ are coupling constants.\footnote{Note that at the tree level the $k=1$ term can be removed by using the equation of motion. So, we will ignore the $k=1$ interaction completely. } The AdS radius $R_{\rm AdS}$ is large but finite.  Our goal is to derive constraints on the coefficients $\mu_k$. In the process, the form of the effective action (\ref{intro:EFT}) will emerge automatically. Note that different higher derivative interactions, in general, can be suppressed by different scales. However, we will assume that all interactions are suppressed by some small coupling $0<\mu\ll 1$:
\be\label{weak}
G_N, \alpha_3^2, \mu_k \sim \mu\ .
\ee 
We intend to impose constraint on the weakly coupled effective theory and hence we work in the leading order in $\mu$.\footnote{One can think of $\mu$ as the analog of the string coupling in string theory. Similarly, the cut-off scale $M$ in the effective action (\ref{intro:EFT}) can be regarded as the string scale.} This will be implemented by keeping only tree level processes.

It should be noted that there are higher derivative 4-$\phi$ interactions (with 12 or more derivatives) that cannot be written as $\phi^2 \Box^k \phi^2$ even when we apply the equation of motion. However, these other higher-derivative interactions are not bounded by the argument of this paper, provided $\mu_2$ is non-zero. Hence, we ignore these other higher derivative 4-$\phi$ interactions since they will not affect any of the bounds obtained in this paper. We will discuss this again in section \ref{sec:others} in detail.

\subsection{Dual CFT}
We will impose constraints on this action from the consistency of the dual CFT$_d$, where $D=d+1$. The AdS theory (\ref{action}) is dual to an interacting CFT in $d$-dimensions. The bulk field $\phi$ is dual to a scalar primary operator $\O$ with dimension $m^2 R_{\rm AdS}^2=\Delta(\Delta-d)$. The two-point function is completely fixed by conformal invariance\footnote{For a review see appendix \ref{gkpw}.} 
\be\label{eq:2pt}
\langle \O(x_1)\O(x_2)\rangle=\frac{(2\Delta-d)C_\Delta}{x_{12}^{2\Delta}}\ , \qquad C_\Delta=\frac{\ \Gamma[\Delta]}{\pi^{d/2}\Gamma[\Delta-d/2]}\ .
\ee
Of course, the graviton $h_{\mu\nu}$ is dual to the CFT stress tensor $T_{\mu\nu}$. The EFT (\ref{action}) is a well behaved theory at energies below the cut-off scale $M$. Our goal is to impose constraints on the coupling constants by requiring that the EFT is the low energy description of a UV complete theory. Equivalently, in the CFT side we will assume that the dual CFT is well behaved. Next, we discuss exactly what we mean by a well behaved CFT. 

\subsection{CFT Axioms}
We make the assumption that the dual CFT obeys the Euclidean bootstrap axioms. In particular, we only make use of the following three properties:
\begin{enumerate}
\item[(i)]{{\bf OPE Unitarity}-- All OPE coefficients of real operators are real. }
\item[(ii)]{{\bf Crossing Symmetry}-- CFT  four-point correlators are crossing symmetric.}
\item[(iii)]{{\bf Analyticity}-- Lorentzian CFT four-point correlators are analytic in the usual domain (see figure \ref{dt1}).}
\end{enumerate}
These CFT properties are well-established and they imply rigorous non-perturbative constraints on certain Regge correlators as derived in \cite{Kundu:2020gkz}. We will use these constraints to derive precise bounds on the higher-derivative couplings of the EFT (\ref{action}).

\section{A Review of the Bounds on CFT Regge Correlators}\label{sec:review}
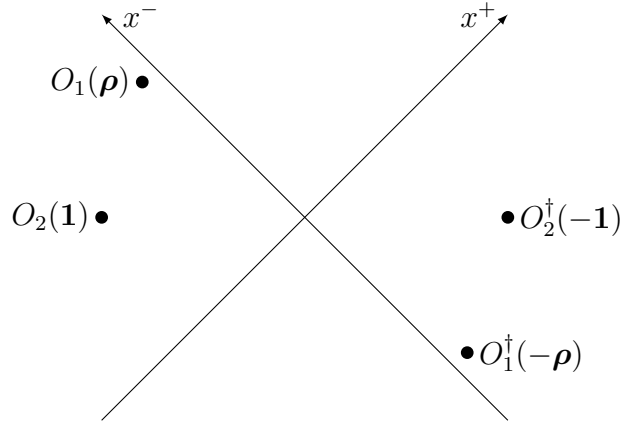
\begin{figure}
\begin{center}
\begin{tikzpicture}[baseline=-3pt, scale=1.80]
\begin{scope}[very thick,shift={(4,0)}]
\coordinate (v1) at (-1.5,-1.5) {};
\coordinate(v2) at (1.5,1.5) {};
\coordinate (v3) at (1.5,-1.5) {};
\coordinate(v4) at (-1.5,1.5) {};

\draw[thin,-latex]  (v1) -- (v2)node[left]{$x^+$};
\draw[thin,-latex]  (v3) -- (v4)node[right]{$\ x^-$};
\draw(-1.5,0)node[left]{ $\ O_2(\mathbf{1})$};
\draw(1.5,0)node[right]{ $ O_2^\dagger(-\mathbf{1})$};
\filldraw[black]  (-1.5,0) circle (1pt);
\filldraw[black]  (1.5,0) circle (1pt);
\coordinate(v5) at (0,0) {};
\def \fac {.6};
\filldraw[black]  (-1.2,1) circle (1 pt);
\filldraw[black]  (1.2,-1) circle (1pt);
\draw(-1.2,1)node[left]{ $ O_1(\rr)$};
\draw(1.2,-1)node[right]{ $ O_1^\dagger(-\rr)$};

\end{scope}
\end{tikzpicture}
\end{center}
\caption{\label{config} \small  Lorentzian four-point correlator (\ref{corr}) where all operators are restricted to a $2$d subspace.}
\end{figure} 

\begin{figure}
\centering
\includegraphics[scale=0.5]{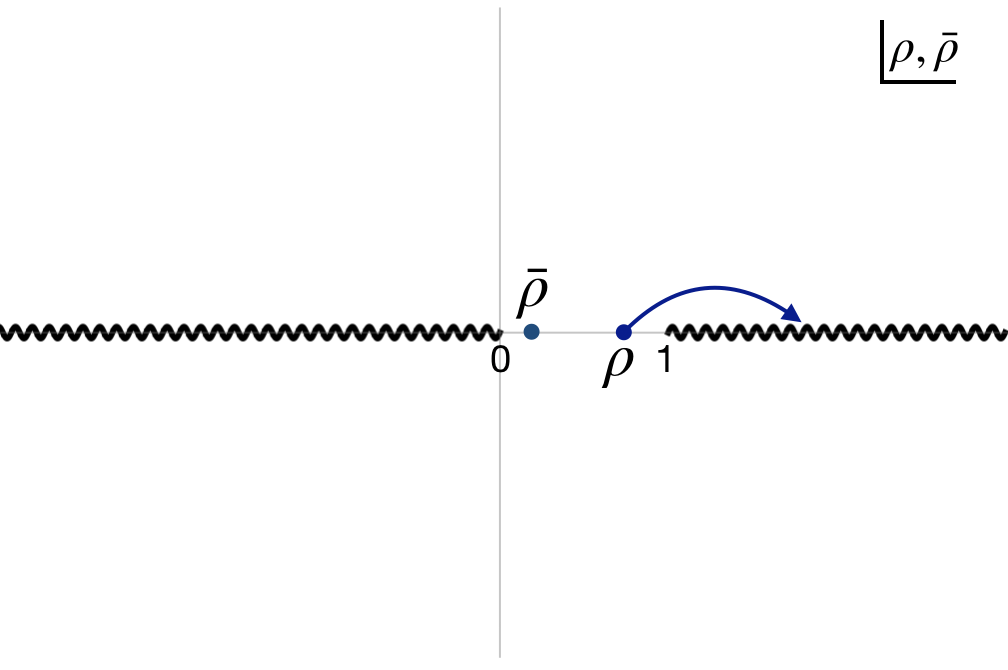}
\caption{ \label{dt1} \small Analytic structure of the correlator (\ref{corr}) – branch cuts appear only when two operators become null separated. The Regge limit is obtained from the Euclidean correlator by analytically continuing $\r$ along the path shown.}
\end{figure}

In this section we review the bounds of \cite{Kundu:2020gkz} on CFT Regge correlators for scalar external operators. Points $x \in \mathbb{R}^{1,d-1}$ in CFT$_d$ are denoted as follows:
\be
x = (t,y,\vec{x})\equiv (x^-,x^+,\vec{x})\ , 
\ee
where, $x^{\pm}=t\pm y$ are lightcone coordinates. We study the Lorentzian CFT correlator\footnote{The Hermitian conjugatation in (\ref{corr}) acts only on operators, not on coordinates. }
\be\label{corr}
G=\frac{\langle O_2(\mathbf{1}) O_1(\rr)O_1^\dagger(-\rr)O_2^\dagger(-\mathbf{1}) \rangle}{ \langle O_2(\mathbf{1}) O_2^\dagger(-\mathbf{1}) \rangle \langle  O_1(\rr)O_1^\dagger(-\rr) \rangle} 
\ee
of two arbitrary CFT scalar operators, where operators inside the correlator are ordered as written. All the points are restricted to be on a 2d subspace:   
\begin{align}\label{points}
\mathbf{1}&\equiv(t=0,y=-1,\vec{0})\ , \qquad \rr\equiv(x^-=\r,x^+=-\br,\vec{0})\ ,\nonumber\\
-\mathbf{1}&\equiv(t=0,y=1,\vec{0})\ , \qquad -\rr\equiv(x^-=-\r,x^+=\br,\vec{0})\
\end{align}
with $1>\br>0$ and $\r>1$, as shown in figure \ref{config}. The operator ordering in (\ref{corr}) is important since some of the operators, as shown in the figure \ref{config}, are timelike separated. This Lorentzian correlator can be obtained from the Euclidean correlator by analytically continuing  $\r$ along the path shown in figure \ref{dt1}.

For later convenience, we parametrize 
\be\label{sigma}
\r=\frac{1}{\sigma}\ , \qquad \br=\sigma \eta  
\ee
with $\eta>0$ and hence $G\equiv G(\eta,\sigma)$. The CFT Regge limit can be reached by taking 
\be\label{regge}
 \sigma\rightarrow 0 \ , \qquad \text{with } \qquad \eta=\text{fixed}>0
\ee
of the correlator $G(\eta,\sigma)$.  The Regge correlator $G(\eta,\sigma)$, as a function of complex $\sigma$, is analytic near $\sigma \sim 0$ (for $0<\eta<1$) on the lower-half $\sigma$-plane \cite{Hartman:2016lgu,Caron-Huot:2017vep,Kundu:2020gkz}.

\subsection{Boundedness of the Regge Correlator}
We can define another Lorentzian correlator 
\be\label{corr0}
G_0(\eta,\sigma)=\frac{\langle O_2(\mathbf{1})   O_1(\rr)O_2^\dagger(-\mathbf{1}) O_1^\dagger(-\rr) \rangle}{\langle O_2(\mathbf{1}) O_2^\dagger(-\mathbf{1}) \rangle \langle  O_1(\rr)O_1^\dagger(-\rr) \rangle}
\ee
which is determined by Euclidean OPE and hence in the limit (\ref{regge})
\be\label{g0approx}
G_0(\eta,\sigma) = 1+\cdots\ , 
\ee
where dots represent terms that are suppressed by positive powers of $\sigma$. The Regge correlator $G(\eta,\sigma)$ is bounded by the ``Euclidean" correlator $G_0(\eta,\sigma)$. In particular, for real $\sigma$ with $|\sigma| < 1$ OPE unitarity and crossing symmetry imply  
\be\label{con3}
|G(\eta,\sigma)| \leq  G_0(\eta,\sigma)\ ,
\ee
where, $G_0(\eta,\sigma)>0$. In the strict Regge limit (\ref{regge}), this simplifies to 
\be\label{bound}
|G(\eta,\sigma)| \leq 1+\O(\sigma^{a})
\ee
with $a>0$.

More generally, the bound (\ref{con3}) follows from Rindler positivity as described in \cite{Hartman:2016lgu,Kundu:2020gkz}. For CFT correlators of scalar operators, Rindler positivity is a consequence of OPE unitarity and crossing symmetry. This is reviewed in appendix \ref{app:rindler} (see also \cite{Caron-Huot:2017vep}).

\subsection{CFT Constraints}\label{sec:cft}
Next, we focus on Regge correlators with a very specific Regge behavior for some range of $\sigma$:
\be\label{con_anc}
G(\eta,\sigma)= 1+ i \sum_{L=1,2,\cdots}  \frac{c_L(\eta)}{ \sigma^{L-1}}\ , \qquad \sigma_*\le |\sigma|\ll \eta<1
\ee
up to terms that decay in the Regge limit. The cut-off $\sigma_*$ dictates the regime of validity of the Regge expansion (\ref{con_anc}). Later we will relate $\sigma_*$ to $1/\Delta_{\rm gap}^2$.

The Regge correlator $G(\eta,\sigma)$, as a function of complex $\sigma$, is analytic near $\sigma \sim 0$ on the lower-half $\sigma$-plane \cite{Hartman:2016lgu,Kundu:2020gkz}. Using this analyticity property we can write a CFT dispersion relation for $c_L(\eta)$ \cite{Kundu:2020gkz}:
\be\label{sum}
c_L(\eta)=\frac{1}{\pi}\int_{-R}^R d\sigma\ \sigma^{L-2}(1- \mbox{Re}\ G(\eta,\sigma))\ , \qquad \sigma_* \le R\ll \eta<1\ ,
\ee
 where $L\ge 2$. The above relation leads to bounds on $c_L(\eta)$ for all $0<\eta<1$. Note that the left hand side does not depend on $R$. This implies $ \mbox{Re}\ G(\eta,\sigma)$ deviates significantly from 1 only when $\sigma \lesssim \sigma_*$. This is closely related to the fact that the tree level 4-pt scattering amplitude for the EFT (\ref{action}) has no imaginary part.

 \subsubsection*{Positivity}
 The boundedness condition (\ref{bound}) immediately implies \cite{Kundu:2020gkz}
 \be\label{bound1}
c_L(\eta)\ge 0\ ,  \qquad \text{for} \qquad \text{even}\ L\ge 2
\ee
for $0<\eta<1$. One can worry whether the  $\O(\sigma^{a})$ correction terms in (\ref{bound}) can affect the above positivity condition for higher $L$. For  the CFT dual to (\ref{action}), we can always take a limit where $R$ is small enough such that these corrections are suppressed.\footnote{At the end of this section we will discuss more about these corrections. } 
 
\subsubsection*{Parametric Separation} 
The fact that $\mbox{Re}\ G(\eta,\sigma)\le 1$ also implies \cite{Kundu:2020gkz}
 \be\label{bound2}
\frac{c_{L+2}(\eta) }{c_L(\eta)} \le \sigma_*^{2}\ , \qquad \frac{|c_{L+1}(\eta)| }{c_L(\eta)} \le \sigma_*
\ee
for all even $L\ge2$ and  $0<\eta<1$. Therefore $|c_L(\eta)|$, as a function of $L$, must decrease monotonically as a power law.  Furthermore, the above bound along with the condition (\ref{bound}) also require that the Regge correlator (\ref{con_anc}) is consistent only if 
\be
c_2 \lesssim \sigma_*\ .
\ee

\subsubsection*{Log-Convexity for Even $L$} 
The Cauchy-Schwarz inequality of integrable functions leads to the log-convexity condition for $c_L(\eta)$ with even $L$ \cite{Kundu:2020gkz}: 
\be\label{bound3}
\(\frac{c_{L+2}(\eta)}{c_L(\eta)}\)^2\le \frac{c_{L+4}(\eta)}{c_{L}(\eta)}\ , \qquad \text{for even} \quad L\ge 2
\ee
and $0<\eta<1$.
\subsubsection*{Boundedness of Odd $L$}
There is no sign constraint on $c_L(\eta)$ with odd $L$. However, the absolute value of $c_L(\eta)$ for odd $L$ is bounded \cite{Kundu:2020gkz}
\be\label{bound4}
|c_L(\eta)|\le \sqrt{c_{L-1}(\eta)c_{L+1}(\eta)}\ , \qquad \text{for odd} \quad L\ge 3\
\ee
and $0<\eta<1$. This also follows from the positivity condition (\ref{bound}) and the Cauchy-Schwarz inequality.

~\\
 Note that the chaos sign and the growth bounds of \cite{Maldacena:2015waa} are contained in the above consistency conditions. The condition (\ref{bound1}) is a generalization of the chaos sign bound. Whereas, the condition (\ref{bound2}) implies  that the Regge correlator (\ref{con_anc}) must not grow faster than $1/\sigma$ within the regime of validity  $\sigma_*\le |\sigma|\ll \eta<1$.

 Finally, let us note that the above constraints hold for arbitrary external operators with or without spins (and not necessarily primary)\footnote{See \cite{Kundu:2020gkz} for details.} as long as the Regge correlator has the form (\ref{con_anc}). For such a general case, the positivity of the integrand in (\ref{sum}) follows from Rindler positivity.

 \subsection{Correction Terms and Validity of the CFT Dispersion Relation}\label{sec:correction}
 All of the above constraints depend on the dispersion relation (\ref{sum}). So, it is only natural to ask whether there are corrections to this dispersion relation. In this section, we argue that any such correction terms do not affect the dispersion relation (\ref{sum}) since they are always suppressed for CFTs that are dual to some EFT in AdS. Casual readers may skip this subsection. 
 
 The first correction comes from the $\O(\sigma^a)$ terms of (\ref{bound}). Moreover, similar $\O(\sigma^a)$ correction terms can be present in the Regge expansion (\ref{con_anc}). So, the leading correction to the dispersion relation (\ref{sum}) comes from a term 
\be\label{eq:correction}
\delta (1- \mbox{Re}\ G(\eta,\sigma)) \sim (\delta c) \sigma^{a} \qquad \text{with} \qquad a>0
\ee
since all terms with negative $a$ have integer $a$ with imaginary coefficients.\footnote{It is important to note that any correction term with integer (positive or negative) power of $\sigma$ and an imaginary coefficient cannot affect the sum-rule (\ref{sum}) \cite{Kundu:2020gkz}.} 

First, let us justify the dispersion relation (\ref{sum}) for the scenario where operators $O_1$ and $O_2$ are different. In this case, especially for CFTs that are dual to some EFT in AdS, it is easy to see that $a\ge d$ since CFT operators that are exchanged are either double trace operators or single trace operators with $a=\Delta\ge d$ from a bulk three-point interaction.\footnote{Let us recall that we are restricting to the case where all fields have $m^2\ge 0$.} In particular,  the contribution of a correction term (\ref{eq:correction}) to the dispersion relation of $c_2$ is given by
 \be
\frac{1}{\pi}\int_{-R}^R d\sigma(1- \mbox{Re}\ G(\eta,\sigma))\sim c_2+  \mbox{Re}\ \delta c\int_{-R}^R d\sigma \sigma^a =c_2+ \O(1)  |\delta c| R^{a+1}
 \ee
 where the line integrals are evaluated just below the real $\sigma$-axis. Now, note that the leading contribution to $c_2 \sim \frac{\mu_2}{R_{\rm AdS}^D}$, whereas, the leading contribution to $\delta c$ comes from  scalar three-point couplings: $\delta c \sim \frac{\alpha_3^2}{R_{\rm AdS}^{D-6}}$.  Therefore, for any non-zero $c_2$, the correction term is suppressed for 
 \be
 R\ll \(\frac{|\mu_2|}{\alpha_3^2}\)^{\frac{1}{D}} \frac{1}{R_{\rm AdS}^{6/D}}\ .
 \ee
 On the other hand, the cut-off $\sigma_*$ scales as $1/R_{\rm AdS}^2$.\footnote{This can be seen easily from the scaling of individual terms of the expansion (\ref{con_anc}) for the bulk theory (\ref{action}). In particular, the expansion (\ref{con_anc}) for the Regge correlator of the dual CFT  is an expansion in the quantity $1/\sigma R_{\rm AdS}^2$, as can be seen from (\ref{regge:k}). The cut-off  $\sigma_*$ is controlled by the relative strength of consecutive terms in the expansion (\ref{con_anc}) and hence $\sigma_*\propto 1/R_{\rm AdS}^2$.} Therefore, for large $R_{\rm AdS}$ we can always choose $1\gg R\ge \sigma_*$  such that the correction term is parametrically suppressed for $D\ge 4$.\footnote{We are also making the mild assumption that $\frac{|\mu_2|}{\alpha_3^2}$ is not parametrically small when measured in the units of the mass cut-off scale (for example $M$, as defined in the introduction) associated with the bulk theory (\ref{action}). } If three-point bulk interactions such as $\alpha_3$ are absent, all other corrections (even from the bulk graviton exchange) to the sum-rule are more suppressed. Hence, the dispersive sum-rule for $c_2$ can always be trusted, at least for $D\ge 4$, for small $ R\rightarrow 0$.

 Let us now analyze the  relation (\ref{sum}) for higher $L$. Note that we can estimate:
 \be \label{eq:estimate}
 1- \mbox{Re}\ G(\eta,\sigma) \sim \frac{\pi c_2}{2\sigma_*}
 \ee 
 which implies
 \be
 \frac{1}{\pi}\int_{-\sigma_*}^{\sigma_*} d\sigma\ \sigma^{L-2}(1- \mbox{Re}\ G(\eta,\sigma))\sim c_L +\O(1)  |\delta c|  \sigma_*^{a+L-1}
 \ee
 where we see from (\ref{eq:estimate}) that $|c_L| \sim c_2 \sigma_*^{L-2}$ and hence the second term can  be ignored just like before even for $L>2$.\footnote{For odd $L$, it is possible that $|c_L| \ll c_2 \sigma_*^{L-2}$ because of cancellations implying that the dispersion relation (\ref{sum}) is not reliable. However, in this case all of the CFT bounds for odd $L$ are satisfied automatically. }

When $O_1=O_2$, there is a loophole in the above argument which we now fix. The disconnected Witten diagrams associated with the  4-pt correlator of the scalar operator of dimension $\Delta$ have the leading correction term $\sim \sigma^{2\Delta}$ with order 1 coefficient. However, one can subtract these contributions without affecting any of the bounds on $c_L$. For example, when $\Delta$ of the external scalar operator is an integer, we can replace $1- \mbox{Re}\ G(\eta,\sigma)$ in the sum-rule (\ref{sum}) by $G_0^{\rm free}(\eta,\sigma)- \mbox{Re}\ G(\eta,\sigma)$, where $G_0^{\rm free}(\eta,\sigma)$ is the correlator (\ref{corr0}) for the AdS theory (\ref{action}) without any interactions. This new sum-rule holds because $G_0^{\rm free}(\eta,\sigma)$ is analytic on the lower-half $\sigma$ plane for integer $\Delta$. Moreover, $G_0^{\rm free}(\eta,\sigma)- \mbox{Re}\ G(\eta,\sigma)$ is positive on the real line up to correction terms that are exactly the same as the above discussion of nonidentical operators. So, we repeat the same argument again to conclude that the modified dispersive sum-rule for $c_L$ is reliable for $D\ge 4$ and integer $\Delta$. This is sufficient for us, since for any fixed $m^2$, we can always tune $R_{\rm AdS}$ such that $\Delta$ is an integer. In any case, for non-integer $\Delta$ one can still write a more general sum-rule for $G(\eta,\sigma)$ by subtracting  contributions from the identity operator in all channels. The procedure is outlined in appendix \ref{app:subtract}.
 
 To summarize, we conclude that the CFT dual to the AdS theory (\ref{action}) must obey the consistency conditions (\ref{bound1}), (\ref{bound2}), (\ref{bound3}), and (\ref{bound4}) for $D\ge 4$. 
 \section{Constraining Scalar EFT in AdS without Gravity}\label{sec:EFT_AdS}
\begin{figure}
\centering
\includegraphics[scale=0.6]{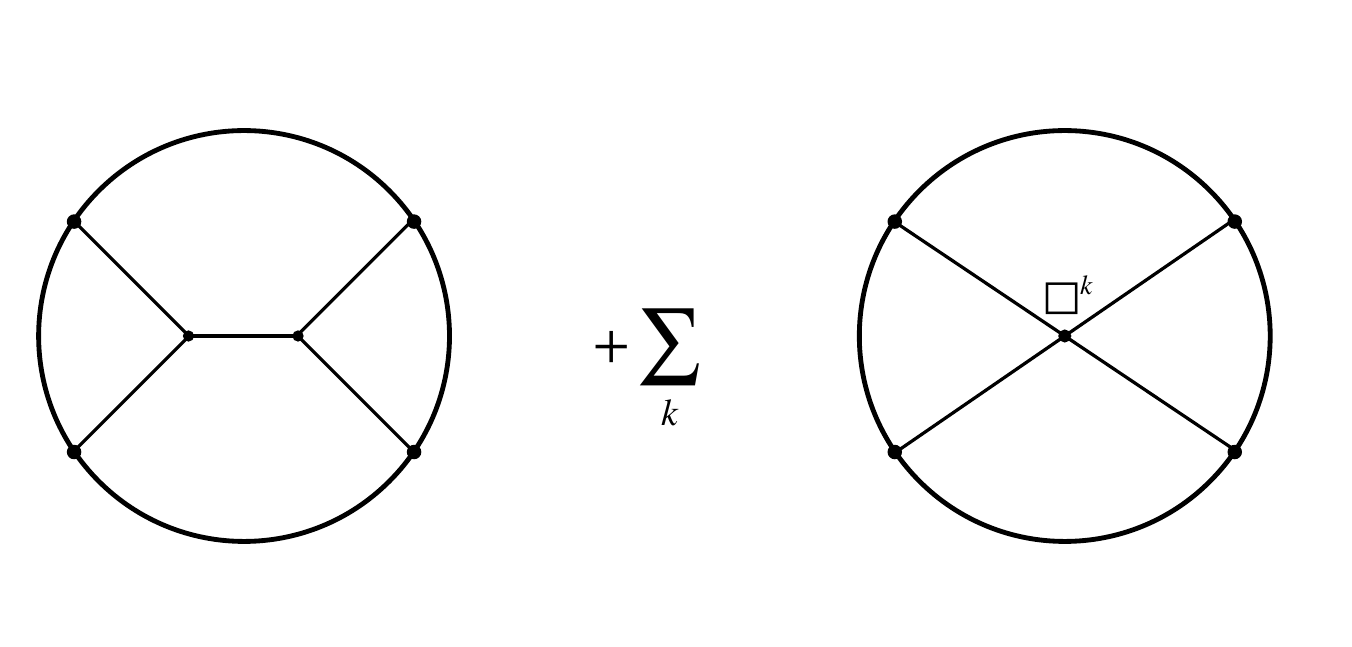}
\caption{ \label{fig:witten} \small  The tree-level Witten diagrams that are relevant in the Regge limit for $G_N=0$. Of course, the exchange diagram should be summed over all channels. }
\end{figure}

The main goal of this section is to impose bounds on the EFT (\ref{action}) from CFT consistency conditions.  To that end, we compute contributions of each EFT interactions to the Lorentzian correlator 
\be\label{eq:corr}
G(\eta,\sigma)=\frac{\langle \O(\mathbf{1}) \O(\rr)\O(-\rr)\O(-\mathbf{1}) \rangle}{ \langle \O(\mathbf{1}) \O (-\mathbf{1}) \rangle \langle  \O (\rr) \O(-\rr) \rangle} 
\ee
in the Regge limit (\ref{regge}), where operator $\O$ is dual to the scalar field $\phi$. First, we consider the purely non-gravitational case by setting $G_N=0$. The leading contribution to the correlator $G(\eta,\sigma)$ comes from the disconnected Witten diagrams. The dominant subleading contribution comes from the tree level Witten diagrams that are shown in figure \ref{fig:witten}. 
 
Before we proceed with the computation, let us review what is already known about the Regge limit. For example, from \cite{Heemskerk:2009pn,Chandorkar:2021viw} we know the scaling of the leading Regge contribution  of each interaction in (\ref{action}):
\begin{align}\label{regge:k}
\alpha_3 \phi^3  \qquad &\Rightarrow \qquad  \sim i \alpha_3^2 R_{\rm AdS}^{6-D} \sigma\ ,\nonumber\\
\mu_{2n} \phi^2 \Box^{2n} \phi^2  \qquad &\Rightarrow \qquad  \sim i \frac{\mu_{2n}}{R_{\rm AdS}^{D+4n-4}}\frac{1}{\sigma^{2n-1}}\ , \nonumber\\
\mu_{2n+1} \phi^2 \Box^{2n+1} \phi^2  \qquad &\Rightarrow \qquad \sim i \frac{\mu_{2n+1}}{R_{\rm AdS}^{D+4n-2}}\frac{1}{\sigma^{2n-1}}\ ,
\end{align}
for integer $n$. From the scaling behavior (\ref{regge:k}) it is clear that the Regge correlator (\ref{eq:corr}) has the desired expansion (\ref{con_anc}) up to terms that are suppressed by positive powers of $\sigma$. Notice that contributions of $\phi^2 \Box^k \phi^2$ for odd $k$ are always suppressed in the large $R_{\rm AdS}$ limit. Hence, the above scaling behavior implies that we can only impose constraints on interactions $\mu_{2n} \phi^2 \Box^{2n} \phi^2$ for $n=1,2,\cdots$ from the CFT consistency conditions of the preceding section.

We observe that the leading contribution to $c_L$ with even $L\ge 2$ comes entirely from the interaction $\phi^2 \Box^{L} \phi^2$. Contributions from $k>L$ interactions to $c_L$ are all suppressed in the large $R_{\rm AdS}$ limit. It is also clear from (\ref{regge:k}) that $c_L$ with odd $L$ are all zero. This simply follows from the fact that $\O$ is a real scalar operator. 

So,  it is sufficient for us to consider each interaction separately 
\be\label{action:k}
S_{L}=\frac{\mu_{L}}{2 }\int d^{D} x \sqrt{-g}  \phi^2 \Box^{L} \phi^2 \ ,
\ee
with $L>0$ being an even integer, where $d^{D} x\equiv dz d^{d}x$.  Note that 
\be
\frac{\mu_{L}}{2 }\phi^2 \Box^{L} \phi^2 \sim 2^{L-1} \mu_{L} \phi^2 (\nabla_{\mu_1}\cdots \nabla_{\mu_{L}}\phi)^2+\cdots\ ,
\ee
where dots represent terms with lower number of derivatives after we impose the free equation of motion.

It is more convenient to first compute the Euclidean correlator and then analytically continue to obtain the Regge correlator. So, the on-shell Euclidean action associated with (\ref{action:k}) is obtained from (\ref{sk1})
\be
S_{\rm on-shell}^{(L)}=- 2^{L-1} \mu_{L}\int dz d^d x \sqrt{g}\   \phi^2 (\nabla_{\mu_1}\cdots \nabla_{\mu_{L}}\phi)^2+\cdots\ ,
\ee
where again dots represent terms with lower number of derivatives that cannot contribute to $c_L$. This on-shell action can be further simplified by using the bulk-to-boundary propagator (\ref{AdS:bdb})
\begin{align}\label{os:eqn1}
S_{\rm on-shell}^{(L)}=- 2^{L-1} \mu_{L}C_\Delta^4& \int_{AdS}\int_{\Phi^4}  \tilde{K}_\Delta(z,x;x_1)\tilde{K}_\Delta(z,x;x_2)\nonumber\\
&\nabla_{\mu_1}\cdots \nabla_{\mu_{L}}\tilde{K}_\Delta(z,x;x_3)\nabla^{\mu_1}\cdots \nabla^{\mu_{L}}\tilde{K}_\Delta(z,x;x_4)\ ,
\end{align}
where the derivatives are taken with respect to the bulk point $\{z,x\}$.\footnote{The coefficient $C_\Delta$ is defined in (\ref{eq:2pt}).} In the above expression, we have utilized the notations of \cite{Kundu:2019zsl}
\be\label{notation}
\int_{AdS}\equiv \int dz d^{d}x \sqrt{g}\ , \qquad \int_{\Phi^4} \equiv \prod_{i=1}^4\int  \Phi(x_i)d^dx_i\ , \qquad x_{ij}^2=(x_i-x_j)^2\ .
\ee
Note that $x$ represents points on the AdS boundary. Moreover, the reduced bulk-to-boundary propagator $ \tilde{K}_{\Delta}(z,x;x')$ is defined in (\ref{reduced}) to reduce clutter. The boundary value of the field $\phi(z,x)$ is given by $\Phi(x)$ which acts as the source for the CFT$_d$ primary operator $\O(x)$ in the usual way.

We can now use the identity (\ref{identity1}) to write $S_{\rm on-shell}^{(L)}$ in terms of the $D$-function which is defined in (\ref{defineD}) in the standard way. We notice from equation (\ref{D:regge}) that all $D$-functions decay in the Regge limit $D_{\Delta_1\Delta_2 \Delta_3 \Delta_4}\sim\sigma$. On the other hand, $x_{ij}^2$ factors  for the kinematics (\ref{points2}) can grow as $\sim 1/\sigma$. Therefore, terms in $S_{\rm on-shell}^{(L)}$ that have the largest factors of $x_{ij}^2$ dominate in the Regge limit (\ref{regge}). This greatly simplifies the analysis since we only care about the growing part of the Regge correlator $G(\eta,\sigma)$. In particular, the leading Regge contribution from the on-shell action (\ref{os:eqn1}) comes from 
\begin{align}\label{os:eqn2}
S_{\rm on-shell}^{(L)}=- 2^{2L-1} \mu_{L} C_\Delta^4\(\frac{\Gamma\(\Delta+L\)}{\Gamma\(\Delta\)}\)^2   \int_{AdS}\int_{\Phi^4}   x_{34}^{2L} \tilde{K}_\Delta(z,x;x_1)\tilde{K}_\Delta(z,x;x_2)\nonumber\\
\times \tilde{K}_{\Delta+L}(z,x;x_3)\tilde{K}_{\Delta+L}(z,x;x_4)+\cdots\ ,
\end{align}
where, dots represent terms that will not contribute to $c_L$. It is now a straightforward exercise to show that the leading Regge contribution of the interaction  (\ref{action:k}) is
\begin{align}\label{eq:corr1}
G(\eta,\sigma)\sim \mu_{L} \frac{(16\eta)^\Delta 2^{2L-1} C_\Delta^2}{ (2\Delta-d)^2 R_{\rm AdS}^{D+2L-4}}\(\frac{\Gamma\(\Delta+L\)}{\Gamma\(\Delta\)}\)^2 \frac{16}{\sigma^{L}}D_{\Delta+L\ \Delta\ \Delta+L\ \Delta}(\eta,\sigma)\ ,
\end{align}
where $D=d+1$. From the above result, we obtain  an expression for $c_L$ for even $L\ge 2$ in the limit of large $R_{\rm AdS}$ (with $\Delta$ fixed):
\be\label{eq:cL}
c_L(\eta)=\frac{\kappa_\Delta \mu_L}{R_{\rm AdS}^{D+2L-4}}F_{2\Delta+L}(\eta)
\ee
where, $\kappa_\Delta$ is a positive coefficient independent of $L$ 
\be\label{def:kappa}
\kappa_\Delta=\frac{4}{\Gamma (\Delta )^2 \Gamma \left(-\frac{D}{2}+\Delta +\frac{3}{2}\right)^2}
\ee
and the $F$-function is given by using (\ref{define:f}):
\be
F_{2\Delta+L}(\eta)=\frac{1}{\eta ^{\frac{L-1}{2}}}f_{\Delta+L\ \Delta\  \Delta+L\ \Delta}\(-\frac{1}{2}\log\(\eta\) \)\ .
\ee 
which is positive for $0<\eta<1$. As we mentioned before, all $c_L(\eta)$ coefficients with odd $L$ vanish exactly.

\subsection{Bounds}
We are now in a position to utilize the CFT constraints from section \ref{sec:cft} to derive bounds on the EFT (\ref{action}).

\subsubsection{Positivity}
First, we impose the condition (\ref{bound1}). The fact that both $\kappa_\Delta$ and $F_{2\Delta+L}(\eta)$ for $0<\eta<1$ are positive immediately implies 
 \begin{bBox}
 \be\label{EFT:bound1}
\mu_k\ge 0\ ,  \qquad \text{for} \qquad \text{even}\ k\ge 2\ .
\ee
\end{bBox}
Moreover, saturation of any one of (\ref{EFT:bound1}) necessarily requires that the all of them are saturated. These bounds are consistent with the flat space bound of \cite{Adams:2006sv} from analyticity and unitarity of $2\rightarrow 2$ scattering amplitudes. Note that there is no such positivity condition on $\mu_k$ with odd $k$ from the CFT consistency conditions.

\subsubsection{Scale suppression of higher derivative interactions}
We now impose the condition (\ref{bound2}). First, let us apply (\ref{bound2}) to $L=2$:
\be\label{scale1}
\frac{\mu_4}{\mu_2}\le \frac{R_{\rm AdS}^4 F_{2\Delta+2}(\eta)}{F_{2\Delta+4}(\eta)}\sigma_*^2
\ee
for all $0<\eta<1$, where we are assuming that the AdS theory is interacting ($\mu_2>0$). First thing we notice that a mass scale $M=\Delta_{\rm gap}/R_{\rm AdS}$ is emerging naturally where we have identified 
\be\label{gap}
\sigma_*\equiv \frac{1}{\Delta_{\rm gap}^2}\sqrt{\frac{\Gamma (2 \Delta +3) \Gamma \left(-\frac{D}{2}+2 \Delta +\frac{9}{2}\right)}{\Gamma (2 \Delta +1) \Gamma \left(-\frac{D}{2}+2 \Delta +\frac{5}{2}\right)}}\ .
\ee
This definition of $\Delta_{\rm gap}$ needs some explanation. It is expected that $\sigma_* \propto 1/\Delta_{\rm gap}^\#$  since  $ \mbox{Re}\ G(\eta,\sigma)$ deviates significantly from 1 when $\sigma \lesssim \sigma_*$ implying a breakdown of (\ref{con_anc}). The exact power in (\ref{gap}) follows from the linear relationship between $M=R_{\rm AdS}\Delta$. The order one numerical factor has been chosen such that in certain scenarios $\Delta_{\rm gap}$ has the physical interpretation of the lightest heavy state exchanged.\footnote{We will make this more precise in section \ref{sec:mass}.} Given a UV complete CFT dual and the low energy Regge behavior (\ref{con_anc}), one can compute $\sigma_*$ from the sum-rule (\ref{sum}) with $R=\sigma_*$. Then (\ref{gap}) should be thought of as a precise definition of $\Delta_{\rm gap}$. The bulk cut-off scale is then given by the relation: $M=\Delta_{\rm gap}/R_{\rm AdS}$. Of course, this definition of $M$ is not unique. This definition is analogous to the definition of $M$ in \cite{Caron-Huot:2021rmr} and in certain cases these two definitions are exactly equivalent, as we show in section \ref{sec:mass}.

The strongest bound from (\ref{scale1}) is obtained for the value of $\eta$ that minimizes the right hand side. One can check that this is achieved in the limit $\eta\rightarrow 0$. Therefore, by using results from appendix \ref{app:F} we obtain a strict bound:
\begin{bBox}
\be\label{EFT:bound2}
\frac{\mu_{k+2}}{\mu_k}\le \frac{n_k(\Delta)}{n_{k+2}(\Delta)}\(\frac{R_{\rm AdS}}{\Delta_{\rm gap}}\)^4 \qquad \text{for} \qquad \text{even}\ k\ge 2
\ee
\end{bBox}
where, $n_k(\Delta)$ is given by
\be\label{eq:gamma}
n_k(\Delta)=\frac{\Gamma\(2\Delta+k-\frac{D-1}{2}\)\Gamma\(2\Delta+k-1\)}{\Gamma\(2\Delta-\frac{D-5}{2}\)\Gamma\(2\Delta+1\)}\left(\frac{\Gamma (2 \Delta +3) \Gamma\(2\Delta-\frac{D-9}{2}\)}{\Gamma (2 \Delta +1)\Gamma\(2\Delta-\frac{D-5}{2}\)}\right)^{1-\frac{k}{2}}
\ee
with $m^2R_{\rm AdS}^2=\Delta(\Delta-D+1)$.  Of course, the bound (\ref{EFT:bound2}) depends heavily on our definition of $\Delta_{\rm gap}$ (\ref{gap}). Notice that $n_k(\Delta)$ is the same coefficient that appears in (\ref{intro:EFT}).

The bound (\ref{EFT:bound2})  validates our expectation that  higher derivative interactions $\phi^2\Box^k\phi^2$ are suppressed by inverse powers of $\Delta_{\rm gap}$ for even $k$. However, CFT consistency conditions of the preceding section do not impose similar constraints on $\phi^2\Box^k\phi^2$ interactions with odd $k$. This perhaps suggests that our bounds are far from being optimal. 

Let us make few comments about the coefficient $n_k(\Delta)$ which is a log-convex function of $k$. From (\ref{eq:gamma}) we find that $n_2(\Delta)=n_4(\Delta)=1$ for any $m^2$ and $D$ implying 
\be\label{eq:scale}
\frac{\mu_4}{\mu_2} \le \frac{1}{M^4}\ .
\ee
We will derive this relation in flat space by assuming the Regge boundedness condition: ${\mathcal A}(s,t=0)<|s|^2$ for large $s$. In that case, $M$ is the mass of the lightest massive state exchanged. This explains the choice (\ref{gap}). 

Furthermore, note that $n_k(\Delta)$ is non-trivial only in the massless limit $m\rightarrow 0$. In particular, if we take $R_{\rm AdS}\rightarrow \infty $ with fixed $0<m\ll M$, we obtain  
\be
n_k(\Delta)\approx 1
\ee
for all finite $k \ll m R_{\rm AdS}$. Therefore, for non-zero $m$, all our bounds simplify greatly. 

\subsubsection{Log-convexity condition}
The CFT condition (\ref{bound3}) leads to a rather strong condition on couplings $\mu_k$ that does not depend on the exact definition of $\Delta_{\rm gap}$ (or equivalently the scale $M$). The optimal bound again is achieved for $\eta\rightarrow 0$, yielding 
\begin{bBox}
\be\label{EFT:bound3}
\frac{\mu_{k+2}^2}{\mu_k \mu_{k+4}}\le \frac{n_k(\Delta)n_{k+4}(\Delta)}{n_{k+2}(\Delta)^2} \qquad \text{for} \qquad \text{even}\ k\ge 2
\ee
\end{bBox}
where, $n_k(\Delta)$ is defined in (\ref{eq:gamma}). The right hand side is exactly 1 for $k \ll m R_{\rm AdS}\ll M R_{\rm AdS}$, as discussed above. 

Therefore, the EFT (\ref{action}), in the absence of gravity, can be UV completed only when it satisfies (\ref{EFT:bound3}), irrespective of how the cut-off scale $M$ is defined.  It would be nice to derive a similar bound for odd $\mu_k$ couplings.

\subsubsection{Odd couplings}
We still can say few things about the odd $\mu_k$ couplings. First of all, if $\mu_2=0$ the  leading contribution to $c_2$ comes from $\mu_3$. However, this contribution to $c_2$ changes sign as we tune $\eta$ within the domain $0<\eta<1$ for $m^2\ge 0$ (see appendix \ref{app:odd}). Hence $\mu_3$ must vanish exactly.  Then the condition (\ref{bound2}) implies that 
\begin{bBox}
\be\label{EFT:bound4}
\mu_2=0 \qquad \Rightarrow  \qquad \mu_k=0 
\ee
\end{bBox}
for even or odd  $k>2$.\footnote{We assume that $m^2\ge 0$. For negative mass$^2$, see appendix \ref{app:odd} for comments. } When $\mu_2>0$, the bound on odd $\mu_k$ couplings are rather weak. The above argument then implies that $\frac{|\mu_{k}|}{\mu_{k-1}} \lesssim R_{\rm AdS}^2$ for odd $k\ge 3$.

\subsection{Final Effective Action}\label{sec:FEA}
Let us now summarize the results of this section by writing the AdS scalar effective action (\ref{action}) as follows
\begin{align}\label{EFT}
S[\phi]=\frac{1}{2}& \int d^{D} x \sqrt{-g} \left(- g^{\mu\nu}\nabla_\mu \phi \nabla_\nu\phi -m^2 \phi^2+\alpha_3 \phi^3+\mu_0 \phi^4\)\nonumber\\
&+\frac{\mu}{2} \int d^{D} x \sqrt{-g} \sum_{k=2,3,4,\cdots} \frac{\lambda_k}{n_k(\Delta) M^{2(k-2)}}\phi^2 \Box^k \phi^2+\cdots \ ,
\end{align}
where $M=\Delta_{\rm gap}/R_{\rm AdS}$ is the scale of new physics and the numerical factor $n_k(\Delta)$ is defined in (\ref{eq:gamma}). The scalar field can have mass but $0\le m^2 \ll M^2$. Note that $n_2(\Delta)=n_4(\Delta)=1$ for any $m^2$ and $D$. For $k>4$,  $n_k(\Delta)$, in the large $R_{\rm AdS}$ limit, differs from 1 only for $m=0$.\footnote{More generally, in the large $R_{\rm AdS}$ limit, $n_k(\Delta)$ with $k>4$ differs from 1 only for $m\rightarrow 0$ and $R_{\rm AdS}\rightarrow \infty$ with $R_{\rm AdS}m$ fixed.} 

So far gravity is non-dynamical $G_N=0$. We have defined a positive coupling constant $\mu\ge 0$ which has the dimension $1/M^{D}$. The $\lambda$-coefficients are dimensionless, however, to begin with we do not assume that they are $\O(1)$. We do assume the theory is weakly coupled $\mu M^D\sim |\mu_0|M^{D-4}\sim \alpha_3^2 M^{D-6}\ll 1$ and hence analyze the theory at tree level.

The main goal of this paper is to address the question: when can this EFT be UV completed? Or equivalently what are the necessary conditions for this EFT to be embedded into a UV theory that is dual to a CFT with  $\Delta_{\rm gap}\gg 1$ obeying the CFT axioms? In this section, we conclude that the EFT (\ref{EFT}), with $\mu\ge 0$, must have the following properties ($D\ge 4$):
\begin{enumerate}[label=(\roman*)]
\item \label{condition1}{$ \lambda_k >0$ for all even $k\ge 2$\ ,}
\item \label{condition3} {$\lambda_{k+2}\le \lambda_k$ for all even $k\ge 2$ }\ ,
\item \label{condition5} {$\frac{1}{k_2-k_1}\ln \frac{\lambda_{k_1}}{\lambda_{k_2}}\ge \frac{1}{k_3-k_1}\ln \frac{\lambda_{k_1}}{\lambda_{k_3}}$ for all even $k_3>k_2>k_1\ge 2$.}
\end{enumerate}
The last  condition follow directly from the local log-convexity condition (\ref{EFT:bound3}). 
It should be noted again that the condition \ref{condition3} depends on the exact definition of $\Delta_{\rm gap}$ and hence the scale $M$. On the other hand, other two conditions do not depend on the exact definition of the scale $M$. For an arbitrary definition of $M$, the condition \ref{condition3} should be thought of in the following way. There must always exist a rescaling $M\rightarrow XM$, with order one $X$, which makes the EFT consistent with the condition \ref{condition3}. 

It should also be emphasized that \ref{condition1}-\ref{condition5} are necessary conditions but we believe they are far from being sufficient. For example, it is expected that similar bounds exist even for odd $k$.\footnote{Note that the condition (\ref{EFT:bound4}) has been implemented by assuming the EFT has the form (\ref{EFT}) along with $\lambda_2>0$. } However, our argument does not impose any restriction on the odd $\lambda_k$ couplings other than $|\lambda_k|\lesssim \Delta_{\rm gap}^2$.

\subsection{Flat Space Limit}\label{sec:flat}

 We end this section with some discussion on the flat space limit of the EFT (\ref{EFT}). In this section we restrict to the massless case: $m=0$. The flat space limit should be taken in the following way:
 \be
 R_{\rm AdS}\rightarrow \infty\ , \qquad \Delta_{\rm gap}\rightarrow \infty \qquad \text{with} \qquad \frac{\Delta_{\rm gap}}{R_{\rm AdS}}=M=\text{fixed} \ .
 \ee
 In the massless case
 \be\label{flat:gamma}
n_k^{(0)}=\left(\frac{3}{2}\right)^{1-\frac{k}{2}}\frac{\Gamma \left(\frac{3 (D-1)}{2}+k\right) \Gamma (2 D+k-3)}{\Gamma \left(\frac{3 D}{2}+\frac{1}{2}\right) \Gamma (2 D-1)} (D (D+1) (2 D-1) (3 D+1))^{1-\frac{k}{2}}
\ee
increases fast for $k>4$ as we increase $k$. In this limit, the constraints \ref{condition1}-\ref{condition5} lead to bounds on the flat space EFT of a massless scalar. We can compare these flat space bounds with the results from \cite{Caron-Huot:2021rmr} by relating various coupling constants:
\be
g_2= 4\mu\lambda_2\ , \quad g_3= \frac{12\mu \lambda_3}{n_3^{(0)}M^2}\ , \quad g_4=  \frac{2\mu \lambda_4}{M^4}\ , \quad g_6=  \frac{\mu \lambda_6}{n_6^{(0)}M^8}\ , \quad \cdots\ .
\ee
In particular, in the absence of gravity we obtain
\be
g_2,\ g_4,\ g_6, \cdots \ge 0\ .
\ee
Furthermore, with our definition of $M$, we find that the bound (\ref{eq:scale}) agrees with the bound obtained in \cite{Caron-Huot:2021rmr}.

As we will explain in section \ref{sec:gravity}, all bounds for $g_k$ with $k>2$ remains unaffected even when gravity is dynamical. Furthermore, we obtain a rather interesting inequality by applying \ref{condition5}:
\be\label{convexity:simon}
\frac{g_4^2}{g_2 g_6}\le \frac{(2 D+1) (3 D+5) (3 D+7)}{3 D (2 D-1) (3 D+1)}\ .
\ee
It would be interesting to compare this bound with the analysis of \cite{Caron-Huot:2020cmc}. It is possible to derive an infinite set of such constraints from \ref{condition5}. Note that constraints involving $g_2$ will only be affected when gravity is turned on. Let us stress that there is a discreet difference between the massless case $m=0$ and the massless limit $m\rightarrow 0$ when we take the flat space limit. We will discuss this in the next section.

Unlike \cite{Caron-Huot:2020cmc,Caron-Huot:2021rmr}, our analysis is insensitive to $\phi^3$ and $\phi^4$ interactions of (\ref{EFT}).  However, we still need to pay attention to these interactions. In particular,   the coupling constant $\alpha_3$ for the $\phi^3$ interaction has positive mass dimension for $D\le 5$. So, this coupling can lead to large mixing effects in the dual CFT when we take the flat space limit $R_{\rm AdS}\rightarrow \infty$ \cite{Kaplan:2019soo}. In particular, when 
 \be
 |\alpha_3| R_{\rm AdS}^{3-D/2} \sim 1
 \ee
 there is a large mixing between the naive generalized free field operator $\O$ and $[\O\O]_{n,0}$ because of the decay channel $\phi\rightarrow \phi\phi$. It is unclear whether the flat space bounds are reliable when the mixing effect is large. Nevertheless,  we can avoid this issue for $D\le 5$ by giving the bulk field $\phi$ some $\mathbb{Z}_2$ symmetry. Or we can take the flat space limit of the AdS theory (\ref{action}) such that 
 \be
R_{\rm AdS}M\gg 1\  \qquad  \text{with} \qquad  |\alpha_3| R_{\rm AdS}^{3-D/2}\ll 1. 
 \ee

There is one more subtlety that we must address. When the $\phi^2\Box^2\phi^2$ interaction is absent in the AdS EFT (\ref{EFT}), all the higher derivative interactions $\phi^2\Box^k\phi^2$ must also vanish. However, our analysis does not require this to be true in the exact flat space limit. For example, the coefficient of the $\phi^2\Box^2\phi^2$ interaction can be suppressed by $R_{\rm AdS}$ in such a way that the dual CFT is well behaved. Moreover, the EFT can have a $\phi^2\Box^3\phi^2$ interaction which is not suppressed by $R_{\rm AdS}$ but fine-tuned such that  $c_2(\eta)$ is still positive. In this scenario,  the $\phi^2\Box^2\phi^2$ interaction goes to zero  in the flat space limit with a non-vanishing $\phi^2\Box^3\phi^2$ interaction. It has been recently conjectured that such EFTs emerge naturally in the IR from 6D supersymmetric RG flows on to the Higgs branch \cite{Heckman:2021nwg}. So, such RG flows in AdS$_6$ with finite radius are expected to generate a $\phi^2\Box^2\phi^2$ interaction for the dilaton which is suppressed by $1/R_{\rm AdS}^2$.

\subsection{Other Higher-Derivative Interactions}\label{sec:others}
The observant reader may have noticed that the effective action (\ref{EFT}) can have other higher derivative 4-$\phi$ interactions. For example, even in flat space there are multiple inequivalent 4-$\phi$ interactions with 12 or more derivatives. In this section, we will argue that these other higher-derivative interactions are not bounded by the argument of this paper  (provided $\mu$ is non-zero). So, these other higher derivative 4-$\phi$ interactions will not affect any of the bounds obtained in this paper.

First, let us consider the flat space EFT (massive or massless) with 4-$\phi$ interactions. At the $2k$-derivative level, there are exactly two types of interactions:
\be\label{eq:others}
a_k \phi^2 \Box^k \phi^2+\sum_{a,b,c>0}^{a+b+c=k} b_k^{\{a,b,c\}} \(\p^{\mu_1}\cdots \p^{\mu_k}\phi \)\(\p_{\mu_1}\cdots \p_{\mu_a}\phi \)\(\p_{\mu_1}\cdots \p_{\mu_b}\phi \)\(\p_{\mu_1}\cdots \p_{\mu_c}\phi \)\ .
\ee
All other possible interactions can be written in the above form by utilizing the free equation of motion and integration by parts. Note that the second term can only be non-zero and indepedent (when we use the equation of motion) for $k\ge 6$.\footnote{For example, for $k=3$ the second term can be equivalently written as $ \phi^2 \Box^3 \phi^2$ plus terms with 4 or less derivatives.} 

Now we move on to the AdS case and replace $\p_\mu \rightarrow \nabla_\mu$. In AdS, derivatives do not commute in general. So, one may construct several more terms from the second term of (\ref{eq:others}) by choosing different ordering of derivatives. However, different derivative orderings differ only by factors of $1/R_{\rm AdS}^2$
\be
\(\cdots\nabla_\mu \nabla_\nu \cdots \) \phi - \(\cdots\nabla_\nu \nabla_\mu \cdots \) \phi\sim \frac{1}{R_{\rm AdS}^2}  \( \cdots \) \phi
\ee
 and hence derivative ordering is not important in the large $R_{\rm AdS}M$ limit. Let us now figure out the Regge contribution of the second term of (\ref{eq:others}) in AdS. One can easily check that the leading Regge contribution of the second term comes from the on-shell action 
\be
 \int_{AdS}\int_{\Phi^4}   x_{34}^{2c}x_{24}^{2b}x_{14}^{2a} \tilde{K}_{\Delta+c}(z,x;x_1)\tilde{K}_{\Delta+b}(z,x;x_2) \tilde{K}_{\Delta+a}(z,x;x_3)\tilde{K}_{\Delta+k}(z,x;x_4)
\ee
which grows slower than $\frac{1}{\sigma^{k-1}}$ since $a,b,c>0$. Hence, for even $k$, the leading Regge contribution always comes from the first term of (\ref{eq:others}). Therefore, we conclude that these other higher derivative 4-$\phi$ interactions do not affect any of the bounds obtained in this paper.

\section{Flat Space Limit: Massless \& Massive Scalars}\label{sec:mass}
In this section, we compare bounds from the previous section with bounds obtained by studying flat space scattering amplitudes. We again start with the effective action (\ref{action}) without dynamical gravity $G_N=0$. For simplicity we take the three-point coupling $\alpha_3=0$, so that the tree level 4-point scattering amplitude does not have poles at $m^2$. The forward limit ($t=0$) of the tree level 4-point scattering amplitude associated with the effective action (\ref{action}) is given by
\be\label{tree:amplitude}
\mathcal{A}(s,t=0)=8\sum_{I=0}^\infty\mu_{2I} s^{2I}\ .
\ee
At this point we make four assumptions: (1) the forward amplitude is bounded for large $s$:
\be\label{bounded:regge}
{\mathcal A}(s,t=0)<|s|^2\ ,
\ee
(2) the amplitude is analytic in the upper-half complex $s$-plane, (3) the amplitude obeys partial-wave unitarity implying $\mbox{Im}\ \mathcal{A}(s,t=0)>0$ for real $s$, (4) the amplitude is crossing symmetric. 

These are the key assumptions which allow us to write a dispersive sum-rule for $\mu_{2I}$. In particular, repeating the argument of \cite{Adams:2006sv}, we can write
\be\label{dispersion1}
\mu_k=\frac{1}{4\pi}\int_{M_*^2}^\infty ds \frac{\mbox{Im}\ \mathcal{A}(s,t=0)}{s^{k+1}}>0\ ,
\ee
for all even $k\ge 2$, where $M_*$ is the cut-off scale at which $\mbox{Im} \mathcal{A}(s=M_*^2,t=0)$ becomes non-zero. The cut-off scale $M_*\propto M$, however, the two scales can be different in general by some order one proportionality constant.  

From (\ref{dispersion1}), we can also derive a monotonicity and a log-convexity conditions for even $k\ge 2$:
\begin{align}\label{dispersion2}
\frac{\mu_{k+2}}{\mu_k}\le \frac{1}{M_*^4}\ , \qquad \frac{\mu_{k+2}^2}{\mu_k \mu_{k+4}}\le 1\ .
\end{align}
The second inequality can be used to derive a global log-convexity condition (\ref{intro:condition3}) for even $\mu_k$.

Thus, under the above assumptions we showed that {\it the tree level amplitude, in the forward limit, has a polynomial expansion in $s^2$ with coefficients obeying (i) positivity, (ii) monotonicity, and (iii) log-convexity conditions.  } At first sight, these conditions seem to be stronger than the flat space limit of the AdS conditions (\ref{intro:condition1})-(\ref{intro:condition3}). For the remainder of this section we will address whether, and in what sense, the above bounds are related to the AdS bounds.

\subsection{Massive Scalars}
An important feature of our AdS bounds is that they differ significantly for massive and massless scalars, especially when we take the flat space limit. First, we consider the massive case $0<m\ll M$, where $M$ is the cut-off scale defined in the previous section. We take the flat space limit by $M R_{\rm AdS}\gg 1$, keeping $m$ fixed.  So, in this limit $\Delta\approx m R_{\rm AdS}\gg 1$ and hence 
\be\label{eq:massive}
n_k(\Delta)= 1
\ee 
for all $k\ll m R_{\rm AdS}$. Therefore, in this case, for all $m>0$ and $D\ge 4$ the AdS conditions (\ref{intro:condition1})-(\ref{intro:condition3}) are identical to conditions (\ref{dispersion1}) and (\ref{dispersion2}) that were derived from the flat space sum-rule, provided we identify $M=M_*$. This provides compelling evidence in favor of the assumptions that were used to derive the sum-rule (\ref{dispersion1}) for massive external scalars. Moreover, for massive scalars, as we will explain in the next section,   conditions (\ref{dispersion1}) and (\ref{dispersion2}) remain valid for even $k\ge 4$ even when there is dynamical gravity.  This is rather non-trivial since the validity of the Regge boundedness condition (\ref{bounded:regge}) is not obvious in the presence of the graviton. 

\subsection{Massless Scalars}
The situation is a lot more subtle for massless external scalars. We can start with $m=0$ and then take the large $R_{\rm AdS}$ limit. One can also take a massless limit in which we first take the  large $R_{\rm AdS}$ limit (with fixed $m$) and then $m\rightarrow 0$. Clearly, our bounds are different in these two limits. In the latter case, we again obtain (\ref{eq:massive}). Hence, conditions (\ref{dispersion1}) and (\ref{dispersion2}) still hold. For example, in this case (\ref{convexity:simon}) becomes 
\be
\frac{g_4^2}{g_2 g_6}\le 1\ .
\ee

On the other hand, if we take $m=0$ first, we obtain bounds from section \ref{sec:flat}. In particular, now we have conditions (\ref{intro:condition1})-(\ref{intro:condition3}) with $n_k(\Delta=D-1)\equiv n_k^{(0)}$ given by (\ref{flat:gamma}). These bounds are weaker than the conditions (\ref{dispersion1}) and (\ref{dispersion2}). 

Furthermore, the massless limit can also be taken  in a more general way: $R_{\rm AdS}\rightarrow \infty$, $m\rightarrow 0$ with $\Delta=\ $fixed. In this case, we again obtain the weaker set of bounds (\ref{intro:condition1})-(\ref{intro:condition3}) with $n_k(\Delta)$ given by (\ref{eq:gamma}). Therefore, depending on how we take the massless limit (or equivalently the value of $\Delta$), we obtain a different set of constraints. We recover the flat space conditions (\ref{dispersion1}) and (\ref{dispersion2}) only for $\Delta\gg 1$. This suggests that in general some of the assumptions that were used to derive the sum-rule (\ref{dispersion1}) are not valid for massless scalars. This is perhaps not surprising since the Regge boundedness condition (\ref{bounded:regge}) can break down in the presence of massless states. 

Nonetheless, we can still provide a general condition on the  tree level amplitude of massless scalars which does not require any assumption other than the usual CFT-axioms. {\it The tree level amplitude of massless scalars, in the forward limit, has a polynomial expansion in $s^2$
\be
\mathcal{A}(s,t=0)=\sum_{I=0}^\infty\frac{c_{2I} }{n_{2I}(\Delta)}\(\frac{s}{M^2}\)^{2I}
\ee
with coefficients $c_{2I}$ obeying (i) positivity, (ii) monotonicity, and (iii) log-convexity conditions (\ref{intro:condition1})-(\ref{intro:condition3}) for $I\ge 1$.  } Of course, $M$ and $\Delta$ are theory dependent but fixed for a specific four-point amplitude.\footnote{The cut-off scale $M=X M_*$ is proportional to the mass $M_*$ of the lightest particle exchanged. The proportionality factor $X\sim \O(1)$, however, it may differ from 1 in general. The parameter $\Delta$ should be regarded as a measure of the breakdown of the Regge boundedness condition (\ref{bounded:regge}). } Whereas, the numerical coefficient $n_{2I}(\Delta)$ is theory independent and given by (\ref{eq:gamma}).  

In the presence of gravity, $\mathcal{A}(s,t)$ has a pole at $t=0$. However, $\mathcal{A}(s,t\rightarrow 0)$ still must satisfy the above condition for $I\ge 2$. 

\section{Scalar EFT with Gravity}\label{sec:gravity}
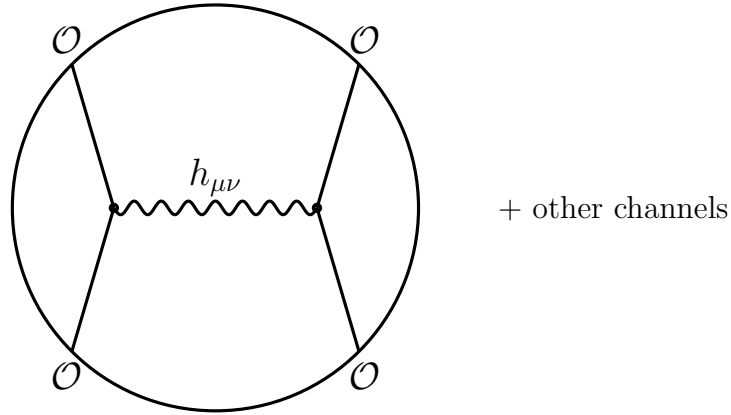
\begin{figure}[h]
\begin{center}
\usetikzlibrary{decorations.markings}    
\usetikzlibrary{decorations.markings}    
\begin{tikzpicture}[baseline=-3pt,scale=0.9]
\begin{scope}[very thick,shift={(4,0)}]
\draw (0,0) circle [radius=3];
\draw  (1.5,0) circle (1.5pt)-- (2.12,2.12);
\draw  (-1.5,0) circle (1.5pt) -- (-2.12,-2.12);
\draw  (1.5,0) -- (2.12,-2.12);
\draw  (-1.5,0) -- (-2.12,2.12);
\draw [domain=-1.5:1.5, samples=500] plot (\x, {0.1* cos(5*pi*\x r)});
\draw(2.2,2.12)node[above]{\large $\O$};
\draw(2.2,-2.12)node[below]{\large $\O$};
\draw(-2.2,2.12)node[above]{\large $\O$};
\draw(-2.2,-2.12)node[below]{\large $\O$};
\draw(4,0)node[right]{+ other channels };
\draw(0,0)node[above]{\large $h_{\mu \nu}$};
\end{scope}
\end{tikzpicture}
\end{center}
\caption{\label{witten2} \small The leading  gravitational contribution to  the Lorentzian correlator (\ref{eq:corr}) comes from the Witten diagram with a single graviton exchange.}
\end{figure}

We now discuss the effects of gravity on the bounds on the EFT (\ref{EFT}) by turning on $G_N\neq 0$:
\begin{align}\label{EFT:gravity}
S[\phi,g]=\frac{1}{16\pi G_N} \int d^{D} x \sqrt{-g}\(R+\frac{(D-1)(D-2)}{R_{\rm AdS}^2}\)+S[\phi] \ ,
\end{align} 
where, $S[\phi]$ is given by (\ref{EFT}).  We analyze the EFT at tree level, so we assume that the theory is weakly coupled as described by (\ref{weak}). Now the central charge of the dual CFT is large $c_T\gg \Delta_{\rm gap}\gg 1$ but finite. We again compute the Lorentzian correlator (\ref{eq:corr}) in the Regge limit (\ref{regge}), where operator $\O$ is dual to the scalar field $\phi$. The leading contribution to the connected part of the correlator $G(\eta,\sigma)$ comes from Witten diagrams \ref{fig:witten} plus the graviton exchange Witten diagram as shown in figure \ref{witten2}.

In the Regge limit (\ref{regge}), contribution from the channel $\O(\rr)\O(-\rr)\rightarrow h_{\mu\nu}\rightarrow  \O(\mathbf{1}) \O(-\mathbf{1})$ grows as $1/\sigma$.  The other channels do not contribute at all to the Regge growth. So, in the presence of gravity $c_L(\eta)$ for $L>2$ remains unaffected. On the other hand, $c_2(\eta)$ receives a contribution from gravity. In particular, the gravitational contribution to $c_2(\eta)$ can be obtained from \cite{JoaoRegge,Costa:2017twz,Kulaxizi:2017ixa}
\be
c_2(\eta)|_{\rm gravity}=\frac{\pi G_N }{R_{\rm AdS}^{D-2}}\kappa_\Delta \tilde{F}_g(\eta)\ ,
\ee
where, the numerical factor $\kappa_\Delta$ is defined in (\ref{def:kappa}). The function $\tilde{F}_g(\eta)$ is given by an integral of the harmonic functions $\Omega_{i\nu}$  in the hyperbolic space (see (\ref{eq:harmonic}))
\be
\tilde{F}_g(\eta)=\frac{1}{\sqrt{\eta}}\int_{-\infty}^\infty d\nu\frac{\Gamma\left(\frac{2\Delta+2-d/2+i\nu}{2} \right)^2 \Gamma\left(\frac{2\Delta+2-d/2-i\nu}{2} \right)^2}{\nu^2+\(\frac{d}{2}\)^2} \Omega_{i\nu}\(-\frac{1}{2}\log\(\eta\) \)
\ee
where $D=d+1$. Therefore, the full $c_2(\eta)$ is given by
\be
c_2(\eta)= \frac{\kappa_\Delta }{R_{\rm AdS}^{D}}F_{2\Delta+2}(\eta)\(\mu \lambda_2+ \pi G_N R_{\rm AdS}^2\frac{\tilde{F}_g(\eta)}{F_{2\Delta+2}(\eta)}\)>0\ ,
\ee
where the positivity follows from  condition (\ref{bound1}) for $0<\eta<1$. One can check that the optimal bound in this case is obtained in the limit $\eta\rightarrow 1$. In this limit, we find that $\lambda_2$ is now allowed to have negative values:
\begin{bBox}
\be\label{gravity:bound1}
\lambda_2 > -\frac{\pi G_N R_{\rm AdS}^2}{\mu}N_D(\Delta)\ ,
\ee
\end{bBox}
where $N_D(\Delta)$ is an $\O(1)$ numerical factor given in appendix \ref{App:N}. In particular, for the massless case we find
\be
N_4=0.1775\ , \quad N_5=0.0882\ , \quad N_6=0.0525\ , \quad N_7=0.0348\ , \quad N_8=0.0247\ , \quad \cdots\ .
\ee
Note that the bound (\ref{gravity:bound1}) cannot be saturated in a way which is consistent with the sum-rule (\ref{sum}).  

So, we conclude that in the presence of gravity $\lambda_2 $ is not required to be positive. This is consistent with the results of  \cite{Caron-Huot:2021rmr}. On the other hand, the bounds \ref{condition1}-\ref{condition5} are still valid for all even $k\ge 4$. Before we proceed, we must note that $\lambda_2$, if negative, cannot be arbitrarily large even in the large $R_{\rm AdS}$ limit. To see that, we write (\ref{gravity:bound1}) as:
\be
\mu \lambda_2 M^D > -\frac{ \Delta_{\rm gap}^D}{c_T}\O(1)\ ,
\ee
where $c_T$ is the CFT central charge. Validity of our analysis requires that we take $c_T \rightarrow \infty$ first and then $\Delta_{\rm gap}\rightarrow \infty$. This implies that we should use caution when we take the flat space limit. In particular, we must take $R_{\rm AdS}$ to be large such that $\frac{1}{\sqrt{G_N M^{D-2}}}\gg MR_{\rm AdS}\gg 1$. Hence, the right hand side of the above expression remains small even in the flat space limit. 

We now analyze the bound (\ref{EFT:bound2}) in the presence of gravity. Since, $c_k(\eta)$ for $k\ge 4$ remains unchanged, we only need to analyze the $k=2$ case. We assume that $\frac{G_N}{\mu M^2}\sim \O(1)$ so that the gravity effects are significant. The optimal bound, in the presence of gravity, is now obtained at the limit $\eta\rightarrow 1$ yielding 
\begin{bBox}
\be\label{gravity:bound2}
0\le \lambda_4 \le \tilde{N}_D(\Delta) \(\lambda_2 +\frac{\pi G_N R_{\rm AdS}^2}{\mu}N_D(\Delta)\)\ ,
\ee
\end{bBox}
where $\tilde{N}_D(\Delta)>1$ is given in appendix \ref{App:N}. One may wish to recover the $G_N=0$ result \ref{condition3} from the above inequality. The above bound is still valid when $G_N=0$, however, it is not optimal. This is simply because of the order of limits. As we take $\frac{G_N}{\mu M^2}\rightarrow 0$, the optimal bound is obtained for a value of $\eta$ which is close to zero and hence the upper bound of $\lambda_4$ approaches $\lambda_2$. The correction term from finite but small $\frac{G_N}{\mu M^2}$ now can be computed numerically, though we will have to leave this for the future.

Finally, we focus on the log-convexity condition $c_4(\eta)^2 \le c_2(\eta)c_6(\eta)$ in the presence of gravity. We again assume that there is no parametric separation between $G_N$ and $\mu$ in units of $M$: $\frac{G_N}{\mu M^2}\sim \O(1)$. Repeating the argument of the preceding section, however for $\eta\rightarrow 1$, we obtain 
\begin{bBox}
\be\label{gravity:bound3}
\frac{\lambda_4^2}{\lambda_6}\le \frac{\tilde{N}_D(\Delta)}{ \tilde{N}_D(\Delta+1)} \(\lambda_2 +\frac{\pi G_N R_{\rm AdS}^2}{\mu}N_D(\Delta)\)
\ee
\end{bBox}
where, $N$-coefficients are given in appendix \ref{App:N}. One can check that the pre-factor $\frac{\tilde{N}_D(\Delta)}{ \tilde{N}_D(\Delta+1)}>1$ and asymptotes to 1 for large $\Delta$. Interestingly the ratio $\frac{\tilde{N}_D(\Delta)}{ \tilde{N}_D(\Delta+1)}$ is independent of the spacetime dimension $D$. 

\subsection{Summary of Bounds}
Let us now summarize the results of this section. The EFT (\ref{EFT:gravity}) has a well behaved CFT dual with  $\Delta_{\rm gap}\gg 1$, if and only the EFT, with $\mu\ge 0$, has  the following properties ($D\ge 4$):
\begin{enumerate}
\item\label{condition:gravity1} {Conditions \ref{condition1}-\ref{condition5} are satisfied for all even $k\ge 4$ , }
\item \label{condition:gravity2}{$ \lambda_2$ is bounded from below by the relation (\ref{gravity:bound1}),}
\item \label{condition:gravity3} {$\lambda_4$ is bounded from above  by the relation (\ref{gravity:bound2}), }
\item \label{condition:gravity4}{$\lambda_2 , \lambda_4$, and $\lambda_6$ must satisfy the convexity condition (\ref{gravity:bound3}).}
\end{enumerate}
Therefore, presence of gravity makes the EFT bounds weaker. 

\subsection{Flat Space Limit}

Finally, let us make a few comments about the flat space limit of the above bounds. Clearly, the constraint (\ref{condition:gravity1}) persists even in the flat space limit. These constraints, in flat space, are consistent with the bounds of \cite{Caron-Huot:2021rmr}.\footnote{It would be interesting to extend our analysis and compare with more recent results (such as \cite{Hatefi:2021czu}) on all order higher derivative couplings in different string theories.}

On the other hand, constraints (\ref{condition:gravity2})-(\ref{condition:gravity4}) do not produce precise bounds for the flat space EFT. For example, consider the  condition (\ref{condition:gravity2}) in the flat space limit. As discussed before, the flat space limit should be taken such that $\frac{1}{\sqrt{G_N M^{D-2}}}\gg MR_{\rm AdS}\gg 1$. Therefore, the condition (\ref{condition:gravity2}), in the flat space limit, suggests that $\mu \lambda_2 M^D > -\varepsilon$, where $\varepsilon$ is some small number.  This is certainly consistent with the results of \cite{Caron-Huot:2021rmr} (for $D\ge 5$), however, we do not have a precise definition of $\varepsilon$. This perhaps suggests that $\varepsilon$ is theory dependent. Nevertheless, the important point is that $\varepsilon$ is strictly positive.

\section{Multiple Scalar Fields in AdS}\label{sec:twofields}
In this section, we analyze EFTs of multiple scalars in AdS. The main motivation for this section is to demonstrate that there are additional constraints from the same CFT consistency conditions  that must be satisfied when there are multiple fields. As we showed earlier, odd $k$ interactions with a single scalar field are not constrained from our CFT analysis. However, in this section we will consider higher derivative interactions with multiple scalar fields to demonstrate that some odd $k$ interactions are constrained from the CFT consistency conditions of section \ref{sec:cft}. 

Furthermore, for multiple fields there are interference effects that are also constrained by the same CFT consistency conditions.  These interference effects have been utilized in \cite{Kundu:2020bdn} to derive non-linear bounds on the dilaton-axion effective action associated with 4D RG flows with global symmetry breaking. In this section, we will derive such interference bounds in a systematic way.   

For the purpose of demonstration of the general idea, we choose a simple theory: an EFT of two scalar fields (with the same mass) in AdS with $\mathbb{Z}_2$ symmetry and without gravity $G_N=0$. We follow the convention of section \ref{sec:FEA} and start with the effective action\footnote{Note that we are ignoring $k=0,1$-interactions. These interactions as well as any other interaction that cannot be written in the form (\ref{EFT:2phi}), if present, will not affect the bounds obtained in this section.}
\begin{align}\label{EFT:2phi}
S[\phi_1,\phi_2]&=\frac{1}{2} \int d^{D} x \sqrt{-g} \left(- g^{\mu\nu}\nabla_\mu \phi_1 \nabla_\nu\phi_1- g^{\mu\nu}\nabla_\mu \phi_2 \nabla_\nu\phi_2 -m^2 (\phi_1^2+\phi_2^2)\)\nonumber\\
&+\frac{\mu}{2}\sum_{k=2}^\infty \int \frac{d^{D} x \sqrt{-g}}{n_k(\Delta) M^{2(k-2)}} \(\lambda_k^{(1)}\phi_1^2 \Box^k \phi_1^2+\lambda_k^{(2)}\phi_2^2 \Box^k \phi_2^2+g_k\phi_1^2 \Box^k \phi_2^2+\tilde{g}_k\phi_1\phi_2 \Box^k \phi_1\phi_2\right) \nonumber\\
&+\cdots\ ,
\end{align}
where $M=\Delta_{\rm gap}/R_{\rm AdS}$ is the scale of new physics, $\mu\ge 0$, and the numerical factor $n_k(\Delta)$ is defined in (\ref{eq:gamma}).\footnote{Let us recall that in our convention $n_2(\Delta)=n_4(\Delta)=1$ for all $\Delta$ and $D$. Moreover, for $\Delta_{\rm gap}\gg m R_{\rm AdS}\gg k$, we have $n_k=1$. } Note that $\lambda$ and $g$ coefficients are dimensionless.
The argument of the previous sections still holds implying that both $\lambda_k^{(1)}$ and $\lambda_k^{(2)}$ must satisfy conditions \ref{condition1}-\ref{condition5} independently. However, as we will show in this section, there are additional non-trivial constraints that involve $g_k$ and $\tilde{g}_k$ couplings.

The AdS theory (\ref{EFT:2phi}) is dual to an interacting CFT in $d=D-1$ dimensions. The bulk fields $\phi_1$ and $\phi_2$ are dual to two scalar operators $\O_1$ and $\O_2$ respectively, with dimensions $m^2 R_{\rm AdS}^2=\Delta(\Delta-d)$. The two point functions are given by (\ref{eq:2pt}). Let us now consider a general four-point CFT correlator 
\be
G(\eta,\sigma)=\frac{\langle \O_B(\mathbf{1}) \O_A(\rr)\O_A^\dagger(-\rr)\O_B^\dagger(-\mathbf{1}) \rangle}{ \langle \O_B(\mathbf{1}) \O_B^\dagger (-\mathbf{1}) \rangle \langle  \O_A (\rr) \O_A^\dagger(-\rr) \rangle} 
\ee
in the Regge limit (\ref{regge}). The operators are defined as
\be
\O_A=\O_1+ a \O_2\ , \qquad \O_B=\O_1+ b \O_2\ ,
\ee
where, $a$ and $b$ are arbitrary complex numbers. We repeat the calculation of section \ref{sec:EFT_AdS} and obtain  an expression for $c_L(\eta)$ in the limit of large $MR_{\rm AdS}$ (with $mR_{\rm AdS}$ fixed):
\begin{align}\label{2field:cL}
c_L(\eta)=&\frac{\kappa_\Delta \mu F_{2\Delta+L}(\eta)}{n_L(\Delta)\Delta_{\rm gap}^{D+2L-4}(1+|a|)^2(1+|b|)^2}\nonumber\\
&\times\(\lambda_L^{(1)}+\lambda_L^{(2)}|a|^2|b|^2+\frac{1}{4}\tilde{g}_L\(|a|^2+|b|^2\)+\frac{1}{8}(2g_L+\tilde{g}_L)(a+a^*)(b+b^*)\)
\end{align}
for even $L\ge 2$, where $\kappa_\Delta$ is a positive coefficient independent of $L$, as defined in equation (\ref{def:kappa}). On the other hand, $c_L(\eta)$ for odd $L$ is non-zero. In particular for odd $L\ge 3$ we obtain
\begin{align}\label{2field:cL2}
c_L(\eta)=&\frac{\kappa_\Delta \mu F_{2\Delta+L}(\eta)}{n_L(\Delta)\Delta_{\rm gap}^{D+2L-4}(1+|a|)^2(1+|b|)^2}\frac{1}{8}(\tilde{g}_L-2g_L)(a-a^*)(b-b^*)\ .
\end{align}

\subsection{Bounds}
We are now in a position to derive bounds by utilizing the CFT consistency conditions of section \ref{sec:cft}. All CFT conditions of section \ref{sec:cft} apply to $c_L(\eta)$ obtained in this section for $0<\eta<1$ and all choices of $a$ and $b$. As we have discussed before, it is sufficient to derive constraints at the limit $\eta\rightarrow 0$. However, now the bounds will also depend on the particular choice of $a$ and $b$. 

Our CFT setup, as we discussed before, is probing local high energy scattering deep in the bulk. Since the local high energy scattering is insensitive to the spacetime curvature, the AdS bounds of this section remain valid even in the flat space limit. 

\subsubsection{Positivity for even $k$}
The condition (\ref{bound1}) now imposes 
\be
\lambda_k^{(1)}> 0\ , \qquad \lambda_k^{(2)}> 0\ , \qquad \tilde{g}_k>0
\ee
for even $k\ge 2$ generalizing the bound \ref{condition1}. Furthermore, now we can derive a non-linear interference bound by choosing $a$ and $b$ that minimize (\ref{2field:cL}), yielding 
\be
|\tilde{g}_k+2g_k|\le 4 \sqrt{\lambda_{k}^{(1)} \lambda_{k}^{(2)}}+\tilde{g}_{k}
\ee
for all even $k\ge 2$. Note that the above bounds are consistent with bounds obtained in \cite{Kundu:2020bdn} on the dilaton-axion effective action.

\subsubsection{Monotonicity for even $k$}
The condition (\ref{bound2}) leads to the following monotonicity conditions:
\be
\lambda_k^{(1)}\ge \lambda_{k+2}^{(1)}\ , \qquad \lambda_k^{(2)}\ge \lambda_{k+2}^{(2)}\ , \qquad \tilde{g}_k\ge \tilde{g}_{k+2}
\ee
for all even $k\ge 2$ generalizing the bound \ref{condition3}. We can again derive a non-linear interference bound by optimizing with respect to $a$ and $b$:
\be
|(\tilde{g}_k-\tilde{g}_{k+2})+2(g_k-g_{k+2})|\le 4\sqrt{(\lambda_k^{(1)}-\lambda_{k+2}^{(1)})(\lambda_k^{(2)}-\lambda_{k+2}^{(2)})}+(\tilde{g}_k-\tilde{g}_{k+2})
\ee
for all even $k\ge 2$.

\subsubsection{Boundedness for odd $k$}
The condition (\ref{bound2}) now imposes bounds also on odd $k$ coupling constants. By optimizing with respect to $a$ and $b$, we find that 
\be
|\tilde{g}_k-2g_k|\le 4 \sqrt{\lambda_{k-1}^{(1)} \lambda_{k-1}^{(2)}}+\tilde{g}_{k-1}
\ee
for all odd $k\ge 3$. Note that there is a particular combination of interactions for any odd $k$ which is not bounded from our argument.

\subsubsection{Log-convexity for even $k$}
To begin with, we can utilize (\ref{bound3}) for different limits of $a$ and $b$ to obtain local log-convexity conditions: $\lambda_{k+2}^{(1)}\le \sqrt{\lambda_{k}^{(1)}\lambda_{k+4}^{(1)}}$, $\lambda_{k+2}^{(2)}\le \sqrt{\lambda_{k}^{(2)}\lambda_{k+4}^{(2)}}$, and $\tilde{g}_{k+2}\le \sqrt{\tilde{g}_k \tilde{g}_{k+4}}$ for all even $k\ge 2$. These local conditions lead to the global log-convexity condition \ref{condition5} for $\lambda_{k}^{(1)}$, $\lambda_{k}^{(2)}$, and $\tilde{g}_k$ individually for even $k\ge 2$. Furthermore, there is again a more general local log-convexity condition:
\be\label{C:bound}
C_{k+2}(a,b)^2\le C_{k}(a,b)C_{k+4}(a,b) \ , \qquad \text{even}\ k\ge 2
\ee
for all real $a$ and $b$, where 
\be
C_{k}(a,b)=4\lambda_k^{(1)}+4\lambda_k^{(2)}a^2 b^2+\tilde{g}_k\(a+b\)^2+4g_ka b>0\ .
\ee
Of course, we can again write a global log-convexity condition for $C_{k}(a,b)$ as before. 

Note that the strongest bound can be obtained by optimizing (\ref{C:bound}) with respect to $a$ and $b$. The actual expression is not very illuminating and hence we will not transcribe it here.

\subsubsection{Log-convexity for odd $k$}
Odd $k$-interactions also obey a local (but not global) log-convexity condition. This can be obtained by using (\ref{bound4}):
\be
\(\tilde{g}_k-2g_k\)^2 \le \frac{1}{y^2}\(2\lambda_{k-1}^{(1)}+2\lambda_{k-1}^{(2)} y^2+ \tilde{g}_{k-1} y\)\(2\lambda_{k+1}^{(1)}+2\lambda_{k+1}^{(2)} y^2+ \tilde{g}_{k+1} y\)
\ee
for  all odd $k\ge 3$ and $0<y<\infty$. Of course, the optimal bound is obtained by minimizing the right hand side with respect to $y$. 

~\\

Finally, we wish to note that now the fields can couple to a massive or a massless gauge field. However, that will not alter equations (\ref{2field:cL}) or (\ref{2field:cL2}) and hence the bounds remain unchanged. On the other hand, when we couple the theory (\ref{EFT:2phi}) to gravity, as we discussed in the previous section, it will contribute to $c_2(\eta)$. So, all bounds for $k\ge 4$ (even or odd) are valid even when $G_N\neq 0$. 

\subsection{Application: Complex Scalar Field}
Let us now consider a complex scalar field
\begin{align}\label{action_sc}
S= \frac{1}{2} \int d^{D} x \sqrt{-g}\left(- \nabla_\mu \phi^\dagger \nabla^\mu\phi+m^2 \phi\phi^\dagger+\sum_{k=0}^\infty \frac{\mu}{n_k(\Delta) M^{2(k-2)}}\left(\alpha_k\phi^2 \Box^k {\phi^\dagger}^2+\beta_k\phi\phi^\dagger \Box^k \phi\phi^\dagger\right)\right)\ .
\end{align}
Results of this section apply to this EFT as well. In particular, bounds on this EFT can be obtained easily once we identify 
\be
\lambda_k^{(1)}=\lambda_k^{(2)}=\alpha_k+\beta_k\ , \qquad g_k=2(\beta_k-\alpha_k)\ , \qquad \tilde{g}_k=4\alpha_k\ .
\ee 


\section{Conclusions \& Comments}\label{sec:conclusions}
In this paper we addressed the question of what EFTs in AdS$_D$ cannot be embedded into a UV theory that is dual to a CFT$_{D-1}$ obeying the usual CFT axioms. We considered EFTs of scalar fields in AdS spacetime of large radius and derived precise constraints (\ref{intro:condition1})-(\ref{intro:condition3}) on the coupling constants of higher derivative interactions $\phi^2\Box^k\phi^2$ from the dual CFT.  Our
derivation of the bounds does not make any assumptions about the dual CFT beyond the well established  conformal bootstrap axioms. Furthermore, we showed that inclusion of gravity only affects constraints involving the $\phi^2\Box^2\phi^2$ interaction which now can have a negative coupling constant even in $D=4$. It is unclear whether this fact survives in the exact flat space limit. It will be interesting to explore this further since positivity of this interaction  is essential in the proof of the 4D $a$-theorem.

Our CFT setup was a Lorentzian four-point correlator in the Regge limit which was designed to probe local high energy scattering deep in the AdS. We utilized the fact that the growth of this CFT Regge correlator is highly constrained from  the argument of \cite{Kundu:2020gkz}. Conceptually, bounds obtained in this paper are closely related to the CFT Nachtmann theorem of \cite{KomargodskiZhiboedov,Kundu:2020gkz}. In fact, the CFT Nachtmann theorem was derived in \cite{Kundu:2020gkz} by starting from the same four-point correlator, however, in the Lorentzian lightcone limit ($\eta\rightarrow 0$, then $\sigma\rightarrow 0$). Moreover, the condition (\ref{intro:condition1}) can be derived from the CFT Nachtmann theorem (with some caveat, as we explain later) once we identify anomalous dimensions $\gamma_{n,\ell}$ of double-trace operators $[\O\O]_{n,\ell}$ are related to $\lambda_\ell$ (for even $\ell$) as follows  \cite{Heemskerk:2009pn,Kraus:2020nga}
\be
\gamma_{n,\ell} \propto - \lambda_\ell\ .
\ee
On the other hand, constraints (\ref{intro:condition2})-(\ref{intro:condition3}) are strictly stronger than what one obtains from the Nachtmann theorem.\footnote{This is a direct consequence of the fact that the order of limits $\eta,\sigma\rightarrow 0$ is non-trivial.} Furthermore, one should exercise caution while applying the Nachtmann theorem to an ``effective" CFT  which is defined order by order in perturbation theory. Of course, even for such a CFT the Nachtmann theorem of \cite{KomargodskiZhiboedov,Kundu:2020gkz} does hold, however, identifying families of minimal twist operators can be subtle. It is particularly complicated when the family of minimal twist operators consists of different set of operators at different orders in perturbation theory. We emphasize that for ``effective" CFTs constraints obtained from the CFT Regge limit are more reliable since they follow directly from the  CFT sum-rule (\ref{sum}).

Finally, we end with some general comments about the swampland bounds on EFTs in flat space which are obtained by using various properties of 4-point scattering amplitudes. We mainly focus on two types of flat space arguments: (i) based on dispersive sum rules, (ii) based on positivity of the eikonal phase-shift. The first type of arguments, as explained in the introduction, lead to precise bounds, however, require some assumption about the Regge boundedness of the 4-point amplitude. Whereas,   positivity of the eikonal phase-shift seems to be a more rigorous condition \cite{Camanho:2014apa}\footnote{For some subtleties see \cite{Hollowood:2015elj,deRham:2020zyh,Alberte:2020jsk}.} which leads to non-trivial constraints \cite{Camanho:2014apa,Edelstein:2016nml,Camanho:2016opx,Hinterbichler:2017qcl, Bonifacio:2017nnt, Afkhami-Jeddi:2018apj,Chowdhury:2018nfv, Kaplan:2019soo,Kaplan:2020ldi,Kaplan:2020tdz}, however,  these constraints in some sense are parametric in nature.  On the other hand, by now it is known that both types of bounds can be obtained in AdS from the same CFT sum-rule (\ref{sum}). So, roughly speaking our CFT Regge correlator (\ref{corr}) is the AdS analogous of the flat space finite impact parameter scattering amplitude $\mathcal{A}(s,\vec{b})$ of \cite{Caron-Huot:2020cmc}, since both capture two types of constraints described above. It would be interesting to derive the full set of constraints of \cite{Caron-Huot:2021rmr} by viewing a flat space EFT as the flat space limit of the EFT in AdS.

More generally, it would be nice to unify the flat space bounds and the AdS bounds in a more systematic way. This can be achieved at the level of individual bounds, however a more useful goal would be to rigorously derive the Regge boundedness condition (and the closely related CRG condition of \cite{Chowdhury:2019kaq,Chandorkar:2021viw}) of the flat space amplitude directly from the CFT axioms by taking the flat space limit.\footnote{Note that the flat space limit is subtle when gravity is dynamical.} It is tempting to translate the results of this paper in to  a Regge boundedness condition for the flat space finite impact parameter scattering amplitude $\mathcal{A}(s,\vec{b})$  of \cite{Caron-Huot:2021rmr}  (or some variation of it) for arbitrary external states. In particular, CFT conditions  of section \ref{sec:cft} suggest
\begin{quote}
{\it Any finite impact parameter scattering amplitude $\mathcal{A}(s,\vec{b})$ for large $s$ cannot grow faster than $s^2$ within any range of $s$. }
\end{quote}
In other words, there can be terms in $\mathcal{A}(s,\vec{b})$ that grow as $s^3, s^4, \cdots$ for large $s$ but none of them can dominate within any range of $s$. Classical version of this statement is very similar to the CRG conjecture of \cite{Chowdhury:2019kaq}, however, it is not equivalent since $\mathcal{A}(s,\vec{b})$ is in the impact parameter space. Note that this Regge boundedness condition, even if true, is weaker than what is required in \cite{Caron-Huot:2021rmr}. Nevertheless, it is of importance to have a rigorous proof of the above Regge boundedness condition or some stronger version of it. 

\section*{Acknowledgments}

It is my pleasure to thank Simon Caron-Huot, Subham Dutta Chowdhury, Tom Hartman, Jonathan Heckman, Jared Kaplan, Suman Kundu, Shiraz Minwalla, and Hao Zhang for several helpful discussions over the years. I thank  Dalimil Mazac, Joao Penedones, Leonardo Rastelli, and David Simmons-Duffin for helpful correspondence. I would also like to thank Madhuparna Pal for all the support during the Covid pandemic. I was supported in part by the Simons Collaboration Grant on the Non-Perturbative Bootstrap.

\begin{appendix}
\section{Rindler Positivity from Conformal Bootstrap}\label{app:rindler}
In this appendix we will show that Rindler positivity, for scalar external operators, follows from OPE unitarity and crossing symmetry. 
Consider the Euclidean correlator ($0<\rho, \bar{\rho}<1$)
\begin{align}
\langle O_2(-\rr) O_1(\rr) O_1^\dagger(\mathbf{1})O_2^\dagger(-\mathbf{1})\rangle_E=\frac{1}{(16 \rho \bar{\rho})^{\frac{\Delta_1+\Delta_2}{2}}} \left( \frac{(1-\rho)(1-\bar{\rho})}{(1+\rho)(1+\bar{\rho})}\right)^{\Delta_{12}}\nonumber\\
\times\sum_p c_{O_2 O_1 p}c_{O_2^\dagger O_1^\dagger p}(-1)^\ell g^{\Delta_{21},\Delta_{12}}_{\Delta,\ell}(z,\bz)
\end{align}
where $\Delta_{12}=\Delta_1-\Delta_2$ and cross-ratios are
\be
z=\frac{4\rho}{(1+\rho)^2}\ , \qquad \bz=\frac{4\bar{\rho}}{(1+\bar{\rho})^2}\ .
\ee
Unitarity ensures that $c_{O_2 O_1 p}c_{O_2^\dagger O_1^\dagger p}>0$. Moreover, positivity of the conformal block expansion in $\r,\br$ now implies  
\be
\langle O_2(-\rr) O_1(\rr) O_1^\dagger(\mathbf{1})O_2^\dagger(-\mathbf{1})\rangle_E=\frac{1}{(16 \rho \bar{\rho})^{\frac{\Delta_1+\Delta_2}{2}}}\sum_{h,\bh} b_{h,\bh}\rho^h \bar{\rho}^{\bh}\ ,  \qquad b_{h,\bh}\ge 0\ 
\ee
where  $h=\frac{1}{2}(\Delta\pm \ell)$ and $\bh=\frac{1}{2}(\Delta\mp \ell)$. The sum is over all operators both primaries and their descendants. Note that $b_{h,\bh}\ge 0$ also follows from  reflection positivity, as shown in \cite{Hartman:2015lfa}. The above facts immediately implies that for $1>z,\bz>0$
\begin{align}\label{pos_exp}
\sum_p &c_{O_2 O_1 p}c_{O_2^\dagger O_1^\dagger p}(-1)^\ell g^{\Delta_{21},\Delta_{12}}_{\Delta,\ell}(z,\bz)\nonumber\\
&=\left((1-z)(1-\bz)\right)^{\Delta_{21}/2}\sum_{h,\bh} b_{h,\bh}\left(\frac{1-\sqrt{1-z}}{1+\sqrt{1-z}} \right)^h \left(\frac{1-\sqrt{1-\bz}}{1+\sqrt{1-\bz}} \right)^{\bh}\ .
\end{align}

\subsection*{Rindler Positivity}
We now consider the correlator $G$ of equation (\ref{corr}), however, in the Euclidean regime ($0<\rho, \bar{\rho}<1$). In the direct channel expansion
\be
G_E=\sum_p c_{O_1O_1^\dagger p}c_{O_2^\dagger O_2 p}g^{0,0}_{\Delta,\ell}(z,\bz)
\ee
with
\be
z=\frac{4\rho}{(1+\rho)^2}\ , \qquad \bz=\frac{4\bar{\rho}}{(1+\bar{\rho})^2}\ .
\ee
The subscript $E$ is there to remind ourselves that we are in the Euclidean regime. Positivity of this correlator is not obvious from the direct channel expansion. So, we expand in the crossed channel
\begin{align}
G_E=\frac{(16\r\br)^{\Delta_2}}{\left((1-\rho)(1-\bar{\rho}) \right)^{\Delta_1+\Delta_2}} \left( \frac{1}{(1+\rho)(1+\bar{\rho})}\right)^{\Delta_{21}}\sum_p c_{O_2 O_1 p}c_{O_2^\dagger O_1^\dagger p}(-1)^\ell g^{\Delta_{21},\Delta_{12}}_{\Delta,\ell}(z,\bz)
\end{align}
where now cross-ratios are
\be
z=\frac{(1-\rho )^2}{(1+\rho )^2}\ , \qquad \bz=\frac{(1-\bar{\rho })^2}{(1+\bar{\rho })^2}\ .
\ee
Using the positive expansion (\ref{pos_exp}), we can write 
\begin{align}
G_E=\frac{\(16\r\br\)^{\frac{\Delta_1+\Delta_2}{2}}}{\left((1-\rho)(1-\bar{\rho}) \right)^{\Delta_1+\Delta_2}}
\sum_{h,\bh} b_{h,\bh}\left(\frac{1-\sqrt{\rho}}{1+\sqrt{\rho}} \right)^{2h} \left(\frac{1-\sqrt{\bar{\rho}}}{1+\sqrt{\bar{\rho}}} \right)^{2\bh}
\end{align}
which is positive for $0<\rho,\bar{\rho}<1$.

\subsubsection*{Lorentzian Correlators}
Rindler positivity is most useful in the Lorentzian regime $\rho>1$ and $0<\br<1$. So, we now consider this regime where some of the operators are time-like separated and hence operator ordering does matter. The positive ordered correlator $G_0$, as defined in (\ref{corr0}), in the Lorentzian regime ($\rho>1$ and $0<\br<1$) is given directly by the Euclidean correlator and hence 
\be\label{app:eq:G0}
G_0=\frac{\(16\r\br\)^{\frac{\Delta_1+\Delta_2}{2}}}{\left(\rho(1-1/\rho)(1-\bar{\rho}) \right)^{\Delta_1+\Delta_2}}
\sum_{h,\bh} b_{h,\bh}\left(\frac{1-1/\sqrt{\rho}}{1+1/\sqrt{\rho}} \right)^{2h} \left(\frac{1-\sqrt{\bar{\rho}}}{1+\sqrt{\bar{\rho}}} \right)^{2\bh}\ge 0\ .
\ee
This establishes Rindler positivity in the Lorentzian regime. 

This leads to the other Lorentzian correlator $G$, as defined in (\ref{corr}) and another distinct Lorentzian correlator that we can define 
\be\label{corr:app}
G=\frac{\langle O_2(\mathbf{1}) O_1(\rr)O_1^\dagger(-\rr)O_2^\dagger(-\mathbf{1}) \rangle}{ \langle O_2(\mathbf{1}) O_2^\dagger(-\mathbf{1}) \rangle \langle  O_1(\rr)O_1^\dagger(-\rr) \rangle} \ , \qquad \tilde{G}=\frac{\langle  O_1(\rr) O_2(\mathbf{1})O_2^\dagger(-\mathbf{1}) O_1^\dagger(-\rr)\rangle}{ \langle O_2(\mathbf{1}) O_2^\dagger(-\mathbf{1}) \rangle \langle  O_1(\rr)O_1^\dagger(-\rr) \rangle} \ .
\ee
These Lorentzian correlators, in the regime $\rho>1$ and $0<\br<1$, are obtained from analytic continuations of the Euclidean correlator  
\begin{align}
\tilde{G}=\frac{\(16\r\br\)^{\frac{\Delta_1+\Delta_2}{2}}}{\left(\rho(1-1/\rho)(1-\bar{\rho}) \right)^{\Delta_1+\Delta_2}}\sum_{h,\bh} b_{h,\bh}\left(\frac{1-1/\sqrt{\rho}}{1+1/\sqrt{\rho}} \right)^{2h} \left(\frac{1-\sqrt{\bar{\rho}}}{1+\sqrt{\bar{\rho}}} \right)^{2\bh}e^{i\pi(2h-\Delta_\psi-\Delta_O)}
\end{align}
and similarly 
\begin{align}\label{app:eq:G}
G=\frac{\(16\r\br\)^{\frac{\Delta_1+\Delta_2}{2}}}{\left(\rho(1-1/\rho)(1-\bar{\rho}) \right)^{\Delta_1+\Delta_2}}\sum_{h,\bh} b_{h,\bh}\left(\frac{1-1/\sqrt{\rho}}{1+1/\sqrt{\rho}} \right)^{2h} \left(\frac{1-\sqrt{\bar{\rho}}}{1+\sqrt{\bar{\rho}}} \right)^{2\bh}e^{-i\pi(2h-\Delta_\psi-\Delta_O)}\ .
\end{align}
From the above expansions, we conclude that the Lorentzian correlators $G_0$, $G$, and $\tilde{G}$, in the regime $\rho>1$ and $0<\br<1$, obey the following properties: 
\begin{align}
& G_0 \ge 0\ ,\qquad  G=\tilde{G}^*\ ,\\
& |G|\le G_0\ ,\qquad |\tilde{G}|\le G_0\ .
\end{align}
\section{A Sum-Rule by Subtracting the Identity Operator}\label{app:subtract}
In this section, we derive a sum-rule similar to (\ref{sum}) by subtracting the identity operator from all channels.  This discussion is only important when $O_1=O_2=O$ in the correlator (\ref{corr0}) with $\Delta_1=\Delta_2=\Delta$. We will restrict to real scalar operators, however, this discussion can be easily generalized for complex scalars. 

In this case, we write (\ref{app:eq:G0}) and (\ref{app:eq:G}) as
\be\label{app:eq:G0_2}
G_0=\frac{\(16\r\br\)^{\Delta}}{\rho^{2\Delta}\left((1-1/\rho)(1-\bar{\rho}) \right)^{2\Delta}}\(1+
\sum_{h,\bh\neq 0} b_{h,\bh}\left(\frac{1-1/\sqrt{\rho}}{1+1/\sqrt{\rho}} \right)^{2h} \left(\frac{1-\sqrt{\bar{\rho}}}{1+\sqrt{\bar{\rho}}} \right)^{2\bh}\)
\ee
and 
\begin{align}\label{app:eq:G_2}
G=\frac{\(16\r\br\)^{\Delta}e^{2\pi i \Delta }}{\rho^{2\Delta}\left((1-1/\rho)(1-\bar{\rho}) \right)^{2\Delta}}\(1+\sum_{h,\bh\neq 0} b_{h,\bh}\left(\frac{1-1/\sqrt{\rho}}{1+1/\sqrt{\rho}} \right)^{2h} \left(\frac{1-\sqrt{\bar{\rho}}}{1+\sqrt{\bar{\rho}}} \right)^{2\bh}e^{-2\pi i h}\)
\end{align}
by isolating the contribution from the identity operator. We can compare these correlators with correlators for the CFT which is dual to a free scalar theory in AdS. In this generalized free CFT, the corresponding correlators are
 \be\label{app:eq:G0_free}
G_0^{free}=1+\frac{\(16\r\br\)^{\Delta}}{\rho^{2\Delta}\left((1-1/\rho)(1-\bar{\rho}) \right)^{2\Delta}}+\frac{\(16\r\br\)^{\Delta}}{\rho^{2\Delta}\left((1+1/\rho)(1+\bar{\rho}) \right)^{2\Delta}}
\ee
and
\begin{align}\label{app:eq:G_2:free}
G^{free}=1+\frac{\(16\r\br\)^{\Delta}e^{2\pi i \Delta }}{\rho^{2\Delta}\left((1-1/\rho)(1-\bar{\rho}) \right)^{2\Delta}}+\frac{\(16\r\br\)^{\Delta}}{\rho^{2\Delta}\left((1+1/\rho)(1+\bar{\rho}) \right)^{2\Delta}}\ .
\end{align}
These two correlators of the generalized free theory are different only when $\Delta$ is not an integer. We now define subtracted correlators:
\be
\delta G_0(\eta,\sigma)=G_0-G_0^{free}\ , \qquad \delta G(\eta,\sigma)=G-G^{free}\ ,  
\ee
where $\eta$ and $\sigma$ are defined in (\ref{sigma}).  Moreover, note that for positive $|\sigma|<1$ and $0<\eta<1$
\be
\mbox{Re}\(\delta G_0(\eta,\sigma)-\delta G(\eta,\sigma)\)=\sum_{h,\bh\neq 0} b_{h,\bh}\left(\frac{1-\sqrt{\sigma}}{1+\sqrt{\sigma}} \right)^{2h} \left(\frac{1-\sqrt{\eta \sigma}}{1+\sqrt{\eta \sigma}} \right)^{2\bh}\(1-\cos(2\pi h)\)\ge 0
\ee
which follows from $b_{h,\bh}\ge 0$. This positivity is true for all unitary CFTs. For negative $|\sigma|<1$, the same positivity condition can be derived by starting from $\tilde{G}$ correlator at positive $\sigma$.

Now we can perform a contour integral on the complex lower-half $\sigma$-plane, as described in \cite{Kundu:2020gkz}. This now yields a modified sum-rule for the expansion (\ref{con_anc})
\be\label{sum:new}
c_L(\eta)=\frac{1}{\pi}\int_{-R}^R d\sigma\ \sigma^{L-2}\mbox{Re}\(\delta G_0(\eta,\sigma)-\delta G(\eta,\sigma)\)\ , \qquad \sigma_* \le R\ll \eta<1\ ,
\ee
which is a more formal (and precise) version of the sum-rule (\ref{sum}). 

The above sum-rule has one key advantage. In order to illustrate that we focus on CFTs that are dual to some EFT in AdS. Clearly these subtracted correlators come entirely from the interacting part of the AdS EFT.  The possible corrections to the above sum-rule comes from terms 
\be\label{app:correction_sum}
\(\delta G_0(\eta,\sigma)-\delta G(\eta,\sigma)\) \sim (\delta c) \sigma^{a} \qquad \text{with} \qquad a\ge d
\ee
where $\delta c$ is obtained entirely from the interacting part of the EFT. Hence, the entire argument of section \ref{sec:correction} about the correction terms now can be repeated  implying  that the consistency conditions (\ref{bound1}), (\ref{bound2}), (\ref{bound3}), and (\ref{bound4}) are valid even when $\Delta$ is non-integer. 

The observant reader may have noticed that the correlator $\delta G_0(\eta,\sigma)$, in general, is not a well-defined object on the the complex lower-half $\sigma$-plane. However, we can always define a function $\delta G^{(-)}_0(\eta,\sigma)$ which is analytic on the lower half $\sigma$-plane (minus the real line) and has the property $\mbox{Re} \delta G^{(-)}_0(\eta,\sigma)= \delta G_0(\eta,\sigma)$ on the real line ($\mbox{Im}\ \sigma \rightarrow 0_-$). For example, $\delta G_0(\eta,\sigma)$, in the limit $\sigma\rightarrow 0$,  has terms like
\be
\delta G_0(\eta,\sigma)\sim c_a |\sigma|^a
\ee
with positive $a$. We can define $\delta G^{(-)}_0(\eta,\sigma)$ as a function on the lower-half $\sigma$ plane with terms 
\be
\delta G^{(-)}_0(\eta,\sigma) \sim c_a \(1+i \tan\(\frac{\pi a}{2}\)\) \sigma^a
\ee
and derive the sum-rule (\ref{sum:new}) using $\delta G^{(-)}_0(\eta,\sigma)$. Clearly, any additional correction that can occur because of $\delta G^{(-)}_0(\eta,\sigma)$ will also obey (\ref{app:correction_sum}) and hence the sum-rule (\ref{sum:new}) is valid for CFTs dual to any AdS EFT for $D\ge 4$. The sum-rule is valid even for $D=3$ as long as $0\le m^2\ll M^2$ and the $\phi^3$ interaction is absent. 
\section{Correlators of CFTs Dual to EFTs in AdS}\label{gkpw}
Derivation of our bounds depends heavily on determining the exact numerical factors. So, we review the computation of correlators in the AdS/CFT correspondence. The tree level Witten diagrams can  be obtained from the Euclidean on-shell action:
\be
e^{-S_{\rm on-shell}[\Phi]}=\langle e^{\int \Phi \O}\rangle\ ,
\ee
where $\Phi$ is the boundary value of the bulk field $\phi$ with CFT dual $\O$. For simplicity we will work in the Euclidean signature with the metric
\be
ds^2=\frac{dz^2+\delta_{\mu\nu}dx^\mu dx^\nu}{z^2}\ .
\ee
We start with a single scalar field in AdS :
\be
S=\frac{1}{2}\int d^{d+1}x \sqrt{g} \left[g^{\mu \nu }\partial_\mu \phi \partial_\nu \phi+ m^2 \phi^2 \right]
\ee
which leads to the equation of motion
\be
(\Box-m^2)\phi=0\ .
\ee
\subsubsection*{Bulk-to-boundary propagator}
This has the solution
\be\label{AdS:bdb}
\phi(z,x)=C_\Delta \int d^d x' \frac{z^\Delta}{(z^2+|x-x'|^2)^\Delta}\Phi(x')\equiv  \int d^d x' K_\Delta(z,x;x')\Phi(x')
\ee
with $m^2=\Delta(\Delta-d)$ and
\be
C_\Delta=\frac{\ \Gamma[\Delta]}{\pi^{d/2}\Gamma[\Delta-d/2]}\ .
\ee
Note that the bulk to boundary propagator satisfies
\be
(\Box(z,x)-m^2)K_\Delta(z,x;x')=0\ .
\ee
Furthermore, note that
\be
K_\Delta(z\rightarrow 0,x;x')=z^{d-\Delta}\left(\delta^{d}(x-x')+\O(z^2)\right)+ z^\Delta \left( \frac{C_\Delta}{|x-x'|^{2\Delta}}+\O(z^2)\right)\ .
\ee

\subsubsection*{Bulk-to-bulk propagator}
The bulk-to-bulk propagator is defined as the solution of the differential equation
\be 
(\Box(z,x)-m^2)G_\Delta(z,x;z',x')=\frac{1}{\sqrt{g(z,x)}}\delta(z-z')\delta^d(x-x')\ .
\ee
The propagator can  be explicitly written as
\be
G_\Delta(z,x;z',x')=-\frac{2^{\Delta } \xi ^{\Delta } \Gamma (\Delta ) \Gamma \left(-\frac{d}{2}+\Delta +\frac{1}{2}\right) }{(4 \pi )^{\frac{d+1}{2}} \Gamma (-d+2 \Delta +1)}\, _2F_1\left(\frac{\Delta }{2},\frac{\Delta }{2}+\frac{1}{2};-\frac{d}{2}+\Delta +1;\xi ^2\right)
\ee
where,
\be
\xi=\frac{2z  z'}{z^2+{z'}^2+(x-x')^2}\ .
\ee
Let us also note the asymptotic behavior of the propagator 
\be
G_\Delta(z\rightarrow \epsilon,x;z',x')=- \frac{\epsilon^\Delta}{2\Delta -d}K_\Delta(z',x';x)\ .
\ee

\subsection{CFT 2-pt Functions}
The on-shell action is given by
\begin{align}
S_{\rm on-shell}&=-\frac{1}{2}\int_{z=\epsilon} d^d x \frac{1}{z^{d-1}}\phi(z,x) \partial_z \phi(z,x)\nonumber\\
&=-\frac{1}{4 \epsilon^{d-1}}\int_{z=\epsilon} d^d x \partial_z(\phi(z,x))^2 \ .
\end{align}
This on-shell action can be evaluated by using the asymptotic expression for the bulk-to-boundary propagator yielding\footnote{The following identity can be useful:
\be
\int d^d x \left(\frac{z}{z^2+x^2}\right)^{\Delta_1}\left(\frac{z}{z^2+|x-x'|^2}\right)^{\Delta_2}=\frac{\pi^{d/2}\Gamma(\Delta_1+\Delta_2-d/2)}{\Gamma(\Delta_1)\Gamma(\Delta_2)}\int_{0}^1 ds\ \frac{z^{\Delta_1+\Delta_2}s^{\Delta_2-1}(1-s)^{\Delta_1-1}}{\left(s(1-s){x'}^2+z^2\right)^{\Delta_1+\Delta_2-d/2}}\ .
\ee
} 
\be
S_{\rm on-shell}=- \frac{C_\Delta d}{2}\int d^d x d^d x' \frac{\Phi(x')\Phi(x)}{|x-x'|^{2\Delta}}-\frac{d-\Delta}{2 \epsilon^{2\Delta-d}}\int d^d x \Phi(x')^2\ .
\ee
The divergent term can be removed by adding a counter-term at $z=\epsilon$:
\be
S_{ct}=\frac{d-\Delta}{2 \epsilon^{d}}\int_{z=\epsilon} d^d x\ \phi(z,x)^2\ .
\ee
The on-shell counter-term also contributes a finite part
\be
S_{ct}=\frac{d-\Delta}{2 \epsilon^{2\Delta-d}}\int d^d x \Phi(x)^2+C_\Delta (d-\Delta)\int d^d x d^d x' \frac{\Phi(x)\Phi(x')}{|x-x'|^{2\Delta}}
\ee
and hence the total on-shell action becomes
\be
S_{\rm on-shell}=-\frac{(2\Delta-d)C_\Delta}{2}\int d^d x d^d x' \frac{\Phi(x')\Phi(x)}{|x-x'|^{2\Delta}}\ .
\ee
Finally the two-point function is
\be
\langle \O(x_1)\O(x_2) \rangle =\frac{(2\Delta-d)C_\Delta}{|x_1-x_2|^{2\Delta}}\ .
\ee

\subsection{Perturbative Expansion of the Euclidean On-Shell Action}
We now study the following Euclidean bulk action
\be
S=\int d^{d+1}x \sqrt{g} \left( \frac{1}{2}(\partial \phi)^2+ \frac{1}{2} m^2 \phi^2+L_{int}\right)\ .
\ee
The bulk equation of motion is now given by
\be
(\Box-m^2)\phi= \frac{\delta L_{int}}{\delta \phi}\ .
\ee
We can again write down a formal solution of the equation of motion
\be
\phi(x,z)=\int d^d x' K_\Delta(z,x;x')\Phi(x')+ \int d^{d}x' dz' \sqrt{g'} G_\Delta(z,x;z',x') \frac{\delta L_{int}}{\delta \phi}(z',x')
\ee
and the on-shell action is given by
\begin{align}
S_{\rm on-shell}=&-\frac{1}{4 \epsilon^{d-1}}\int_{z=\epsilon} d^d x  \partial_z(\phi(z,x))^2+\frac{d-\Delta}{2 \epsilon^{d}}\int_{z=\epsilon} d^d x\ \phi(z,x)^2\nonumber\\
&+\int d^{d}xdz \sqrt{g}\left(L_{int}-\frac{1}{2}\phi \frac{\delta L_{int}}{\delta \phi}\right)\ \\
&\equiv S_0+S_{ct}+S_{int}\ .
\end{align}
First, we find that
\begin{align}
S_0+S_{ct}=&-\frac{c_{\Delta} (2\Delta-d)}{2 }\int_{z=\epsilon}d^d x_1 d^d x_2  \frac{\Phi(x_1)\Phi(x_2)}{|x_1-x_2|^{2\Delta}}\nonumber\\
&  +\frac{1}{2 }\int_{z=\epsilon}d^d x_1 \int d^{d}x' dz' \sqrt{g} K_\Delta(x_1;z',x')\Phi(\vec{x}_1)\frac{\delta L_{int}}{\delta \phi}(z',x')\ .
\end{align}
So the total Euclidean on-shell action can be written in a nice form
\begin{align}\label{sk1}
S_{\rm on-shell}&=-\frac{c_{\Delta} (2\Delta-d)}{2 }\int_{z=\epsilon}d^d x_1 d^d x_2  \frac{\Phi(x_1)\Phi(x_2)}{|x_1-x_2|^{2\Delta}}+\int d^{d}xdz \sqrt{g} L_{int}(z,x)\nonumber\\
&-\frac{1}{2}\int d^{d}xdz \sqrt{g}\int d^{d}x' dz' \sqrt{g'} G_\Delta(z,x;z',x')\frac{\delta L_{int}}{\delta \phi}(z',x')\frac{\delta L_{int}}{\delta \phi}(z,x)\ ,
\end{align}
where, the bulk field $\phi$ should be understood as
\be\label{sk2}
\phi(x,z)=\int d^d x' K_\Delta(z,x;x')\Phi(x')+ \int d^{d}x' dz' \sqrt{g'} G_\Delta(z,x;z',x') \frac{\delta L_{int}}{\delta \phi}(z',x')\ .
\ee
We can use equation (\ref{sk2}) to perform a perturbative expansion of (\ref{sk1}). Note that contact diagrams receive contributions only from the second term in (\ref{sk1}). On the other hand, both the second and the third term can contribute to an exchange diagram. 

\subsection{Example}
Let us now consider the example
\be
S=\int d^{d+1}x \sqrt{g} \left( \frac{1}{2}(\partial \phi)^2+ \frac{1}{2} m^2 \phi^2+\lambda_3 \phi^3+\lambda_4 \phi^4\right)\ .
\ee
The bulk equation of motion is now given by
\be
(\Box-m^2)\phi=3\lambda_3 \phi^2+4\lambda_4 \phi^3\equiv \frac{\delta L_{int}}{\delta \phi}\ .
\ee

\subsubsection*{Three-point function}
We can now write down the cubic action by using (\ref{sk1}):
\be
S_{(3)}=\lambda_3\int d^{d}xdz \sqrt{g} \int   d^d x_1 d^d x_2 d^d x_3 K_\Delta(z,x;x_1)  K_\Delta(z,x;x_2)  K_\Delta(z,x;x_3)\Phi(x_1) \Phi(x_2) \Phi(x_3)
\ee
and hence the tree-level three-point function is given by
\be
\langle \O(x_1)\O(x_2)\O(x_3)\rangle=-6\lambda_3\int d^{d}xdz \sqrt{g}K_\Delta(z,x;x_1)  K_\Delta(z,x;x_2)  K_\Delta(z,x;x_3)\ .
\ee
For the sake of completeness let us note that \cite{Freedman:1998tz}
\begin{align}
\int d^{d}xdz \sqrt{g} K_{\Delta_1}(z,x;x_1)  K_{\Delta_2}(z,x;x_2)  K_{\Delta_3}(z,x;x_3)=\frac{a_{ijk}}{|x_1-x_2|^{\Delta_{12}}|x_1-x_3|^{\Delta_{13}}|x_3-x_2|^{\Delta_{32}}}
\end{align}
with
\be
a_{ijk}=\frac{\Gamma\left(\frac{\Delta_{12}}{2}\right)\Gamma\left(\frac{\Delta_{32}}{2}\right)\Gamma\left(\frac{\Delta_{13}}{2}\right)\Gamma\left(\frac{\sum_i\Delta_{i}-d}{2}\right)}{2\pi^d \Gamma\left(\Delta_1-\frac{d}{2} \right)\Gamma\left(\Delta_2-\frac{d}{2} \right)\Gamma\left(\Delta_3-\frac{d}{2} \right)}
\ee
and $\Delta_{ij}=\Delta_i+\Delta_j-\Delta_k$.
\subsubsection*{Four-point function}
The four-point function receives contributions from both contact diagrams and exchanged diagrams. In the leading order the quartic on-shell action is given by
\be
S_{(4)}=\lambda_4\int d^{d}xdz \sqrt{g}\phi^4+\frac{9}{2}\lambda_3^2 \int d^{d}xdz \sqrt{g}\int d^{d}x' dz' \sqrt{g'} G_\Delta(z,x;z',x')\phi^2(z,x)\phi^2(z',x')\ .
\ee
So the full four-point function is given by
\begin{align}
\langle \O(x_1)\O(x_2)\O(x_3)\O(x_4)\rangle&=- (4!)\lambda_4 (\text{contact Witten diagram})\nonumber\\
&-(3!)^2\lambda_3^2 (\text{three exchanged Witten diagrams})\ .
\end{align}

\section{Properties of $D$-Functions}
The $D(\eta,\sigma)$-function in AdS$_{d+1}$ is defined as
\be\label{defineD}
D_{\Delta_1\Delta_2 \Delta_3 \Delta_4}(\eta,\sigma)=\int d^{d+1}x \sqrt{g} \prod_{i=1}^4 \tilde{K}_{\Delta_i}(z,x;x_i)
\ee
where boundary $x_i$-points are given by (\ref{points}):
\begin{align}\label{points2}
x_1=-x_2=\rr\ ,  \qquad x_4=-x_3=\mathbf{1}\ .
\end{align}
Note that $\tilde{K}$ is the reduced bulk to boundary propagator 
\be\label{reduced}
\tilde{K}_{\Delta}(x')\equiv \tilde{K}_{\Delta}(z,x;x')=\frac{z^\Delta}{(z^2+|x-x'|^2)^\Delta}\ .
\ee

\subsection{Some Useful Identities}\label{app:identities}
The following identities will be very useful for us.
\subsubsection*{First Identity}
 From  \cite{DHoker:1999kzh}, we write
\begin{align}\label{identity1}
g^{\mu\nu}\partial_\mu \tilde{K}_{\Delta_1}(z,x;x_1)\partial_\nu \tilde{K}_{\Delta_2}(z,x;x_2)=&\Delta_1\Delta_2 \left( \tilde{K}_{\Delta_1}(z,x;x_1) \tilde{K}_{\Delta_2}(z,x;x_2) \right.\nonumber\\
&\left.-2x_{12}^2\tilde{K}_{\Delta_1+1}(z,x;x_1) \tilde{K}_{\Delta_2+1}(z,x;x_2) \right)\ ,
\end{align}
where, derivatives are taken with respect to bulk coordinates. 

\subsubsection*{Second Identity}
From \cite{DHoker:1999kzh}, we can also write
\begin{align}
D_{\Delta+1\ \Delta\ \Delta\ \Delta+1}(\eta,\sigma)=D_{\Delta\ \Delta+1\ \Delta+1\ \Delta}(\eta,\sigma)\ , \\
D_{\Delta+1\ \Delta\ \Delta+1\ \Delta}(\eta,\sigma)=D_{\Delta\ \Delta+1\ \Delta\ \Delta+1}(\eta,\sigma)\ .
\end{align}

\subsubsection*{Third Identity}
Let us now write our $D$-functions in terms of $\mathcal{D}$-functions of \cite{Dolan:2000ut}\footnote{Our $\mathcal{D}$-functions are $\bar{D}$ functions of \cite{Dolan:2000ut}.}
\begin{align}
&\mathcal{D}_{\Delta\Delta+1\Delta+1\Delta}(u,v)=\frac{2\Gamma[\Delta]^2\Gamma[\Delta+1]^2}{\Gamma\(2\Delta+1-h\)}\frac{(16\r\br)^{\Delta+1}}{(1-\r)(1-\br)}D_{\Delta\Delta+1\Delta+1\Delta}(\eta,\sigma)\ ,\\
&\mathcal{D}_{\Delta+1\Delta\Delta+1\Delta}(u,v)=\frac{2\Gamma[\Delta]^2\Gamma[\Delta+1]^2}{\Gamma\(2\Delta+1-h\)}\frac{(16\r\br)^{\Delta+1}}{(1+\r)(1+\br)}D_{\Delta+1\Delta\Delta+1\Delta}(\eta,\sigma)\ ,\\
&\mathcal{D}_{\Delta\Delta\Delta\Delta}(u,v)=\frac{2\Gamma[\Delta]^4}{\Gamma\(2\Delta-h\)}(16\r\br)^{\Delta}D_{\Delta\Delta\Delta\Delta}(\eta,\sigma)\ ,
\end{align}
where,
\begin{align}
&u=\frac{(1+\r)^2(1+\br)^2}{16\r\br}=\frac{(1+\sigma )^2 (1+\eta  \sigma )^2}{16 \eta  \sigma ^2}\ ,\\
&v=\frac{(1-\r)^2(1-\br)^2}{16\r\br}=\frac{(1-\sigma )^2 (1-\eta  \sigma )^2}{16 \eta  \sigma ^2}\ .
\end{align}
From \cite{Dolan:2000ut}, we can relate
\be
\mathcal{D}_{\Delta\Delta+1\Delta+1\Delta}(u,v)=-\p_v \mathcal{D}_{\Delta\Delta\Delta\Delta}(u,v)\ , \quad \mathcal{D}_{\Delta+1\Delta\Delta+1\Delta}(u,v)=-\p_u \mathcal{D}_{\Delta\Delta\Delta\Delta}(u,v)\ .
\ee
Therefore, we can derive the following expression 
\begin{align}\label{eq:app123}
&(1-\r)^3(1-\br)^3D_{\Delta\Delta+1\Delta+1\Delta}(\eta,\sigma)+(1+\r)^3(1+\br)^3D_{\Delta+1\Delta\Delta+1\Delta}(\eta,\sigma)\nonumber\\
&=- \frac{\Gamma\(2\Delta+1-h\)}{2\Gamma[\Delta]^2\Gamma[\Delta+1]^2(16\eta)^{\Delta-1}}\(v^2 \p_v+u^2 \p_u \)\mathcal{D}_{\Delta\Delta\Delta\Delta}(u,v)\nonumber\\
&=- \frac{16\(2\Delta-h\)}{\Delta^2 \eta^{\Delta-1}}\(v^2 \p_v+u^2 \p_u \)\eta^\Delta D_{\Delta\Delta\Delta\Delta}(u,v)\nonumber\\
&=- \frac{16\(2\Delta-h\)}{\Delta^2 }\(\frac{f_1(\eta,\sigma)}{\eta^{\Delta-1}}\p_\eta\eta^{\Delta}+f_2(\eta,\sigma)\p_\sigma \)D_{\Delta\Delta\Delta\Delta}(\eta,\sigma)\ ,
\end{align}
where, $h=d/2$ and 
\begin{align}
&f_1(\eta,\sigma)=\frac{(\eta +1) \left(\sigma ^2 \left(\eta  \left(3 \eta  \sigma ^2+\eta +8\right)+1\right)+3\right)}{16 (\eta -1) \sigma ^2}\ ,\\
&f_2(\eta,\sigma)=\frac{(1-\sigma ) (\sigma +1) \left(\eta  (\eta +2) \sigma ^2+1\right) \left(\eta  \left(\eta  \sigma ^2+2\right)+1\right)}{16 (\eta -1)   \sigma  \left(\eta  \sigma ^2-1\right)}\ .
\end{align}

\subsection{Regge limit of the $D$-functions}
Following \cite{Kundu:2019zsl}, in the Regge limit we obtain
\be\label{D:regge}
D_{\Delta_1\Delta_2 \Delta_3 \Delta_4}(\eta,\sigma)=i \frac{\pi^d2^{1-\sum_i \Delta_i}\sigma}{\eta^{\frac{\Delta_1+\Delta_2-1}{2}}\prod_i \Gamma(\Delta_i) }f_{\Delta_1\Delta_2 \Delta_3 \Delta_4}\(-\frac{1}{2}\log\(\eta\) \)\
\ee
where, 
\begin{align}\label{define:f}
f_{\Delta_1\Delta_2 \Delta_3 \Delta_4}(s)=&\int_{-\infty}^\infty d\nu \Omega_{i\nu}(s)\Gamma\left(\frac{\Delta_3+\Delta_4-d/2+i\nu}{2} \right) \Gamma\left(\frac{\Delta_3+\Delta_4-d/2-i\nu}{2} \right)\nonumber\\
&\times\Gamma\left(\frac{\Delta_1+\Delta_2-d/2+i\nu}{2} \right)\Gamma\left(\frac{\Delta_1+\Delta_2-d/2-i\nu}{2} \right)\ .
\end{align}
Harmonic functions $\Omega_{i\nu}$ on $H_{d-1}$  are known in any dimension \cite{Costa:2017twz}
\begin{align}
\Omega_{E}\left(s\right)=&-\frac{E \sin(\pi E) \Gamma\left(\frac{d-2}{2}+E \right)\Gamma\left(\frac{d-2}{2}-E \right)}{2^{d-1}\pi^{\frac{d+1}{2}}\Gamma\left(\frac{d-1}{2} \right)}\nonumber\\
&\times {}_2F_1\left(\frac{d-2}{2}+E,\frac{d-2}{2}-E,\frac{d-1}{2}, \frac{1-\cosh(s)}{2} \right)\ . \label{eq:harmonic}
\end{align}

\subsection{$F$-function}\label{app:F}
The $F$-function is defined as
\be
F_{2\Delta+L}(\eta)=\frac{1}{\eta ^{\frac{L-1}{2}}}f_{\Delta+L\ \Delta\  \Delta+L\ \Delta}\(-\frac{1}{2}\log\(\eta\) \)\ .
\ee 
In the limit $\eta\rightarrow 0$, we obtain from (\ref{define:f}) (see appendix D of \cite{Kundu:2019zsl}):
\be
F_{2\Delta+L}(\eta\rightarrow 0)= 2 \pi^{1-\frac{d}{2}}\Gamma\(2\Delta+L-\frac{d}{2}\)\Gamma\(2\Delta+L-1\)\eta^\Delta \ln\(\frac{1}{\eta}\)
\ee
where $D=d+1$.

\subsection{Another Identity}
We can also derive an exact identity 
\begin{align}
\p_\eta f_{\Delta\Delta \Delta \Delta}\(-\frac{1}{2}\log\eta \)=\frac{1-\eta}{(4\Delta-d)\eta^{3/2}}f_{\Delta+1\Delta \Delta+1 \Delta}\(-\frac{1}{2}\log\eta \)\ .
\end{align}
This will be useful later. 

\section{Regge Contributions of Odd Couplings}\label{app:odd}
In this appendix, our goal is to establish (\ref{EFT:bound4}). To this end, we first prove (\ref{EFT:bound4}) for $k=3$. This will necessarily imply (\ref{EFT:bound4}) for all odd $k\ge 3$, as we explain at the end. 

Using the explicit form of the dilaton effective action (\ref{action}), we obtain the leading  on-shell Euclidean effective action for $k=3$:
\be\label{eq:onshell1}
S^{(k=3)}_{\rm on-shell}=-\frac{\mu_3}{2}\int d^{D}x\  \sqrt{g}  \phi^2 \Box^3 \phi^2 \ .
\ee
The above on-shell action can be rewritten at the leading order in perturbation theory by using the bulk-to-boundary propagator. We notice from \cite{Kundu:2019zsl} that all $D$-functions decay in the Regge limit $D_{\Delta_1\Delta_2 \Delta_3 \Delta_4}(\r,\br)\sim\frac{1}{\r}$. On the other hand, $x_{ij}^2$ factors can grow as $\sim \r$. Therefore, terms in (\ref{eq:onshell1}) that have at least two factors of $x_{ij}^2$ can grow in the Regge limit (\ref{regge}). This greatly simplifies the analysis since we only care about the growing part of the Regge correlator. In particular, the on-shell four-point interaction (\ref{eq:onshell1}) can be approximated as
\begin{align}\label{eq:onshell3}
S^{(k=3)}_{\rm on-shell}&\propto-\mu_3 \int_{\Phi^4}\int_{AdS}\left(- x_{12}^4 x_{34}^2 \tilde{K}_{\Delta+2}(z,x;x_1)\tilde{K}_{\Delta+2}(z,x;x_2)\tilde{K}_{\Delta+1}(z,x;x_3)\tilde{K}_{\Delta+1}(z,x;x_3)\ \right.\nonumber\\
&+\frac{2-d+4\Delta-2d\Delta+4\Delta^2}{2(\Delta+1)^2} x_{12}^2 x_{34}^2 \tilde{K}_{\Delta+1}(z,x;x_1)\tilde{K}_{\Delta+1}(z,x;x_2)\tilde{K}_{\Delta+1}(z,x;x_3)\tilde{K}_{\Delta+1}(z,x;x_3) \nonumber\\ 
&\left.+ \frac{2\Delta-d}{2\Delta}x_{12}^4 \tilde{K}_{\Delta+2}(z,x;x_1)\tilde{K}_{\Delta+2}(z,x;x_2)\tilde{K}_{\Delta}(z,x;x_3)\tilde{K}_{\Delta}(z,x;x_3)\)+\cdots ,
\end{align}
where dots represent terms that do not contribute to the Regge growth. Note that we are not keeping track of the overall (positive) numerical factor, since our conclusion will not depend on it. It is now a straightforward exercise to compute the  Regge contribution of the $k=3$ term: 
\begin{align}
G(\eta,\sigma)\sim & \mu_3\(\frac{2-d+4\Delta-2d\Delta+4\Delta^2}{2(\Delta+1)^2} \r^2 D_{\Delta+1\ \Delta+1\ \Delta+1\ \Delta+1}(\eta,\sigma)+\frac{2\Delta-d}{2\Delta}\r^2 D_{\Delta+2\ \Delta\ \Delta+2\ \Delta}(\eta,\sigma) \right.\nonumber\\
&-\frac{1}{4}(1-\r)^3(1-\br)^3 D_{\Delta+2\ \Delta+1\ \Delta+1\ \Delta+2}(\eta,\sigma)-\frac{1}{4}(1+\r)^3(1+\br)^3 D_{\Delta+2\ \Delta+1\ \Delta+2\ \Delta+1}(\eta,\sigma)\nonumber\\
&\left.-\frac{1}{4}(1-\r)^3(1-\br)^3 D_{\Delta+1\ \Delta+2\ \Delta+2\ \Delta+1}(\eta,\sigma)-\frac{1}{4}(1+\r)^3(1+\br)^3 D_{\Delta+1\ \Delta+2\ \Delta+1\ \Delta+2}(\eta,\sigma)\right)\nonumber\\
&+\O\(\sigma^0\)
\end{align}
In the above expression, we have also exploited the fact that all $D_{\Delta_1\Delta_2\Delta_3\Delta_4}(\r,\br)$ functions with fixed $\Delta_1+\Delta_2=\Delta_3+\Delta_4$ have the same leading Regge behavior \cite{Kundu:2019zsl}. Moreover, in the Regge limit, one can also relate \cite{Kundu:2019zsl}
\be
D_{\Delta+1\ \Delta+1\ \Delta+1\ \Delta+1}(\eta,\sigma)=\frac{(\Delta+1)^2}{\Delta^2}D_{\Delta+2\ \Delta\ \Delta+2\ \Delta}(\eta,\sigma)+\O\(\sigma^2\)
\ee
We can now use various identities discussed in appendix \ref{app:identities} to obtain 
\begin{align}
G(\eta,\sigma)\sim i \frac{\mu_3}{ \eta^{\Delta+\frac{1}{2}}\sigma}\(\frac{-12 (d-1) \Delta -3 d+24 \Delta ^2+4}{2}f_{\Delta+1\ \Delta+1\ \Delta+1\ \Delta+1}\(-\frac{1}{2}\log\eta \)\right.\nonumber\\
\left.-\frac{3 (\eta +1)}{ \sqrt{\eta }}f_{\Delta+2\ \Delta+1\ \Delta+2\ \Delta+1}\(-\frac{1}{2}\log\eta \)\)+\O\(\sigma^0\)\ ,
\end{align}
where, $f$-functions are given by (\ref{define:f}). One now can check that the quantity inside the parentheses, for $m^2\ge 0$, changes sign as we increase $\eta$. For example,  for $\eta\rightarrow 0$ it is negative. Whereas, for $\eta\rightarrow 1$ it becomes positive for $m^2\ge 0$. Hence, if $\mu_2=0$, then the condition (\ref{bound1}) necessarily requires 
\be
\mu_3=0\ , \qquad m^2\ge 0\ .
\ee
Furthermore, the condition (\ref{bound2}) now also requires that 
\be
\mu_k=0
\ee
for all $k\ge4$. 

Interestingly, for negative $m^2$ there is always a range of $\Delta$ for which the quantity inside the parentheses does not change sign. In such a case, we can only derive a sign constraint on $\mu_3$. The condition (\ref{bound2}) now rules out all even $\mu_k$ with $k\ge 4$, however, odd $\mu_k$ with $k\ge 5$ are not ruled out. It is possible that CFT conditions (\ref{bound1}) and (\ref{bound2}) for $L>2$ might rule out such a scenario. Nonetheless, we will restrict to $m^2\ge 0$ to avoid this possible loophole.  


\section{$N$-Coefficients}\label{App:N}
\subsection{$N_D(\Delta)$ }
$N_D(\Delta)$ is a numerical coefficient that appears in the bound of $\phi^2\Box^2 \phi^2$ interaction in the presence of gravity. First, let us note that the Harmonic function (\ref{eq:harmonic}) function has the following behavior in the limit $\eta\rightarrow 1$:
\be
\tilde{\Omega}_{i \nu}=\Omega_{i \nu}\(-\frac{1}{2}\log\(\eta\) \)_{\eta\rightarrow 1}=\frac{2^{1-d} \pi ^{-\frac{d}{2}-\frac{1}{2}} \nu  \sinh (\pi  \nu ) \Gamma \left(\frac{d}{2}-i \nu -1\right) \Gamma \left(\frac{d}{2}+i \nu -1\right)}{\Gamma \left(\frac{d-1}{2}\right)}\ .
\ee
 The $N_D(\Delta)$ coefficient is now given by the ratio:
\be
N_D(\Delta)=\frac{\int_{-\infty}^\infty d\nu\frac{\Gamma\left(\frac{2\Delta+2-d/2+i\nu}{2} \right)^2 \Gamma\left(\frac{2\Delta+2-d/2-i\nu}{2} \right)^2}{\nu^2+\(\frac{d}{2}\)^2} \tilde{\Omega}_{i\nu}}{\int_{-\infty}^\infty d\nu \Gamma\left(\frac{2\Delta+2-d/2+i\nu}{2} \right)^2 \Gamma\left(\frac{2\Delta+2-d/2-i\nu}{2} \right)^2 \tilde{\Omega}_{i\nu}}
\ee
where $D=d+1$. This factor can be easily computed in Mathematica. In particular, we find that for large $\Delta$ and $D>4$:
\be
N_D(\Delta\gg 1)\approx \frac{1}{(D-4)\Delta}\ .
\ee

\subsection{$\tilde{N}_D(\Delta)$ }
The $\tilde{N}_D(\Delta)$ coefficient is now given by the ratio:
\be
\tilde{N}_D(\Delta)=\frac{\Gamma\(2\Delta-\frac{D-9}{2}\)\Gamma\(2\Delta+3\)}{\Gamma\(2\Delta-\frac{D-5}{2}\)\Gamma\(2\Delta+1\)}\frac{\int_{-\infty}^\infty d\nu \Gamma\left(\frac{2\Delta+2-d/2+i\nu}{2} \right)^2 \Gamma\left(\frac{2\Delta+2-d/2-i\nu}{2} \right)^2 \tilde{\Omega}_{i\nu}}{\int_{-\infty}^\infty d\nu \Gamma\left(\frac{2\Delta+4-d/2+i\nu}{2} \right)^2 \Gamma\left(\frac{2\Delta+4-d/2-i\nu}{2} \right)^2 \tilde{\Omega}_{i\nu}}
\ee
where $D=d+1$. Note that $\tilde{N}_D(\Delta)>1$.

\end{appendix}


\end{spacing}

\bibliographystyle{utphys} 
\bibliography{RGBib}

\providecommand{\href}[2]{#2}\begingroup\raggedright\begin{thebibliography}{100}

\bibitem{Adams:2006sv}
A.~Adams, N.~Arkani-Hamed, S.~Dubovsky, A.~Nicolis, and R.~Rattazzi,
  ``{Causality, analyticity and an IR obstruction to UV completion},''
  \href{http://dx.doi.org/10.1088/1126-6708/2006/10/014}{{\em JHEP} {\bfseries
  0610} (2006) 014},
\href{http://arxiv.org/abs/hep-th/0602178}{{\ttfamily arXiv:hep-th/0602178
  [hep-th]}}.

\bibitem{Komargodski:2011vj}
Z.~Komargodski and A.~Schwimmer, ``{On Renormalization Group Flows in Four
  Dimensions},'' \href{http://dx.doi.org/10.1007/JHEP12(2011)099}{{\em JHEP}
  {\bfseries 1112} (2011) 099},
\href{http://arxiv.org/abs/1107.3987}{{\ttfamily arXiv:1107.3987 [hep-th]}}.

\bibitem{Vafa:2005ui}
C.~Vafa, ``{The String landscape and the swampland},''
  \href{http://arxiv.org/abs/hep-th/0509212}{{\ttfamily arXiv:hep-th/0509212}}.

\bibitem{Ooguri:2006in}
H.~Ooguri and C.~Vafa, ``{On the Geometry of the String Landscape and the
  Swampland},'' \href{http://dx.doi.org/10.1016/j.nuclphysb.2006.10.033}{{\em
  Nucl. Phys. B} {\bfseries 766} (2007) 21--33},
  \href{http://arxiv.org/abs/hep-th/0605264}{{\ttfamily arXiv:hep-th/0605264}}.

\bibitem{Brennan:2017rbf}
T.~D. Brennan, F.~Carta, and C.~Vafa, ``{The String Landscape, the Swampland,
  and the Missing Corner},'' \href{http://dx.doi.org/10.22323/1.305.0015}{{\em
  PoS} {\bfseries TASI2017} (2017) 015},
  \href{http://arxiv.org/abs/1711.00864}{{\ttfamily arXiv:1711.00864
  [hep-th]}}.

\bibitem{Palti:2019pca}
E.~Palti, ``{The Swampland: Introduction and Review},''
  \href{http://dx.doi.org/10.1002/prop.201900037}{{\em Fortsch. Phys.}
  {\bfseries 67} no.~6, (2019) 1900037},
  \href{http://arxiv.org/abs/1903.06239}{{\ttfamily arXiv:1903.06239
  [hep-th]}}.

\bibitem{vanBeest:2021lhn}
M.~van Beest, J.~Calder\'on-Infante, D.~Mirfendereski, and I.~Valenzuela,
  ``{Lectures on the Swampland Program in String Compactifications},''
  \href{http://arxiv.org/abs/2102.01111}{{\ttfamily arXiv:2102.01111
  [hep-th]}}.

\bibitem{ArkaniHamed:2006dz}
N.~Arkani-Hamed, L.~Motl, A.~Nicolis, and C.~Vafa, ``{The String landscape,
  black holes and gravity as the weakest force},''
  \href{http://dx.doi.org/10.1088/1126-6708/2007/06/060}{{\em JHEP} {\bfseries
  06} (2007) 060},
\href{http://arxiv.org/abs/hep-th/0601001}{{\ttfamily arXiv:hep-th/0601001
  [hep-th]}}.

\bibitem{deRham:2017avq}
C.~de~Rham, S.~Melville, A.~J. Tolley, and S.-Y. Zhou, ``{Positivity bounds for
  scalar field theories},''
  \href{http://dx.doi.org/10.1103/PhysRevD.96.081702}{{\em Phys. Rev. D}
  {\bfseries 96} no.~8, (2017) 081702},
  \href{http://arxiv.org/abs/1702.06134}{{\ttfamily arXiv:1702.06134
  [hep-th]}}.

\bibitem{deRham:2017zjm}
C.~de~Rham, S.~Melville, A.~J. Tolley, and S.-Y. Zhou, ``{UV complete me:
  Positivity Bounds for Particles with Spin},''
  \href{http://dx.doi.org/10.1007/JHEP03(2018)011}{{\em JHEP} {\bfseries 03}
  (2018) 011}, \href{http://arxiv.org/abs/1706.02712}{{\ttfamily
  arXiv:1706.02712 [hep-th]}}.

\bibitem{Chandrasekaran:2018qmx}
V.~Chandrasekaran, G.~N. Remmen, and A.~Shahbazi-Moghaddam, ``{Higher-Point
  Positivity},'' \href{http://dx.doi.org/10.1007/JHEP11(2018)015}{{\em JHEP}
  {\bfseries 11} (2018) 015}, \href{http://arxiv.org/abs/1804.03153}{{\ttfamily
  arXiv:1804.03153 [hep-th]}}.

\bibitem{Zhang:2018shp}
C.~Zhang and S.-Y. Zhou, ``{Positivity bounds on vector boson scattering at the
  LHC},'' \href{http://dx.doi.org/10.1103/PhysRevD.100.095003}{{\em Phys. Rev.
  D} {\bfseries 100} no.~9, (2019) 095003},
  \href{http://arxiv.org/abs/1808.00010}{{\ttfamily arXiv:1808.00010
  [hep-ph]}}.

\bibitem{Bi:2019phv}
Q.~Bi, C.~Zhang, and S.-Y. Zhou, ``{Positivity constraints on aQGC: carving out
  the physical parameter space},''
  \href{http://dx.doi.org/10.1007/JHEP06(2019)137}{{\em JHEP} {\bfseries 06}
  (2019) 137}, \href{http://arxiv.org/abs/1902.08977}{{\ttfamily
  arXiv:1902.08977 [hep-ph]}}.

\bibitem{Remmen:2019cyz}
G.~N. Remmen and N.~L. Rodd, ``{Consistency of the Standard Model Effective
  Field Theory},'' \href{http://dx.doi.org/10.1007/JHEP12(2019)032}{{\em JHEP}
  {\bfseries 12} (2019) 032}, \href{http://arxiv.org/abs/1908.09845}{{\ttfamily
  arXiv:1908.09845 [hep-ph]}}.

\bibitem{Remmen:2020vts}
G.~N. Remmen and N.~L. Rodd, ``{Flavor Constraints from Unitarity and
  Analyticity},'' \href{http://dx.doi.org/10.1103/PhysRevLett.125.081601}{{\em
  Phys. Rev. Lett.} {\bfseries 125} no.~8, (2020) 081601},
  \href{http://arxiv.org/abs/2004.02885}{{\ttfamily arXiv:2004.02885
  [hep-ph]}}.

\bibitem{Zhang:2020jyn}
C.~Zhang and S.-Y. Zhou, ``{Convex Geometry Perspective on the (Standard Model)
  Effective Field Theory Space},''
  \href{http://dx.doi.org/10.1103/PhysRevLett.125.201601}{{\em Phys. Rev.
  Lett.} {\bfseries 125} no.~20, (2020) 201601},
  \href{http://arxiv.org/abs/2005.03047}{{\ttfamily arXiv:2005.03047
  [hep-ph]}}.

\bibitem{Yamashita:2020gtt}
K.~Yamashita, C.~Zhang, and S.-Y. Zhou, ``{Elastic positivity vs extremal
  positivity bounds in SMEFT: a case study in transversal electroweak
  gauge-boson scatterings},''
  \href{http://dx.doi.org/10.1007/JHEP01(2021)095}{{\em JHEP} {\bfseries 01}
  (2021) 095}, \href{http://arxiv.org/abs/2009.04490}{{\ttfamily
  arXiv:2009.04490 [hep-ph]}}.

\bibitem{Fuks:2020ujk}
B.~Fuks, Y.~Liu, C.~Zhang, and S.-Y. Zhou, ``{Positivity in electron-positron
  scattering: testing the axiomatic quantum field theory principles and probing
  the existence of UV states},''
  \href{http://dx.doi.org/10.1088/1674-1137/abcd8c}{{\em Chin. Phys. C}
  {\bfseries 45} no.~2, (2021) 023108},
  \href{http://arxiv.org/abs/2009.02212}{{\ttfamily arXiv:2009.02212
  [hep-ph]}}.

\bibitem{Remmen:2020uze}
G.~N. Remmen and N.~L. Rodd, ``{Signs, Spin, SMEFT: Positivity at Dimension
  Six},'' \href{http://arxiv.org/abs/2010.04723}{{\ttfamily arXiv:2010.04723
  [hep-ph]}}.

\bibitem{Caron-Huot:2020cmc}
S.~Caron-Huot and V.~Van~Duong, ``{Extremal Effective Field Theories},''
  \href{http://arxiv.org/abs/2011.02957}{{\ttfamily arXiv:2011.02957
  [hep-th]}}.

\bibitem{Bellazzini:2020cot}
B.~Bellazzini, J.~Elias~Mir\'o, R.~Rattazzi, M.~Riembau, and F.~Riva,
  ``{Positive Moments for Scattering Amplitudes},''
  \href{http://arxiv.org/abs/2011.00037}{{\ttfamily arXiv:2011.00037
  [hep-th]}}.

\bibitem{Tolley:2020gtv}
A.~J. Tolley, Z.-Y. Wang, and S.-Y. Zhou, ``{New positivity bounds from full
  crossing symmetry},'' \href{http://arxiv.org/abs/2011.02400}{{\ttfamily
  arXiv:2011.02400 [hep-th]}}.

\bibitem{Gu:2020ldn}
J.~Gu, L.-T. Wang, and C.~Zhang, ``{An unambiguous test of positivity at lepton
  colliders},'' \href{http://arxiv.org/abs/2011.03055}{{\ttfamily
  arXiv:2011.03055 [hep-ph]}}.

\bibitem{Arkani-Hamed:2020blm}
N.~Arkani-Hamed, T.-C. Huang, and Y.-T. Huang, ``{The EFT-Hedron},''
  \href{http://arxiv.org/abs/2012.15849}{{\ttfamily arXiv:2012.15849
  [hep-th]}}.

\bibitem{Li:2021cjv}
X.~Li, C.~Yang, H.~Xu, C.~Zhang, and S.-Y. Zhou, ``{Positivity in Multi-Field
  EFTs},'' \href{http://arxiv.org/abs/2101.01191}{{\ttfamily arXiv:2101.01191
  [hep-ph]}}.

\bibitem{Paulos:2016but}
M.~F. Paulos, J.~Penedones, J.~Toledo, B.~C. van Rees, and P.~Vieira, ``{The
  S-matrix bootstrap II: two dimensional amplitudes},''
  \href{http://dx.doi.org/10.1007/JHEP11(2017)143}{{\em JHEP} {\bfseries 11}
  (2017) 143}, \href{http://arxiv.org/abs/1607.06110}{{\ttfamily
  arXiv:1607.06110 [hep-th]}}.

\bibitem{Caron-Huot:2016icg}
S.~Caron-Huot, Z.~Komargodski, A.~Sever, and A.~Zhiboedov, ``{Strings from
  Massive Higher Spins: The Asymptotic Uniqueness of the Veneziano
  Amplitude},'' \href{http://dx.doi.org/10.1007/JHEP10(2017)026}{{\em JHEP}
  {\bfseries 10} (2017) 026}, \href{http://arxiv.org/abs/1607.04253}{{\ttfamily
  arXiv:1607.04253 [hep-th]}}.

\bibitem{Caron-Huot:2016owq}
S.~Caron-Huot, L.~J. Dixon, A.~McLeod, and M.~von Hippel, ``{Bootstrapping a
  Five-Loop Amplitude Using Steinmann Relations},''
  \href{http://dx.doi.org/10.1103/PhysRevLett.117.241601}{{\em Phys. Rev.
  Lett.} {\bfseries 117} no.~24, (2016) 241601},
  \href{http://arxiv.org/abs/1609.00669}{{\ttfamily arXiv:1609.00669
  [hep-th]}}.

\bibitem{Paulos:2017fhb}
M.~F. Paulos, J.~Penedones, J.~Toledo, B.~C. van Rees, and P.~Vieira, ``{The
  S-matrix bootstrap. Part III: higher dimensional amplitudes},''
  \href{http://dx.doi.org/10.1007/JHEP12(2019)040}{{\em JHEP} {\bfseries 12}
  (2019) 040}, \href{http://arxiv.org/abs/1708.06765}{{\ttfamily
  arXiv:1708.06765 [hep-th]}}.

\bibitem{Cordova:2018uop}
L.~C\'ordova and P.~Vieira, ``{Adding flavour to the S-matrix bootstrap},''
  \href{http://dx.doi.org/10.1007/JHEP12(2018)063}{{\em JHEP} {\bfseries 12}
  (2018) 063}, \href{http://arxiv.org/abs/1805.11143}{{\ttfamily
  arXiv:1805.11143 [hep-th]}}.

\bibitem{Guerrieri:2018uew}
A.~L. Guerrieri, J.~Penedones, and P.~Vieira, ``{Bootstrapping QCD Using Pion
  Scattering Amplitudes},''
  \href{http://dx.doi.org/10.1103/PhysRevLett.122.241604}{{\em Phys. Rev.
  Lett.} {\bfseries 122} no.~24, (2019) 241604},
  \href{http://arxiv.org/abs/1810.12849}{{\ttfamily arXiv:1810.12849
  [hep-th]}}.

\bibitem{Homrich:2019cbt}
A.~Homrich, J.~Penedones, J.~Toledo, B.~C. van Rees, and P.~Vieira, ``{The
  S-matrix Bootstrap IV: Multiple Amplitudes},''
  \href{http://dx.doi.org/10.1007/JHEP11(2019)076}{{\em JHEP} {\bfseries 11}
  (2019) 076}, \href{http://arxiv.org/abs/1905.06905}{{\ttfamily
  arXiv:1905.06905 [hep-th]}}.

\bibitem{EliasMiro:2019kyf}
J.~Elias~Mir\'o, A.~L. Guerrieri, A.~Hebbar, J.~Penedones, and P.~Vieira,
  ``{Flux Tube S-matrix Bootstrap},''
  \href{http://dx.doi.org/10.1103/PhysRevLett.123.221602}{{\em Phys. Rev.
  Lett.} {\bfseries 123} no.~22, (2019) 221602},
  \href{http://arxiv.org/abs/1906.08098}{{\ttfamily arXiv:1906.08098
  [hep-th]}}.

\bibitem{Karateev:2019ymz}
D.~Karateev, S.~Kuhn, and J.~a. Penedones, ``{Bootstrapping Massive Quantum
  Field Theories},'' \href{http://dx.doi.org/10.1007/JHEP07(2020)035}{{\em
  JHEP} {\bfseries 07} (2020) 035},
  \href{http://arxiv.org/abs/1912.08940}{{\ttfamily arXiv:1912.08940
  [hep-th]}}.

\bibitem{Correia:2020xtr}
M.~Correia, A.~Sever, and A.~Zhiboedov, ``{An Analytical Toolkit for the
  S-matrix Bootstrap},'' \href{http://arxiv.org/abs/2006.08221}{{\ttfamily
  arXiv:2006.08221 [hep-th]}}.

\bibitem{Bose:2020shm}
A.~Bose, P.~Haldar, A.~Sinha, P.~Sinha, and S.~S. Tiwari, ``{Relative entropy
  in scattering and the S-matrix bootstrap},''
  \href{http://dx.doi.org/10.21468/SciPostPhys.9.5.081}{{\em SciPost Phys.}
  {\bfseries 9} (2020) 081}, \href{http://arxiv.org/abs/2006.12213}{{\ttfamily
  arXiv:2006.12213 [hep-th]}}.

\bibitem{Guerrieri:2020bto}
A.~Guerrieri, J.~Penedones, and P.~Vieira, ``{S-matrix Bootstrap for Effective
  Field Theories: Massless Pions},''
  \href{http://arxiv.org/abs/2011.02802}{{\ttfamily arXiv:2011.02802
  [hep-th]}}.

\bibitem{Guerrieri:2020kcs}
A.~L. Guerrieri, A.~Homrich, and P.~Vieira, ``{Dual S-matrix bootstrap. Part I.
  2D theory},'' \href{http://dx.doi.org/10.1007/JHEP11(2020)084}{{\em JHEP}
  {\bfseries 11} (2020) 084}, \href{http://arxiv.org/abs/2008.02770}{{\ttfamily
  arXiv:2008.02770 [hep-th]}}.

\bibitem{Kaplan:2020ldi}
J.~Kaplan and S.~Kundu, ``{Closed Strings and Weak Gravity from Higher-Spin
  Causality},'' \href{http://dx.doi.org/10.1007/JHEP02(2021)145}{{\em JHEP}
  {\bfseries 02} (2021) 145}, \href{http://arxiv.org/abs/2008.05477}{{\ttfamily
  arXiv:2008.05477 [hep-th]}}.

\bibitem{Bose:2020cod}
A.~Bose, A.~Sinha, and S.~S. Tiwari, ``{Selection rules for the S-Matrix
  bootstrap},'' \href{http://arxiv.org/abs/2011.07944}{{\ttfamily
  arXiv:2011.07944 [hep-th]}}.

\bibitem{Huang:2020nqy}
Y.-t. Huang, J.-Y. Liu, L.~Rodina, and Y.~Wang, ``{Carving out the Space of
  Open-String S-matrix},'' \href{http://arxiv.org/abs/2008.02293}{{\ttfamily
  arXiv:2008.02293 [hep-th]}}.

\bibitem{Hebbar:2020ukp}
A.~Hebbar, D.~Karateev, and J.~Penedones, ``{Spinning S-matrix Bootstrap in
  4d},'' \href{http://arxiv.org/abs/2011.11708}{{\ttfamily arXiv:2011.11708
  [hep-th]}}.

\bibitem{Sinha:2020win}
A.~Sinha and A.~Zahed, ``{Crossing Symmetric Dispersion Relations in QFTs},''
  \href{http://arxiv.org/abs/2012.04877}{{\ttfamily arXiv:2012.04877
  [hep-th]}}.

\bibitem{Tourkine:2021fqh}
P.~Tourkine and A.~Zhiboedov, ``{Scattering from production in 2d},''
  \href{http://arxiv.org/abs/2101.05211}{{\ttfamily arXiv:2101.05211
  [hep-th]}}.

\bibitem{Haldar:2021rri}
P.~Haldar, A.~Sinha, and A.~Zahed, ``{Quantum field theory and the Bieberbach
  conjecture},'' \href{http://arxiv.org/abs/2103.12108}{{\ttfamily
  arXiv:2103.12108 [hep-th]}}.

\bibitem{He:2021eqn}
Y.~He and M.~Kruczenski, ``{S-matrix bootstrap in 3+1 dimensions:
  regularization and dual convex problem},''
  \href{http://arxiv.org/abs/2103.11484}{{\ttfamily arXiv:2103.11484
  [hep-th]}}.

\bibitem{Caron-Huot:2021rmr}
S.~Caron-Huot, D.~Mazac, L.~Rastelli, and D.~Simmons-Duffin, ``{Sharp
  Boundaries for the Swampland},''
  \href{http://arxiv.org/abs/2102.08951}{{\ttfamily arXiv:2102.08951
  [hep-th]}}.

\bibitem{Noumi:2021uuv}
T.~Noumi and J.~Tokuda, ``{Gravitational Positivity Bounds on Scalar
  Potentials},'' \href{http://arxiv.org/abs/2105.01436}{{\ttfamily
  arXiv:2105.01436 [hep-th]}}.

\bibitem{Froissart:1961ux}
M.~Froissart, ``{Asymptotic behavior and subtractions in the Mandelstam
  representation},'' \href{http://dx.doi.org/10.1103/PhysRev.123.1053}{{\em
  Phys. Rev.} {\bfseries 123} (1961) 1053--1057}.

\bibitem{Martin:1962rt}
A.~Martin, ``{Unitarity and high-energy behavior of scattering amplitudes},''
  \href{http://dx.doi.org/10.1103/PhysRev.129.1432}{{\em Phys. Rev.} {\bfseries
  129} (1963) 1432--1436}.

\bibitem{CHAICHIAN1992151}
M.~Chaichian, J.~Fischer, and Y.~Vernov, ``{Generalization of the
  Froissart-Martin bounds to scattering in a space-time of general
  dimension},''
  \href{http://dx.doi.org/https://doi.org/10.1016/0550-3213(92)90674-Z}{{\em
  Nuclear Physics B} {\bfseries 383} no.~1, (1992) 151 -- 172}.

\bibitem{Komargodski:2011xv}
Z.~Komargodski, ``{The Constraints of Conformal Symmetry on RG Flows},''
  \href{http://dx.doi.org/10.1007/JHEP07(2012)069}{{\em JHEP} {\bfseries 07}
  (2012) 069}, \href{http://arxiv.org/abs/1112.4538}{{\ttfamily arXiv:1112.4538
  [hep-th]}}.

\bibitem{Luty:2012ww}
M.~A. Luty, J.~Polchinski, and R.~Rattazzi, ``{The $a$-theorem and the
  Asymptotics of 4D Quantum Field Theory},''
\href{http://arxiv.org/abs/1204.5221}{{\ttfamily arXiv:1204.5221 [hep-th]}}.

\bibitem{Elvang:2012st}
H.~Elvang, D.~Z. Freedman, L.-Y. Hung, M.~Kiermaier, R.~C. Myers, and
  S.~Theisen, ``{On renormalization group flows and the a-theorem in 6d},''
  \href{http://dx.doi.org/10.1007/JHEP10(2012)011}{{\em JHEP} {\bfseries 10}
  (2012) 011}, \href{http://arxiv.org/abs/1205.3994}{{\ttfamily arXiv:1205.3994
  [hep-th]}}.

\bibitem{Heckman:2021nwg}
J.~J. Heckman, S.~Kundu, and H.~Y. Zhang, ``{EFT of 6D SUSY RG Flows},''
  \href{http://arxiv.org/abs/2103.13395}{{\ttfamily arXiv:2103.13395
  [hep-th]}}.

\bibitem{Poland:2018epd}
D.~Poland, S.~Rychkov, and A.~Vichi, ``{The Conformal Bootstrap: Theory,
  Numerical Techniques, and Applications},''
  \href{http://dx.doi.org/10.1103/RevModPhys.91.015002}{{\em Rev. Mod. Phys.}
  {\bfseries 91} (2019) 015002},
  \href{http://arxiv.org/abs/1805.04405}{{\ttfamily arXiv:1805.04405
  [hep-th]}}.

\bibitem{Hofman:2008ar}
D.~M. Hofman and J.~Maldacena, ``{Conformal collider physics: Energy and charge
  correlations},'' \href{http://dx.doi.org/10.1088/1126-6708/2008/05/012}{{\em
  JHEP} {\bfseries 0805} (2008) 012},
\href{http://arxiv.org/abs/0803.1467}{{\ttfamily arXiv:0803.1467 [hep-th]}}.

\bibitem{Hartman:2015lfa}
T.~Hartman, S.~Jain, and S.~Kundu, ``{Causality Constraints in Conformal Field
  Theory},'' \href{http://dx.doi.org/10.1007/JHEP05(2016)099}{{\em JHEP}
  {\bfseries 05} (2016) 099},
\href{http://arxiv.org/abs/1509.00014}{{\ttfamily arXiv:1509.00014 [hep-th]}}.

\bibitem{Afkhami-Jeddi:2016ntf}
N.~Afkhami-Jeddi, T.~Hartman, S.~Kundu, and A.~Tajdini, ``{Einstein gravity
  3-point functions from conformal field theory},''
  \href{http://dx.doi.org/10.1007/JHEP12(2017)049}{{\em JHEP} {\bfseries 12}
  (2017) 049},
\href{http://arxiv.org/abs/1610.09378}{{\ttfamily arXiv:1610.09378 [hep-th]}}.

\bibitem{Paulos:2016fap}
M.~F. Paulos, J.~Penedones, J.~Toledo, B.~C. van Rees, and P.~Vieira, ``{The
  S-matrix Bootstrap I: QFT in AdS},''
\href{http://arxiv.org/abs/1607.06109}{{\ttfamily arXiv:1607.06109 [hep-th]}}.

\bibitem{Kulaxizi:2017ixa}
M.~Kulaxizi, A.~Parnachev, and A.~Zhiboedov, ``{Bulk Phase Shift, CFT Regge
  Limit and Einstein Gravity},''
  \href{http://dx.doi.org/10.1007/JHEP06(2018)121}{{\em JHEP} {\bfseries 06}
  (2018) 121}, \href{http://arxiv.org/abs/1705.02934}{{\ttfamily
  arXiv:1705.02934 [hep-th]}}.

\bibitem{Costa:2017twz}
M.~S. Costa, T.~Hansen, and J.~Penedones, ``{Bounds for OPE coefficients on the
  Regge trajectory},'' \href{http://dx.doi.org/10.1007/JHEP10(2017)197}{{\em
  JHEP} {\bfseries 10} (2017) 197},
  \href{http://arxiv.org/abs/1707.07689}{{\ttfamily arXiv:1707.07689
  [hep-th]}}.

\bibitem{Afkhami-Jeddi:2017rmx}
N.~Afkhami-Jeddi, T.~Hartman, S.~Kundu, and A.~Tajdini, ``{Shockwaves from the
  Operator Product Expansion},''
  \href{http://dx.doi.org/10.1007/JHEP03(2019)201}{{\em JHEP} {\bfseries 03}
  (2019) 201},
\href{http://arxiv.org/abs/1709.03597}{{\ttfamily arXiv:1709.03597 [hep-th]}}.

\bibitem{Cordova:2017zej}
C.~Cordova, J.~Maldacena, and G.~J. Turiaci, ``{Bounds on OPE Coefficients from
  Interference Effects in the Conformal Collider},''
  \href{http://dx.doi.org/10.1007/JHEP11(2017)032}{{\em JHEP} {\bfseries 11}
  (2017) 032}, \href{http://arxiv.org/abs/1710.03199}{{\ttfamily
  arXiv:1710.03199 [hep-th]}}.

\bibitem{Meltzer:2017rtf}
D.~Meltzer and E.~Perlmutter, ``{Beyond $a = c$: gravitational couplings to
  matter and the stress tensor OPE},''
  \href{http://dx.doi.org/10.1007/JHEP07(2018)157}{{\em JHEP} {\bfseries 07}
  (2018) 157}, \href{http://arxiv.org/abs/1712.04861}{{\ttfamily
  arXiv:1712.04861 [hep-th]}}.

\bibitem{Afkhami-Jeddi:2018own}
N.~Afkhami-Jeddi, S.~Kundu, and A.~Tajdini, ``{A Conformal Collider for
  Holographic CFTs},'' \href{http://dx.doi.org/10.1007/JHEP10(2018)156}{{\em
  JHEP} {\bfseries 10} (2018) 156},
\href{http://arxiv.org/abs/1805.07393}{{\ttfamily arXiv:1805.07393 [hep-th]}}.

\bibitem{Afkhami-Jeddi:2018apj}
N.~Afkhami-Jeddi, S.~Kundu, and A.~Tajdini, ``{A Bound on Massive Higher Spin
  Particles},'' \href{http://dx.doi.org/10.1007/JHEP04(2019)056}{{\em JHEP}
  {\bfseries 04} (2019) 056},
\href{http://arxiv.org/abs/1811.01952}{{\ttfamily arXiv:1811.01952 [hep-th]}}.

\bibitem{Kaplan:2019soo}
J.~Kaplan and S.~Kundu, ``{A Species or Weak-Gravity Bound for Large $N$ Gauge
  Theories Coupled to Gravity},''
  \href{http://dx.doi.org/10.1007/JHEP11(2019)142}{{\em JHEP} {\bfseries 11}
  (2019) 142}, \href{http://arxiv.org/abs/1904.09294}{{\ttfamily
  arXiv:1904.09294 [hep-th]}}.

\bibitem{Haldar:2019prg}
P.~Haldar and A.~Sinha, ``{Froissart bound for/from CFT Mellin amplitudes},''
  \href{http://dx.doi.org/10.21468/SciPostPhys.8.6.095}{{\em SciPost Phys.}
  {\bfseries 8} (2020) 095}, \href{http://arxiv.org/abs/1911.05974}{{\ttfamily
  arXiv:1911.05974 [hep-th]}}.

\bibitem{Conlon:2020wmc}
J.~P. Conlon and F.~Revello, ``{Moduli Stabilisation and the Holographic
  Swampland},'' \href{http://dx.doi.org/10.31526/LHEP.2020.171}{{\em LHEP}
  {\bfseries 2020} (2020) 171},
  \href{http://arxiv.org/abs/2006.01021}{{\ttfamily arXiv:2006.01021
  [hep-th]}}.

\bibitem{Komatsu:2020sag}
S.~Komatsu, M.~F. Paulos, B.~C. Van~Rees, and X.~Zhao, ``{Landau diagrams in
  AdS and S-matrices from conformal correlators},''
  \href{http://dx.doi.org/10.1007/JHEP11(2020)046}{{\em JHEP} {\bfseries 11}
  (2020) 046}, \href{http://arxiv.org/abs/2007.13745}{{\ttfamily
  arXiv:2007.13745 [hep-th]}}.

\bibitem{Kundu:2020bdn}
S.~Kundu, ``{RG Flows with Global Symmetry Breaking and Bounds from Chaos},''
  \href{http://arxiv.org/abs/2012.10450}{{\ttfamily arXiv:2012.10450
  [hep-th]}}.

\bibitem{Hijano:2020szl}
E.~Hijano and D.~Neuenfeld, ``{Soft photon theorems from CFT Ward identites in
  the flat limit of AdS/CFT},''
  \href{http://dx.doi.org/10.1007/JHEP11(2020)009}{{\em JHEP} {\bfseries 11}
  (2020) 009}, \href{http://arxiv.org/abs/2005.03667}{{\ttfamily
  arXiv:2005.03667 [hep-th]}}.

\bibitem{Alday:2021odx}
L.~F. Alday, C.~Behan, P.~Ferrero, and X.~Zhou, ``{Gluon Scattering in AdS from
  CFT},'' \href{http://arxiv.org/abs/2103.15830}{{\ttfamily arXiv:2103.15830
  [hep-th]}}.

\bibitem{Heemskerk:2009pn}
I.~Heemskerk, J.~Penedones, J.~Polchinski, and J.~Sully, ``{Holography from
  Conformal Field Theory},''
  \href{http://dx.doi.org/10.1088/1126-6708/2009/10/079}{{\em JHEP} {\bfseries
  0910} (2009) 079},
\href{http://arxiv.org/abs/0907.0151}{{\ttfamily arXiv:0907.0151 [hep-th]}}.

\bibitem{Heemskerk:2010ty}
I.~Heemskerk and J.~Sully, ``{More Holography from Conformal Field Theory},''
  \href{http://dx.doi.org/10.1007/JHEP09(2010)099}{{\em JHEP} {\bfseries 1009}
  (2010) 099},
\href{http://arxiv.org/abs/1006.0976}{{\ttfamily arXiv:1006.0976 [hep-th]}}.

\bibitem{Fitzpatrick:2010zm}
A.~Fitzpatrick, E.~Katz, D.~Poland, and D.~Simmons-Duffin, ``{Effective
  Conformal Theory and the Flat-Space Limit of AdS},''
  \href{http://dx.doi.org/10.1007/JHEP07(2011)023}{{\em JHEP} {\bfseries 1107}
  (2011) 023},
\href{http://arxiv.org/abs/1007.2412}{{\ttfamily arXiv:1007.2412 [hep-th]}}.

\bibitem{Penedones:2010ue}
J.~Penedones, ``{Writing CFT correlation functions as AdS scattering
  amplitudes},'' \href{http://dx.doi.org/10.1007/JHEP03(2011)025}{{\em JHEP}
  {\bfseries 1103} (2011) 025},
\href{http://arxiv.org/abs/1011.1485}{{\ttfamily arXiv:1011.1485 [hep-th]}}.

\bibitem{ElShowk:2011ag}
S.~El-Showk and K.~Papadodimas, ``{Emergent Spacetime and Holographic CFTs},''
  \href{http://dx.doi.org/10.1007/JHEP10(2012)106}{{\em JHEP} {\bfseries 1210}
  (2012) 106},
\href{http://arxiv.org/abs/1101.4163}{{\ttfamily arXiv:1101.4163 [hep-th]}}.

\bibitem{Fitzpatrick:2011ia}
A.~L. Fitzpatrick, J.~Kaplan, J.~Penedones, S.~Raju, and B.~C. van Rees, ``{A
  Natural Language for AdS/CFT Correlators},''
  \href{http://dx.doi.org/10.1007/JHEP11(2011)095}{{\em JHEP} {\bfseries 1111}
  (2011) 095},
\href{http://arxiv.org/abs/1107.1499}{{\ttfamily arXiv:1107.1499 [hep-th]}}.

\bibitem{Fitzpatrick:2011hu}
A.~L. Fitzpatrick and J.~Kaplan, ``{Analyticity and the Holographic
  S-Matrix},'' \href{http://dx.doi.org/10.1007/JHEP10(2012)127}{{\em JHEP}
  {\bfseries 1210} (2012) 127},
\href{http://arxiv.org/abs/1111.6972}{{\ttfamily arXiv:1111.6972 [hep-th]}}.

\bibitem{Fitzpatrick:2011dm}
A.~L. Fitzpatrick and J.~Kaplan, ``{Unitarity and the Holographic S-Matrix},''
  \href{http://dx.doi.org/10.1007/JHEP10(2012)032}{{\em JHEP} {\bfseries 1210}
  (2012) 032},
\href{http://arxiv.org/abs/1112.4845}{{\ttfamily arXiv:1112.4845 [hep-th]}}.

\bibitem{Fitzpatrick:2012cg}
A.~Fitzpatrick and J.~Kaplan, ``{AdS Field Theory from Conformal Field
  Theory},'' \href{http://dx.doi.org/10.1007/JHEP02(2013)054}{{\em JHEP}
  {\bfseries 02} (2013) 054}, \href{http://arxiv.org/abs/1208.0337}{{\ttfamily
  arXiv:1208.0337 [hep-th]}}.

\bibitem{Goncalves:2014rfa}
V.~Goncalves, J.~Penedones, and E.~Trevisani, ``{Factorization of Mellin
  amplitudes},'' \href{http://dx.doi.org/10.1007/JHEP10(2015)040}{{\em JHEP}
  {\bfseries 10} (2015) 040}, \href{http://arxiv.org/abs/1410.4185}{{\ttfamily
  arXiv:1410.4185 [hep-th]}}.

\bibitem{Alday:2014tsa}
L.~F. Alday, A.~Bissi, and T.~Lukowski, ``{Lessons from crossing symmetry at
  large N},''
\href{http://arxiv.org/abs/1410.4717}{{\ttfamily arXiv:1410.4717 [hep-th]}}.

\bibitem{Hijano:2015zsa}
E.~Hijano, P.~Kraus, E.~Perlmutter, and R.~Snively, ``{Witten Diagrams
  Revisited: The AdS Geometry of Conformal Blocks},''
\href{http://arxiv.org/abs/1508.00501}{{\ttfamily arXiv:1508.00501 [hep-th]}}.

\bibitem{Aharony:2016dwx}
O.~Aharony, L.~F. Alday, A.~Bissi, and E.~Perlmutter, ``{Loops in AdS from
  Conformal Field Theory},''
  \href{http://dx.doi.org/10.1007/JHEP07(2017)036}{{\em JHEP} {\bfseries 07}
  (2017) 036}, \href{http://arxiv.org/abs/1612.03891}{{\ttfamily
  arXiv:1612.03891 [hep-th]}}.

\bibitem{Kundu:2020gkz}
S.~Kundu, ``{A Generalized Nachtmann Theorem in CFT},''
  \href{http://dx.doi.org/10.1007/JHEP11(2020)138}{{\em JHEP} {\bfseries 11}
  (2020) 138}, \href{http://arxiv.org/abs/2002.12390}{{\ttfamily
  arXiv:2002.12390 [hep-th]}}.

\bibitem{Chowdhury:2019kaq}
S.~D. Chowdhury, A.~Gadde, T.~Gopalka, I.~Halder, L.~Janagal, and S.~Minwalla,
  ``{Classifying and constraining local four photon and four graviton
  S-matrices},'' \href{http://dx.doi.org/10.1007/JHEP02(2020)114}{{\em JHEP}
  {\bfseries 02} (2020) 114}, \href{http://arxiv.org/abs/1910.14392}{{\ttfamily
  arXiv:1910.14392 [hep-th]}}.

\bibitem{Bern:2021ppb}
Z.~Bern, D.~Kosmopoulos, and A.~Zhiboedov, ``{Gravitational Effective Field
  Theory Islands, Low-Spin Dominance, and the Four-Graviton Amplitude},''
  \href{http://arxiv.org/abs/2103.12728}{{\ttfamily arXiv:2103.12728
  [hep-th]}}.

\bibitem{Guerrieri:2021ivu}
A.~Guerrieri, J.~Penedones, and P.~Vieira, ``{Where is String Theory?},''
  \href{http://arxiv.org/abs/2102.02847}{{\ttfamily arXiv:2102.02847
  [hep-th]}}.

\bibitem{Hartman:2016lgu}
T.~Hartman, S.~Kundu, and A.~Tajdini, ``{Averaged Null Energy Condition from
  Causality},'' \href{http://dx.doi.org/10.1007/JHEP07(2017)066}{{\em JHEP}
  {\bfseries 07} (2017) 066},
\href{http://arxiv.org/abs/1610.05308}{{\ttfamily arXiv:1610.05308 [hep-th]}}.

\bibitem{Caron-Huot:2017vep}
S.~Caron-Huot, ``{Analyticity in Spin in Conformal Theories},''
  \href{http://dx.doi.org/10.1007/JHEP09(2017)078}{{\em JHEP} {\bfseries 09}
  (2017) 078}, \href{http://arxiv.org/abs/1703.00278}{{\ttfamily
  arXiv:1703.00278 [hep-th]}}.

\bibitem{Maldacena:2015waa}
J.~Maldacena, S.~H. Shenker, and D.~Stanford, ``{A bound on chaos},''
\href{http://arxiv.org/abs/1503.01409}{{\ttfamily arXiv:1503.01409 [hep-th]}}.

\bibitem{Chandorkar:2021viw}
D.~Chandorkar, S.~D. Chowdhury, S.~Kundu, and S.~Minwalla, ``{Bounds on Regge
  growth of flat space scattering from bounds on chaos},''
  \href{http://arxiv.org/abs/2102.03122}{{\ttfamily arXiv:2102.03122
  [hep-th]}}.

\bibitem{Kundu:2019zsl}
S.~Kundu, ``{Renormalization Group Flows, the $a$-Theorem and Conformal
  Bootstrap},'' \href{http://dx.doi.org/10.1007/JHEP05(2020)014}{{\em JHEP}
  {\bfseries 05} (2020) 014}, \href{http://arxiv.org/abs/1912.09479}{{\ttfamily
  arXiv:1912.09479 [hep-th]}}.

\bibitem{JoaoRegge}
M.~S. Costa, V.~Goncalves, and J.~Penedones, ``{Conformal Regge theory},''
\href{http://arxiv.org/abs/1209.4355}{{\ttfamily arXiv:1209.4355 [hep-th]}}.

\bibitem{Hatefi:2021czu}
E.~Hatefi and P.~Sundell, ``{All-Order Quartic Couplings in Highly Symmetric
  D-brane-Anti-D-brane Systems},''
  \href{http://arxiv.org/abs/2103.06302}{{\ttfamily arXiv:2103.06302
  [hep-th]}}.

\bibitem{KomargodskiZhiboedov}
Z.~Komargodski and A.~Zhiboedov, ``{Convexity and Liberation at Large Spin},''
  \href{http://dx.doi.org/10.1007/JHEP11(2013)140}{{\em JHEP} {\bfseries 1311}
  (2013) 140},
\href{http://arxiv.org/abs/1212.4103}{{\ttfamily arXiv:1212.4103 [hep-th]}}.

\bibitem{Kraus:2020nga}
P.~Kraus, S.~Megas, and A.~Sivaramakrishnan, ``{Anomalous dimensions from
  thermal AdS partition functions},''
  \href{http://dx.doi.org/10.1007/JHEP10(2020)149}{{\em JHEP} {\bfseries 10}
  (2020) 149}, \href{http://arxiv.org/abs/2004.08635}{{\ttfamily
  arXiv:2004.08635 [hep-th]}}.

\bibitem{Camanho:2014apa}
X.~O. Camanho, J.~D. Edelstein, J.~Maldacena, and A.~Zhiboedov, ``{Causality
  Constraints on Corrections to the Graviton Three-Point Coupling},''
  \href{http://dx.doi.org/10.1007/JHEP02(2016)020}{{\em JHEP} {\bfseries 02}
  (2016) 020},
\href{http://arxiv.org/abs/1407.5597}{{\ttfamily arXiv:1407.5597 [hep-th]}}.

\bibitem{Hollowood:2015elj}
T.~J. Hollowood and G.~M. Shore, ``{Causality Violation, Gravitational
  Shockwaves and UV Completion},''
  \href{http://dx.doi.org/10.1007/JHEP03(2016)129}{{\em JHEP} {\bfseries 03}
  (2016) 129}, \href{http://arxiv.org/abs/1512.04952}{{\ttfamily
  arXiv:1512.04952 [hep-th]}}.

\bibitem{deRham:2020zyh}
C.~de~Rham and A.~J. Tolley, ``{Causality in curved spacetimes: The speed of
  light and gravity},''
  \href{http://dx.doi.org/10.1103/PhysRevD.102.084048}{{\em Phys. Rev. D}
  {\bfseries 102} no.~8, (2020) 084048},
  \href{http://arxiv.org/abs/2007.01847}{{\ttfamily arXiv:2007.01847
  [hep-th]}}.

\bibitem{Alberte:2020jsk}
L.~Alberte, C.~de~Rham, S.~Jaitly, and A.~J. Tolley, ``{Positivity Bounds and
  the Massless Spin-2 Pole},''
  \href{http://dx.doi.org/10.1103/PhysRevD.102.125023}{{\em Phys. Rev. D}
  {\bfseries 102} no.~12, (2020) 125023},
  \href{http://arxiv.org/abs/2007.12667}{{\ttfamily arXiv:2007.12667
  [hep-th]}}.

\bibitem{Edelstein:2016nml}
J.~D. Edelstein, G.~Giribet, C.~Gomez, E.~Kilicarslan, M.~Leoni, and B.~Tekin,
  ``{Causality in 3D Massive Gravity Theories},''
  \href{http://dx.doi.org/10.1103/PhysRevD.95.104016}{{\em Phys. Rev. D}
  {\bfseries 95} no.~10, (2017) 104016},
  \href{http://arxiv.org/abs/1602.03376}{{\ttfamily arXiv:1602.03376
  [hep-th]}}.

\bibitem{Camanho:2016opx}
X.~O. Camanho, G.~Lucena~G\'omez, and R.~Rahman, ``{Causality Constraints on
  Massive Gravity},'' \href{http://dx.doi.org/10.1103/PhysRevD.96.084007}{{\em
  Phys. Rev. D} {\bfseries 96} no.~8, (2017) 084007},
  \href{http://arxiv.org/abs/1610.02033}{{\ttfamily arXiv:1610.02033
  [hep-th]}}.

\bibitem{Hinterbichler:2017qcl}
K.~Hinterbichler, A.~Joyce, and R.~A. Rosen, ``{Eikonal scattering and
  asymptotic superluminality of massless higher spin fields},''
  \href{http://dx.doi.org/10.1103/PhysRevD.97.125019}{{\em Phys. Rev. D}
  {\bfseries 97} no.~12, (2018) 125019},
  \href{http://arxiv.org/abs/1712.10021}{{\ttfamily arXiv:1712.10021
  [hep-th]}}.

\bibitem{Bonifacio:2017nnt}
J.~Bonifacio, K.~Hinterbichler, A.~Joyce, and R.~A. Rosen, ``{Massive and
  Massless Spin-2 Scattering and Asymptotic Superluminality},''
  \href{http://dx.doi.org/10.1007/JHEP06(2018)075}{{\em JHEP} {\bfseries 06}
  (2018) 075}, \href{http://arxiv.org/abs/1712.10020}{{\ttfamily
  arXiv:1712.10020 [hep-th]}}.

\bibitem{Chowdhury:2018nfv}
T.~A. Chowdhury, R.~Rahman, and Z.~A. Sabuj, ``{Gravitational Properties of the
  Proca Field},'' \href{http://dx.doi.org/10.1016/j.nuclphysb.2018.09.009}{{\em
  Nucl. Phys. B} {\bfseries 936} (2018) 364--382},
  \href{http://arxiv.org/abs/1807.10284}{{\ttfamily arXiv:1807.10284
  [hep-th]}}.

\bibitem{Kaplan:2020tdz}
J.~Kaplan and S.~Kundu, ``{Causality Constraints in Large $N$ QCD Coupled to
  Gravity},'' \href{http://arxiv.org/abs/2009.08460}{{\ttfamily
  arXiv:2009.08460 [hep-th]}}.

\bibitem{Freedman:1998tz}
D.~Z. Freedman, S.~D. Mathur, A.~Matusis, and L.~Rastelli, ``{Correlation
  functions in the CFT($d$)/AdS($d+1$) correspondence},''
  \href{http://dx.doi.org/10.1016/S0550-3213(99)00053-X}{{\em Nucl. Phys.}
  {\bfseries B546} (1999) 96--118},
\href{http://arxiv.org/abs/hep-th/9804058}{{\ttfamily arXiv:hep-th/9804058}}.

\bibitem{DHoker:1999kzh}
E.~D'Hoker, D.~Z. Freedman, S.~D. Mathur, A.~Matusis, and L.~Rastelli,
  ``{Graviton exchange and complete four point functions in the AdS / CFT
  correspondence},''
  \href{http://dx.doi.org/10.1016/S0550-3213(99)00525-8}{{\em Nucl. Phys. B}
  {\bfseries 562} (1999) 353--394},
  \href{http://arxiv.org/abs/hep-th/9903196}{{\ttfamily arXiv:hep-th/9903196}}.

\bibitem{Dolan:2000ut}
F.~A. Dolan and H.~Osborn, ``{Conformal four point functions and the operator
  product expansion},''
  \href{http://dx.doi.org/10.1016/S0550-3213(01)00013-X}{{\em Nucl. Phys.}
  {\bfseries B599} (2001) 459--496},
\href{http://arxiv.org/abs/hep-th/0011040}{{\ttfamily arXiv:hep-th/0011040}}.

\end{thebibliography}\endgroup

\end{document}